\documentclass{article}
\usepackage[utf8]{inputenc}
\usepackage[a4paper, total={6in, 8in},margin=1in]{geometry}
\usepackage{amsmath}
\usepackage{amssymb}
\usepackage{amsfonts}
\usepackage{comment}
\usepackage{graphicx}
\graphicspath{{./images/}}
\usepackage{tikz}
\usepackage{bbm}
\usepackage{amsthm}
\usepackage[english]{babel}
\usepackage{appendix}
\usepackage{bbm}
\usepackage{authblk}
\usepackage{natbib}
\usepackage{appendix}
\usepackage{hyperref}
\newtheorem*{theorem*}{Theorem}
\newtheorem{theorem}{Theorem}
\newtheorem*{corollary*}{Corollary}
\newtheorem{corollary}{Corollary}
\newtheorem*{definition*}{Definition}

\newtheorem{assumption}{Assumption}
\newtheorem*{setup*}{Setup}
\newtheorem{proposition}{Proposition}
\newtheorem*{proposition*}{Proposition}
\newtheorem*{claim*}{Claim}
\newtheorem{lemma}{Lemma}
\usepackage{url}

\providecommand{\keywords}[1]
{
  \small	
  \textbf{Keywords:} #1
}
\title{Asymptotic Inference for Constrained Regression}
\author[1]{Madhav Sankaranarayanan \thanks{E-mail: \texttt{madhav\_sankaranarayanan@g.harvard.edu}}}
\author[2]{Yana Hrytsenko \thanks{E-mail: \texttt{yhrytsenko@gmail.com}}}
\author[3]{Jerome I. Rotter \thanks{E-mail: \texttt{jrotter@lundquist.org}}}
\author[2]{Tamar Sofer \thanks{E-mail: \texttt{tsofer@bidmc.harvard.edu}}}
\author[1]{Rajarshi Mukherjee \thanks{E-mail: \texttt{rmukherj@hsph.harvard.edu}}}

\affil[1]{Department of Biostatistics, Harvard T.H. Chan School of Public Health}
\affil[2]{Cardiovascular Institute, Beth Israel Deaconess Medical Center}
\affil[3]{The Institute for Translational Genomics and Population Sciences, Department of Pediatrics, The Lundquist Institute for Biomedical Innovation at Harbor-UCLA Medical Center}
\date{\today}
\begin{document}
\maketitle
\begin{abstract}
    We consider statistical inference in high-dimensional regression problems under affine constraints on the parameter space. The theoretical study of this is motivated by the study of genetic determinants of diseases, such as diabetes, using external information from mediating protein expression levels. Specifically, we develop rigorous methods for estimating genetic effects on diabetes-related continuous outcomes when these associations are constrained based on external information about genetic determinants of proteins, and genetic relationships between proteins and the outcome of interest. In this regard, we discuss multiple candidate estimators and study their theoretical properties, sharp large sample optimality, and numerical qualities under a high-dimensional proportional asymptotic framework. 
Finally, we apply the developed methods to study the genetic determinants of BMI, fasting insulin and HbA1c, leveraging their genetic correlation with protein expression obtained from an external study. 

\end{abstract}
\keywords{Asymptotic inference; Constrained linear regression; Genetic correlation; Genome wide association study; High dimensional inference; Metabolic trait.}

\section{Introduction}
\subsection{Motivation}

In this paper, we consider improved statistical inference in high-dimensional linear regression problems by leveraging known affine constraints on the underlying unknown parameter vector drawn from external sources. Specifically, we operate under a proportional asymptotic framework that allows us to derive precise efficiency gains implied by the number of independent constraints. 
Our main goal is to calibrate the precise statistical gains implied by the assumed constraints, to develop estimation methods to efficiently extract additional information, and provide asymptotically valid uncertainty quantification. In practical applications, these constraints are reliant on domain knowledge or external data. 
A specific instance of this arises quite naturally in the context of genetic disease modeling, and we will be using these methodologies to improve inference on the genetic pathways for diabetes. Diabetes is a chronic disease that affects an individual's ability to effectively utilize insulin and regulate blood sugar levels. While the onset of diabetes, in particular, type 2 diabetes or adult onset diabetes, is influenced by external factors, there is a marked genetic component of risk \citep{Cole_Florez_2020}. As a result, there is a pressing need for tools and techniques that leverage genetics for the early detection and prevention of diabetes. As diabetes diagnostics are based on glycemic traits, investigating genetic factors that impact glycemic and related metabolic traits can shed crucial light on the relationship between one’s genetic profile and risk of diabetes. Traditional analyses rely on genome-wide association studies (GWAS) to provide evidence of associations between genetic loci and phenotypes of interest. However, proteins mediate the effect of genetic variation on disease, and thus following trends in recent research, we leverage the advent and popularization of protein quantification technologies \citep{He_Shi_Wang_Jiang_Zhu_2020}, to utilize intermediary data and improve the effect estimation of genetic mutation on disease traits. 

We operate under this very framework and incorporate pleiotropic information, i.e., use effect sizes from a genetic variant's association with a non-target trait, to enhance the efficiency of genetic effect estimation on metabolic traits of interest. Specifically, our methodological paradigm aims to leverage pleiotropic information through \textit{genetic correlations}, the measure of correlation between traits that are dependent on genetic contribution alone. In this regard, genetic correlations between protein expression levels and metabolic traits obtained from a reference population allow us to posit a \textit{constraint} on the effect of involved genetic variants of interest, and thereby increase statistical accuracy of effect estimation in a target population. The main results of this paper can be summarized as follows: (i) we develop theoretically rigorous procedures for incorporating protein expression data into the pipeline of genetic effect estimation; (ii) we demonstrate statistical efficiency of our methods under moderately high dimensional settings; (iii) in higher dimensional settings, we provide novel methods that provide valid inference under some additional regularity conditions; (iv) we apply these methods to real genetic datasets for causal variant discovery.

\subsection{Literature}

The work in this paper builds on a well studied problem in classical literature of constrained linear regressions \citep{Aitchison_Silvey_1958,amemiya1985advanced}, more recent literature on high dimensional regression with random designs \citep{Hsu_Kakade_Zhang_2014, Mourtada_2022}, and established results in large random matrix theory in an proportional asymptotic regime \citep{Vaart_1998,Dobriban_Sheng_2022,Dobriban_Wager_2018}. Indeed, the problem of estimating parameters in constrained spaces arises in various settings, thus there are many avenues that have been explored in prior works. In this regard, the affine constraints setup is explored in \cite{Yu_2020} under a fixed design model to develop an algorithmic framework for constructing estimators under sparsity in high dimensions. Further, \cite{Shi_Zhang_Li_2016} allows for simplicial constraints in high dimensions with sparsity, a natural consideration in microbiome data. Aside from linear constraints, elliptical constraints have been studied in fixed design contexts \citep{Donoho_1994} and more recently, in general random designs \citep{Pathak_Wainwright_Xiao_2023}. Also, conical constraints, arising from higher-order constraints are investigated in \cite{Yu_Gupta_Kolar_2019}. The techniques used for these constraints are fundamentally different from the techniques required for affine constraints, but provide insights on the difficulty of estimation in constrained parameter spaces. More generally, \cite{Han_2023} provides exact risk asymptotics for recovery under convex constraints with Gaussian design. We leverage the additional structure of affine constraints and general universality results from random matrix theory to obtain more precise, interpretable, and minimax optimal error analysis in our work along with ways to quantify uncertainty. Although we do not directly work under a more general setup, the theme of the paper naturally connects to a larger body of work on more general semiparametric inference in low dimensional setups with constraints \citep{Klaassen_Susyanto_2016}, 
and estimation under inequality constraints \citep{Shapiro_1989}.


Finally, the question of interest in this work  arise partially from the field of genetic disease etiology. Large-scale genetic data analysis to study disease biology is an ever-growing field of research that addresses prediction and elucidation of causal relationships between risk factors and health and outcomes \citep{Khoury_Newill_Chase_1985,Holtzman_Marteau_2000,Gondro_Van_Der_Werf_Hayes_2013, Clerget-Darpoux_Elston_2013}. Particularly, polygenic risk scores have been used for disease prognostics, particularly in the case of diabetes \citep{Hahn_Kim_Choi_Lee_Kang_2022, Pemmasani_Atmakuri_Acharya_2023}. Moreover, ongoing research has also contributed to the literature on genetic studies of estimation and inference of quantities such as genetic correlations \citep{Elgart_Goodman_Isasi_Chen_De_Vries_Xu_Manichaikul_Guo_Franceschini_Psaty_etal._2021} and heritability \citep{Sofer_2017}. This literature informs the constraints we operate under and provides context and impetus for our research questions and statistical methods. 

\subsection{Notations}
We use the following notations in the remainder of the paper. First, we use bold symbols (e.g. $\boldsymbol{u}$) to denote vectors, non-bold capital letters (e.g. $W$) to denote matrices and non-bold lowercase letters (e.g. $e$) to denote scalars. For a square matrix $W\in\mathbb{R}^{d\times d}$, we denote $\lambda_{\max}(W)=\lambda_1(W)\geq\lambda_2(W)\geq\ldots\geq\lambda_d(W)=\lambda_{\min}(W)$ as the eigenvalues of $W$, and subsequently $\mathrm{Tr}(W):=\sum_{i=1}^d\lambda_i(W)$ as the trace of the matrix. For a matrix $W$, we denote the rank of the matrix as $\mathrm{rank}(W)$. The identity matrix in $\mathbb{R}^{d\times d}$ is denoted by $I_d$ and the zero vector in $\mathbb{R}^d$ is denoted as $\boldsymbol{0}_d$. The symbols 
$\|\cdot \|$ and $ \|\cdot\|_\infty$ denotes the $\ell_2$ and $\ell_\infty$ norm for a vector/matrix, respectively. For two symmetric $d\times d$ matrices $U$ and $V$, we write $U \succ V$ and $U \succeq V$ when $U - V$ is positive definite and positive semi-definite, respectively.

We utilize the following asymptotic notation in $n$. We use the standard Bachmann-Landau notations $o(\cdot), \mathcal{O}(\cdot)$ for deterministic sequences. For deterministic sequences, $\rightarrow$ denotes the limit of the sequence as $n$ tends to infinity. For two deterministic sequences $\{a_n\}_{n\geq 1},\{b_n\}_{n\geq 1}$, we say $a_n$ is asymptotic to $b_n$ denoted by $a_n\asymp b_n$, if $\frac{a_n}{b_n}\rightarrow 1$. The symbols $\stackrel{\mathbb{P}}{\rightarrow}$ and $\stackrel{d}{\rightarrow}$ denote convergence in probability and in distribution, respectively. The symbol $\stackrel{d}{=}$ denotes equivalence in distribution.

For a sequence of random variables $\{Y_n\}_{n \ge 1}$ and a deterministic positive sequence $\{a_n\}_{n \ge 1}$, we write $Y_n = o_{\mathbb{P}}(a_n)$ when $\frac{Y_n}{a_n} \stackrel{\mathbb{P}}{\rightarrow} 0$, and $Y_n = \mathcal{O}_p(a_n)$ when $\underset{K \to \infty}{\lim} \underset{n \to \infty}{\lim} \mathbb{P}\left(\frac{|Y_n|}{a_n} \le K\right) = 1$.

\subsection{Organization}
The following sections contain details of the results summarized earlier and are organized as follows. We present the mathematical setup and the technical background required for the analysis in Section \ref{setup}. Then, we outline the construction of the proposed estimators in Section \ref{methods}. Subsequently, we present our main theoretical analyses in Section \ref{theory}. Finally, we validate the performance of estimators through a suite of simulations with synthetic and real-life data, along with a real data exploration, in Section \ref{s:experiments}.

\section{Setup}
\label{setup}

We consider observing data 
on $n$ individuals as $\left(y_i,\boldsymbol{X}_i\right)_{i=1}^{n}\stackrel{i.i.d.}{\sim}\mathbb{P}$, where $y_i\in\mathbb{R}$ is an outcome of interest, $\boldsymbol{X}_i\in\mathbb{R}^p$ contains baseline covariates. In this setup, we are interested in the precise quantification of the association between $\boldsymbol{X}$ on $y$ under a constrained linear regression model as follows, where
\begin{equation}
\begin{split}
  \label{eq:main_setup}
y_i &= \boldsymbol{X}_i^T\boldsymbol{\beta}^{*}+\epsilon_i,\;\epsilon_i\text{ independent of } \boldsymbol{X}_i,\\ 
\boldsymbol{A}_j^T\boldsymbol{\beta}^{*}&=c_j,\;j=1,\ldots,q\\\mathbb{E}[\epsilon_i]&=0,\mathrm{Var}(\epsilon_i)=:\sigma^2<\infty, \exists\; \delta>0,\;\mathbb{E}\left[\epsilon_i^{2+\delta}\right]<\infty
\end{split}
\end{equation}
We include parallel discussions on generalized linear models in Appendix \ref{app:glm_est}. Under this model, $\boldsymbol{\beta}^{*}$ will be referred to as \textit{effect vector}, which, in this model, belongs to the affine space defined by known vectors $\boldsymbol{A}_j\in\mathbb{R}^p$ and constants $c_j\in\mathbb{R}$ for $j=1,\ldots,q$. For subsequent discussions, we shall refer to $X:=\left[\boldsymbol{X}_1 \ldots \boldsymbol{X}_n\right]^T$ as the \textit{design matrix}, $\boldsymbol{y}:=\left[y_1\ldots y_n\right]^T$ the \textit{outcome vector}, $A:=\left[\boldsymbol{A}_1\ldots\boldsymbol{A}_q\right]^T$ the \textit{constraint matrix}, and $\boldsymbol{c}:=\left[c_1\ldots c_q\right]^T$ the \textit{constraint vector}.

Although we operate under generic known constraints, in our real data applications they arise from borrowing information from an external {reference population}. We briefly elaborate on this to explain the nature of the linear constraints that arise in our data setting. Specifically, suppose we have access to a reference population where one has  $(\boldsymbol{Y},y_{r},\tilde{\boldsymbol{X}})\sim\mathbb{P}_r$ with $\boldsymbol{Y}\in \mathbb{R}^q$ contains $q<p$ auxiliary outcomes, $y_{r}$ is the outcome of interest in the reference population, and $\tilde{\boldsymbol{X}}$ contains baseline covariates. Suppose now that a regression model holds in the reference population designating $y_{r} = \tilde{\boldsymbol{X}}^T\boldsymbol{\beta}^{*}+\tilde{\epsilon},\;\mathbb{E}[\tilde{\epsilon}_i]=0$ and $Y_{j} = \tilde{\boldsymbol{X}}^T\boldsymbol{\beta}_j+\tilde{\epsilon}_{j},\;\mathbb{E}[\tilde{\epsilon}_{j}]=0,\;j=1,\ldots,q$, with a common variance-covariance matrix profile of covariates under both the reference and our study population (i.e. $\Sigma:=\mathbb{E}\left[\tilde{\boldsymbol{X}}\tilde{\boldsymbol{X}}^T\right]=\mathbb{E}\left[\boldsymbol{X}\boldsymbol{X}^T\right]$). Drawing motivation from genetic association studies, we will connect our population to the reference population by defining constraint matrices and vectors as $\boldsymbol{A}_j = \boldsymbol{\beta}_j^T\Sigma, c_i = \boldsymbol{A}_j\boldsymbol{\beta}^{*},\;j=1,\ldots,q$. The precise reason behind these definitions can be understood from the background of genetic correlation analysis that we provide in Appendix \ref{app:bio_info}. For now, especially for our main data analysis (see Section \ref{subsec:data_analysis} for details), the relevance of such target and reference populations and their implied connections through constraints on parameter space can be roughly described as follows. Suppose we focus on genetic variants known as single-nucleotide polymorphisms (SNPs), particularly those responsible for expression levels of certain proteins, known as protein quantitative trait loci (pQTLs). To draw information on this relationship, suppose we have a reference and target population as above with genotypic and metabolic data, with the reference population additionally possessing protein quantification data. Mathematically,  one can consider $q$ proteins for our analysis of $p$ SNPs on a specific metabolic trait in a reference population of individuals with genotype data  $\tilde{\mathbf{X}}$, $\boldsymbol{\beta}_1,\ldots, \boldsymbol{\beta}_q$  the additive allelic effect vectors corresponding to each protein,  $\boldsymbol{\beta}^{*}$ the effect vector for the metabolic trait, ${Y}_1,\ldots,{Y}_q$ the protein expression outcomes in the reference population, and $\boldsymbol{y}_{r}$ the metabolic trait outcomes in the reference population. Similarly, in the target population, suppose $\boldsymbol{y}_t$ are the metabolic trait outcomes in the target population, arising from the same corresponding additive allelic effect vector $\boldsymbol{\beta}^{*}$. The dispersion matrix $\Sigma$ is the linkage disequilibrium (LD) matrix, which contains information on the non-random association of alleles at different loci in a population. In this framework, the above described constraints arise naturally from genetic correlation perspectives between $Y_j$'s and $y$'s. We will not operating under the randomness of $Y_j$ in this work. This paper can be thought of as working through oracle knowledge of the reference population and the main goal is to precise quantification of the gain in efficiency based on this external knowledge in a proportional asymptotic framework and provide easy uncertainty quantification of the optimal estimation procedures. in particular, since the general construction of our methodology will not depend on the precise manner in which the constraints were obtained, we will not refer to the reference population in the theoretical parlance of the paper.

\begin{figure}
    \centering
    \includegraphics[height=17em]{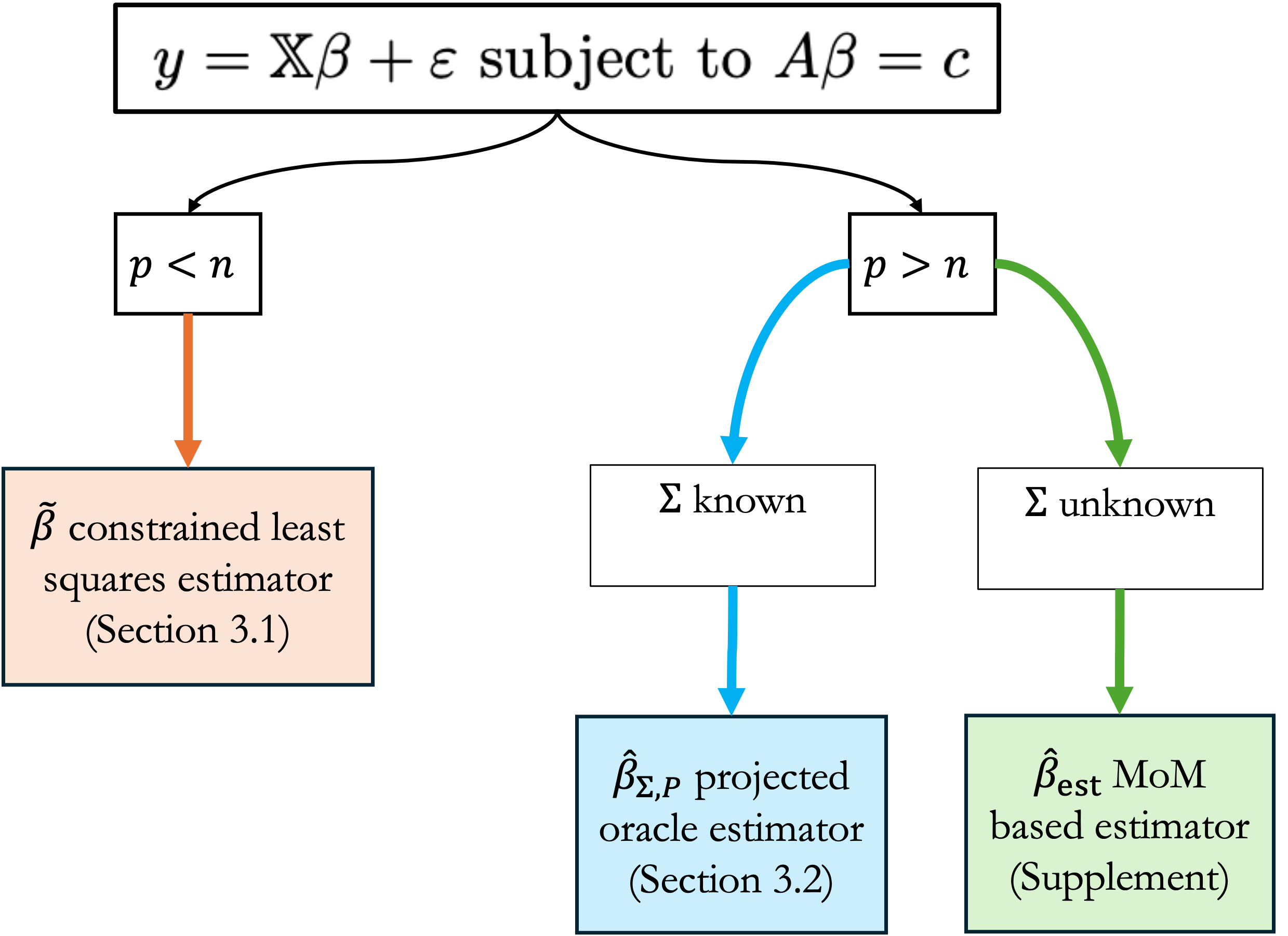}
    \caption{Schematic describing different estimators described in this paper}
    \label{fig:schematic}
\end{figure}
\section{Methods}
\label{methods}
In this section, we describe estimators we will study under scenarios based on different regimes of the dimension $p$ and knowledge of $\Sigma$. Throughout we will assume a proportional asymptotic regime where $p/n\rightarrow \alpha\in (0,\infty)$ and divide our discussions according to moderately high dimensions $(\alpha<1)$ and high dimensions $(\alpha\geq1)$. In moderately high dimensions, we provide sharp minimax optimal estimation of the vector $\boldsymbol{\beta}$ to optimally quantify the gain from external information. Beyond this we also explore estimation of low dimensional functionals of $\boldsymbol{\beta}$ and there $\sqrt{n}$-consistent asymptotically normal inference. We will explicitly specify the targets of estimation in each section and conditions we require therein.

\subsection{Moderately High Dimension ($\alpha<1$)} 
\label{mod_known}

We incorporate the  constraint $A\boldsymbol{\beta}^{*}=\boldsymbol{c}$ into ordinary least squares optimization problem whose unique solution exists with high probability under suitable assumptions on $X$ to get an optimal estimator for $\boldsymbol{\beta}$. A feasible solution, that will be referred to as the \textit{standard projected estimator}, is defined as 
$$\hat{\boldsymbol{\beta}}_{\mathcal{P}}:=\mathcal{P}_{A^\perp}\hat{\boldsymbol{\beta}}_{\mathrm{LS}}+A^T\left(AA^T\right)^{-1}\boldsymbol{c}$$ where $\mathcal{P}_{A^\perp}:=\mathcal{P}_{\perp}(A)=I_p-A^T\left(AA^T\right)^{-1}A$ and $\hat{\boldsymbol{\beta}}_{\mathrm{LS}}$ is the \textit{ordinary least squares} (OLS) estimator of $\boldsymbol{\beta}^{*}$. Although this estimator turns out to be asymptotically sharp minimax optimal under isotropic covariance for $X$,  
this estimator can be improved under more general covariance structures on $X$ by utilizing random projections that preserve the constraint. To that end, we derive an optimal procedure through the Lagrangian optimization based \textit{constrained least squares} (CLS) estimator \citep{amemiya1985advanced}. With the sample estimate of the covariance matrix $\hat{\Sigma}_n=\frac{1}{n}X^TX$, we define the CLS estimator as 
$$\tilde{\boldsymbol{\beta}}:=C_{A^\perp,inv}\hat{\boldsymbol{\beta}}_{\mathrm{LS}}+\hat{\Sigma}_{n,inv}^{-1}A^T\left(A\hat{\Sigma}_{n,inv}^{-1}A^T\right)^{-1}\boldsymbol{c}$$ where $C_{A^\perp,inv}=I_p-\hat{\Sigma}_{n,inv}^{-1}A^T\left(A\hat{\Sigma}_{n,inv}^{-1}A^T\right)^{-1}A$ and $$\hat{\Sigma}_{n,inv}=\begin{cases}
    \hat{\Sigma}_n\;\text{if }\hat{\Sigma}_n\text{ is invertible}\\
    I_p\;\text{otherwise}
\end{cases}$$The matrix $C_{A^\perp,inv}$  is the row space projector on the eigenbasis of $\hat{\Sigma}_{n,inv}$. In our analysis, we consider sub-Gaussian distributions for the independent rows of $X$, which means that $\hat{\Sigma}_n$ is invertible with probability converging to $1$ exponentially fast in the sample size under the proportional asymptotic regime \citep{Eaton_Perlman_1973,vershynin2010introduction}. 
For notational convenience, under the assumptions on the rows of $X$, we will replace $\hat{\Sigma}_{n,inv}$ with $\hat{\Sigma}_n$ and $C_{A^\perp,inv}$ with $C_{A^\perp}$, implicitly conditioning on the good event of invertibility. We study the theoretical and numerical properties of these estimators in Section \ref{theory} and Section \ref{s:experiments} respectively to demonstrate the gains in efficiency as a function of the number of constraints $q$. In the case where $q\ll p $ or even $q=\mathcal{O}(1)$, which will be relevant to our data application, the efficiency gain does not reflect itself in the leading order. To motivate the use of constraints over ordinary least squares, in Sections \ref{s:theory_mod} and \ref{s:experiments} we also discuss precise efficiency gain in the second-order term of asymptotic expansions. 
Finally, we also include a discussion of the construction of estimators for generalized linear models in this moderately high dimensional case in Appendix \ref{app:glm_est}. 

\subsection{High Dimension ($\alpha\geq 1$)}
\label{high_known}

In high dimensions, our approach will depend on the knowledge we possess of the underlying dispersion matrix $\Sigma$. This is because in an ill-specified regime, without oracle information about $\Sigma$, estimation of functionals of $\boldsymbol{\beta}^{*}$ becomes challenging while targeting $\sqrt{n}$-consistent asymptotically normal estimation \citep{verzelen2018adaptive, kong2018estimating, Chen_Liu_Mukherjee_2024}. Within this scope, we present a simple estimator that demonstrates desirable properties of gaining efficiency under constraints. 
This estimator will utilize $\Sigma$ as a stand-in for the gram matrix in a traditional least squares estimator. Following nomenclature given in \cite{Frostig_Heller_2022,Knight_Duan_2024}, we define an \textit{oracle} estimator $\hat{\boldsymbol{\beta}}_{\Sigma}:=\frac{1}{n}\Sigma^{-1}X^T\boldsymbol{y}$ as our high-dimensional estimator of $\boldsymbol{\beta}^{*}$, which is the analogue least squares estimator for $\boldsymbol{\beta}$ using information on the true population variance-covariance matrix of $\boldsymbol{X}_i$'s. Subsequently, we define the \textit{projected oracle estimator} as follows
$$\hat{\boldsymbol{\beta}}_{\Sigma,\mathcal{P}}:=\frac{1}{n}\mathcal{P}_{A^\perp}\Sigma^{-1}X^T\boldsymbol{y}+A^T\left(AA^T\right)^{-1}\boldsymbol{c}$$
incorporating the constraint information into the estimation procedure. We show in Section \ref{theory} that such an estimator attains a notion of asymptotic normality, allowing us to use this estimator for testing purposes. 

Without the knowledge of $\Sigma$ and any additional structure that allows its consistent estimation, the question of optimal $\sqrt{n}$-consistent estimation of functionals of $\boldsymbol{\beta}$ remains an area of active research. Although the modifications of debiased Lasso under proportional asymptotics have been explored in recent literature \citep{celentano2023challenges,bellec2023debiasing, celentano2023lasso, song2024hede, celentano2024correlation, bellec2025observable}, they overwhelmingly rely on the knowledge of $\Sigma$. To address this, some recent work has proposed a method-of-moments type estimator of functionals of $\boldsymbol{\beta}$ in the unknown $\Sigma$ regime -- only demonstrating consistency with a logarithmic rate of convergence \citep{kong2018estimating, Chen_Liu_Mukherjee_2024}. We build on this \citep{kong2018estimating, Chen_Liu_Mukherjee_2024} by
incorporating the constraint information. Similar to \cite{kong2018estimating, Chen_Liu_Mukherjee_2024},  we operate by approximating functions of $\Sigma^{-1}$
using Chebyshev polynomials \citep{DeVore_Lorentz_1993}, and subsequently using higher-order U-statistics
to estimate these approximations. We provide this algorithmic procedure in Appendix \ref{app:high_dim_est}, along with simulations that validate the performance of these estimators



\section{Properties of Estimators}
\label{theory}
To discuss the theoretical properties of the estimators introducted in Section \ref{methods}, we will work with the following set of assumptions:
\begin{assumption}
\label{as:dim}
(Dimensionality and Asymptotics)

       $p,q,n\rightarrow\infty$ such that 
    $\frac{p}{n}\rightarrow\alpha\in[0,\infty)$ (\textit{aspect ratio}) and 
       $\frac{q}{p}\rightarrow\gamma\in[0,1)$ (\textit{constraint ratio})
        
\end{assumption}
\begin{assumption}
\label{as:rd}
(Random Design)
\begin{itemize}
    \item[(a)]\begin{enumerate}
    \item[(i)] $\boldsymbol{X}_i \stackrel{d}{=} \Sigma^{1/2} \boldsymbol{Z}_i$, where entries of $\boldsymbol{Z}$ are mean zero, unit variance and have finite $(8+c)$th moment where $c>0$ is fixed and arbitrary.
    \item[(ii)] $\exists C>0$ such that $\frac{1}{C}\leq\lambda_{k}(\Sigma)\leq C$ for $k=1,\ldots,p$. Additionally, the support of $\mu_{\infty}$ lies within $\left(\frac{1}{C},C\right)$, where $\mu_{\infty}$ denotes the limiting empirical spectral distribution of $\Sigma$.
\end{enumerate}
\item[(b)]\begin{enumerate}
    \item[(i)] For any 1-Lipschitz function $f$ of $\boldsymbol{X}_i$, $\exists C$ such that we have the following concentration inequality:
    $$\mathbb{P}(|f(\boldsymbol{X}_i)-\mathbb{E}\left[f(\boldsymbol{X}_i)\right]|>t)\leq C\exp\left\{-\frac{t^2}{C}\right\}$$
    \item[(ii)] $\exists C^{'}>0$ such that $\frac{1}{C^{'}}\leq \frac{p}{n}\leq C^{'}$ for all $n>0$
    \item[(iii)] $\exists C^{''}>0$ such that $\frac{1}{C^{''}}\leq\lambda_{k}(\Sigma)\leq C^{''}$ for all $k=1,\ldots,p$. Additionally, the support of $\mu_{\infty}$ lies within $\left(\frac{1}{C^{''}},C^{''}\right)$, where $\mu_{\infty}$ denotes the limiting empirical spectral distribution of $\Sigma$.
\end{enumerate}
\end{itemize}

\end{assumption}
\begin{assumption}
\label{as:err}
    (Gaussian Error) $y_i\sim\mathcal{N}\left(\boldsymbol{X}_i^T\boldsymbol{\beta}^{*},\sigma^2\right)$ for $i=1,\ldots,n$
        
\end{assumption}

    \textbf{Remark:} The stipulations on $\boldsymbol{X}$ for Assumption \ref{as:rd}(a) are stronger than those for Assumption \ref{as:rd}(b), but independent control of each $\boldsymbol{X}$ requires the aspect ratio to be bounded away from $0$, which is not required in Assumption \ref{as:rd}(a). Also, the assumptions on the limiting empirical spectral distribution of $\Sigma$ guarantee the existence of certain trace functionals that arise in our results. Additionally, from \cite{Bai_Zhou_2008}, we have the convergence of the empirical spectral measure of $\hat{\Sigma}_n$ to a free multiplicative convolution of a Marchenko-Pastur law $\mathrm{MP}(\alpha)$ and $\mu_{\infty}$, which is necessary for many subsequent convergence in probability results. These assumptions arise from prior work focused on the construction of deterministic equivalents for sample covariance matrices \cite{Hachem_Loubaton_Najim_2007,Louart_Couillet_2021,Chouard_2022}. In subsequent results, we will only refer to the random design assumptions as Assumption \ref{as:rd}, as either assumption will be sufficient, and we discuss these conditions further in Appendix \ref{app:asymp_minimax}.

\subsection{Moderately High Dimension}
\label{s:theory_mod}
For the CLS estimator $\tilde{\boldsymbol{\beta}}$, defined through the matrices $\hat{\Sigma}_n$ and $C_{A^\perp}$, we have the following theorem describing the exact minimax error of estimation under a linear constraint, and the asymptotics of the same error.

\begin{theorem}
\label{thm:minimax}
For the setup given in (\ref{eq:main_setup}), under Assumption \ref{as:err}, the minimax risk of estimation of $\boldsymbol{\beta}^*$ in squared error loss is 
    $$\inf_{\hat{\boldsymbol{\beta}}}\sup_{\boldsymbol{\beta}^*}\mathbb{E}\left|\left|\hat{\boldsymbol{\beta}}-\boldsymbol{\beta}^*\right|\right|^2 = \frac{\sigma^2}{n}\mathbb{E}\left[\mathrm{Tr}\left(C_{A^\perp}\hat{\Sigma}_n^{-1}\right)\right]$$where $$\hat{\Sigma}_n=\frac{X^TX}{n},C_{A^\perp} = I-\hat{\Sigma}_n^{-1}\left(A^T\left(A\hat{\Sigma}_n^{-1}A^T\right)^{-1}A\right)$$Additionally, the estimator 
    $$\tilde{\boldsymbol{\beta}}=C_{A^\perp}\hat{\Sigma}_n^{-1}\left(\frac{X^T\boldsymbol{y}}{n}\right)+\hat{\Sigma}_n^{-1}A^T(A\hat{\Sigma}_n^{-1}A^T)^{-1}\boldsymbol{c}$$ achieves the minimax risk   
\end{theorem}

We would also like to categorize the asymptotics of such an error formulation, which is put forth in the following corollary.

\begin{corollary}
\label{thm:asymp_minimax}
    For the setup given in (\ref{eq:main_setup}), under Assumptions \ref{as:dim},\ref{as:rd},\ref{as:err}, the constrained least squares estimator $\tilde{\boldsymbol{\beta}}$ achieves the following asymptotic squared error loss
    $$\lim_{n\rightarrow\infty}\mathbb{E}\left|\left|\tilde{\boldsymbol{\beta}}-\boldsymbol{\beta}^*\right|\right|^2=\frac{\sigma^2}{(1-(1-\gamma)\alpha)}\lim_{n\rightarrow\infty}\frac{1}{n}\mathrm{Tr}\left(\Sigma^{-1}-\Sigma^{-1}\left(A^T\left(A\Sigma^{-1}A^T\right)^{-1}A\right)\Sigma^{-1}\right)$$
    Under an isotropic $(\Sigma=I_p)$ generation process, this reduces to 
    $$\lim_{n\rightarrow\infty}\mathbb{E}\left|\left|\tilde{\boldsymbol{\beta}}-\boldsymbol{\beta}^*\right|\right|^2=\frac{\sigma^2(1-\gamma)\alpha}{1-(1-\gamma)\alpha}$$
\end{corollary}

\textbf{Remark:} We assume the existence of the limit in the first equation. This is a reasonable assumption based on the random design restrictions, and the second statement of the corollary demonstrates it in the isotropic case. 

This result describes the precise ``difficulty'' of recovering the true effect vector. In the context of mean-squared error, we provide a justification for projection, even in the case where $q\ll p $. Let us denote $G_n:=\|\hat{\boldsymbol{\beta}}_{LS}-\boldsymbol{\beta}^{*}\|^2-\|\hat{\boldsymbol{\beta}}_{\mathcal{P}}-\boldsymbol{\beta}^{*}\|^2$ as the ``gain'' in squared error loss of projecting.

\begin{proposition}
\label{thm:low_q}
Under Assumptions \ref{as:dim} and \ref{as:rd}, with an isotropic $(\Sigma = I_p)$ generation process, we have     $$\mathbb{E}\left[G_n\right] = \frac{q\sigma^2}{n(1-\alpha)}(1+o(1))$$Additionally
\begin{itemize}
    \item[(a)] For $q\rightarrow\infty$, we have

    $$\frac{nG_n}{q}\stackrel{\mathbb{P}}{\rightarrow}\frac{\sigma^2}{1-\alpha}$$
    \item[(b)] Additionally, under Assumption \ref{as:err}, for $q=\mathcal{O}(1)$, we have $$\frac{nG_n}{\sigma^2}\stackrel{d}{=}\sum_{i=1}^qw_i\chi_1^{(i)}$$where $\chi_1^{(i)}$ are i.i.d. central chi-squared random variables with degrees of freedom $1$ and $w_1\geq w_2\geq\ldots\geq w_q$ are the non-zero eigenvalues of $nX\hat{\Sigma}_n^{-1}\mathcal{P}_A\hat{\Sigma}_n^{-1}X^T$.
\end{itemize}
    \label{eq:q_benefit}
\end{proposition}
This proposition, in tandem with Theorem \ref{thm:minimax}, demonstrates the benefit of utilizing even a fixed number of constraints over the OLS, which is relevant to our data example in Section \ref{subsec:data_analysis}. We additionally verify this benefit in Section \ref{s:experiments}. 
Our next result is about the asymptotic distribution of coordinates of $\tilde{\boldsymbol{\beta}}$ at a parametric rate.
\begin{proposition}
\label{thm:asymp_norm}
    Under Assumptions \ref{as:dim} and \ref{as:rd}, we have  
    $$\sqrt{n}\left(\tilde{\boldsymbol{\beta}}_j-\boldsymbol{\beta}^*_j\right)\stackrel{d}{\rightarrow}\mathcal{N}\left(0,\frac{\sigma^2s_{C,j}}{1-(1-\gamma)\alpha}\right)$$
    where $s_{C,j} = \underset{n\rightarrow\infty}{\lim}\boldsymbol{e}^T_j\left(\Sigma^{-1}-\Sigma^{-1}\left(A^T\left(A\Sigma^{-1}A^T\right)^{-1}A\right)\Sigma^{-1}\right)\boldsymbol{e}_j$

    \label{eq:asymp_norm}
\end{proposition}
\textbf{Remark:} We assume the existence of the limit $s_{C,j}$, similar to the remark on Corollary \ref{thm:asymp_minimax}. Also, since
$$\frac{\sigma^2}{1-\alpha}\Sigma^{-1}\succeq\frac{\sigma^2}{1-(1-\gamma)\alpha}\left(\Sigma^{-1}-\Sigma^{-1}\left(A^T\left(A\Sigma^{-1}A^T\right)^{-1}A\right)\Sigma^{-1}\right)$$for any fixed $p$, where $\frac{\sigma^2}{1-\alpha}\Sigma^{-1}$ is the variance matrix of the OLS estimator, the total of the variances of each coordinate of the CLS estimator is lower than that of the OLS estimator. We cannot make any direct comparison of the asymptotic variance for a given coordinate. 

The asymptotic variance is a function of $\Sigma^{-1}$, which requires careful estimation while constructing confidence intervals. We will use a corrected jackknife estimator for the variance, elucidated in \citep{Karoui_Purdom_2016a}, for our numerical experiments. We have the following result, which follows from Theorem 4.1 in \citep{Karoui_Purdom_2016a}:

\begin{proposition}
\label{thm:jackknife}
    Under Assumptions \ref{as:dim} and \ref{as:rd}, given that every row of $\boldsymbol{X}_i$ is drawn from a multivariate normal distribution with mean $\boldsymbol{0}$ and variance $\Sigma$, we have  
    $$\frac{\mathbb{E}\left[\widehat{\mathrm{Var}}_j\right]}{\mathrm{Var}(\tilde{\beta}_j)}\rightarrow\frac{1}{\left(1-\left(1-\gamma\right)\alpha\right)}$$
    where $\widehat{\mathrm{Var}}_j=\frac{n-1}{n}\sum_{i=1}^n\left(\tilde{\beta}_{(i),j}-\tilde{\beta}_j\right)^2$, $\tilde{\boldsymbol{\beta}}_{(i)},\tilde{\boldsymbol{\beta}}$ are the CLS estimators with the $i$th observation removed and for the whole dataset respectively. This holds for any contrast $\boldsymbol{v}^T\tilde{\boldsymbol{\beta}}$ with $\|\boldsymbol{v}\|^2_2=1$
\end{proposition}
Using Propositions \ref{eq:asymp_norm} and \ref{thm:jackknife}, we can conduct inference on any constrast of our regression vector in this proportional regime. 


\subsection{High Dimension}

We demonstrate that the projected oracle estimator has a notion of asymptotic normality, akin to the moderately-high dimension estimator, allowing for inference in this regression problem in high dimensions.

\begin{proposition}
\label{thm:asymp_norm_high}
    Given the setup in (\ref{eq:main_setup}), under Assumptions \ref{as:dim},\ref{as:rd},\ref{as:err}, we have 
    $$\sqrt{n}\left(\hat{\beta}_{\Sigma,\mathcal{P},j}-\beta^*_j\right)\stackrel{d}{\rightarrow}\mathcal{N}\left(0,\left(\beta^{*}_j\right)^2+\left(\sigma^2+\ell_{\boldsymbol{\beta}^{*},\Sigma}\right)s_{\mathcal{P},j}\right)$$where $\ell_{\boldsymbol{\beta}^{*},\Sigma}=\underset{n\rightarrow\infty}{\lim}\|\Sigma^{1/2}\boldsymbol{\beta}^{*}\|^2$ and $s_{\mathcal{P},j} = \underset{n\rightarrow\infty}{\lim}\boldsymbol{e}_j^T\left(\mathcal{P}_{A^\perp}\Sigma^{-1}\mathcal{P}_{A^\perp}\right)\boldsymbol{e}_j$
    \label{eq:hd_asymp_norm}
\end{proposition}

\textbf{Remark:} We assume the existence of the limit $s_{\mathcal{P},j}$, similar to Proposition \ref{eq:asymp_norm}. Also, since 
$\Sigma^{-1}\succeq\mathcal{P}_{A^\perp}\Sigma^{-1}\mathcal{P}_{A^\perp}$ for any projection matrix $\mathcal{P}_{A^{\perp}}$ in fixed $p$, the total variance of each coordinate of this projected oracle estimator is less than that of a standard oracle estimator.

Note that the asymptotic variance is a function of the signal strength and the magnitude of the regression coordinate. Additional information on this signal strength along with variance stabilization will allow us to invert the variance form to construct a confidence interval for $\beta^{*}_j$. We demonstrate the efficacy of this projected oracle estimator in utilizing the constraints over a range of constraint ratios in Section \ref{s:experiments}.

\section{Experimental Efficacy}
\label{s:experiments}
We look to evaluate the performance of our estimators through a set of simulations. We will first demonstrate the properties detailed in the previous section through simulations using synthetic data, and we will apply the proposed estimators to genetic datasets for the purpose of genetic variant discovery.

\subsection{Synthetic Simulations}
For the purposes of our simulations, we refer to the formulation of the reference population from Section \ref{setup}. We use a reference population to construct our constraints, which we use for our target population, and all subsequent analyses are conditional on this reference population. We fix our sample size $n=200$ and $N=1000$, with $1000$ iterations for repeated experiments, and vary the aspect and constraint ratio as described below. 

\begin{enumerate}
    \item[(\textbf{s1})] $p=(10,20,\ldots,190), q=\lfloor\frac{p}{2}\rfloor+1$ (Moderately high dimension, varying aspect ratio)
    \item[(\textbf{s2})] $p=100, q=(0,5,\ldots,100)$ (Moderately high dimension, varying constraint ratio)
    \item[(\textbf{s3})] $p=300, q=(0,10,\ldots,300)$ (High dimension, varying constraint ratio)
\end{enumerate}
Additionally, we consider two cases of the dispersion matrix; (\textbf{m1}) $\Sigma$ isotropic, generated as $\Sigma = I_p$, (\textbf{m2}) $\Sigma$ anisotropic, generated as $\Sigma=0.5I_p+0.5J_p$. Moreover, we generate $\boldsymbol{X}_i\sim \mathcal{N}_p(0,\Sigma)$ for $i$ in $1,\ldots,n$ and $\tilde{\boldsymbol{X}}_{\tilde{i}}\sim \mathcal{N}_p(0,\Sigma)$ for $\tilde{i}$ in $1,\ldots,N$, and the sample covariance matrices $\hat{\Sigma}_n=\frac{X^TX}{n}$ and $\tilde{\Sigma}_N=\frac{\tilde{X}^T\tilde{X}}{N}$. We use $B=[I_q:0_{p-q}], \beta^*_j\sim\mathcal{N}(5,5)$ for $j$ in $1,\ldots,p$, and define $A=B\hat{\Sigma}_N$ and $\boldsymbol{c}=A\boldsymbol{\beta}^*$. Finally, for the outcome, we consider $y=X\boldsymbol{\beta}^*+\boldsymbol{\epsilon},\epsilon_i\sim\mathcal{N}(0,1)\text{ for }i\text{ in }1,\ldots,n$. 
Below we highlight a few results surrounding these setups and defer additional results to Appendix \ref{s:add_exp}. 

\begin{figure}
\begin{minipage}{.47\textwidth}
\centering
  \includegraphics[width = 0.9\textwidth, height = 13em]{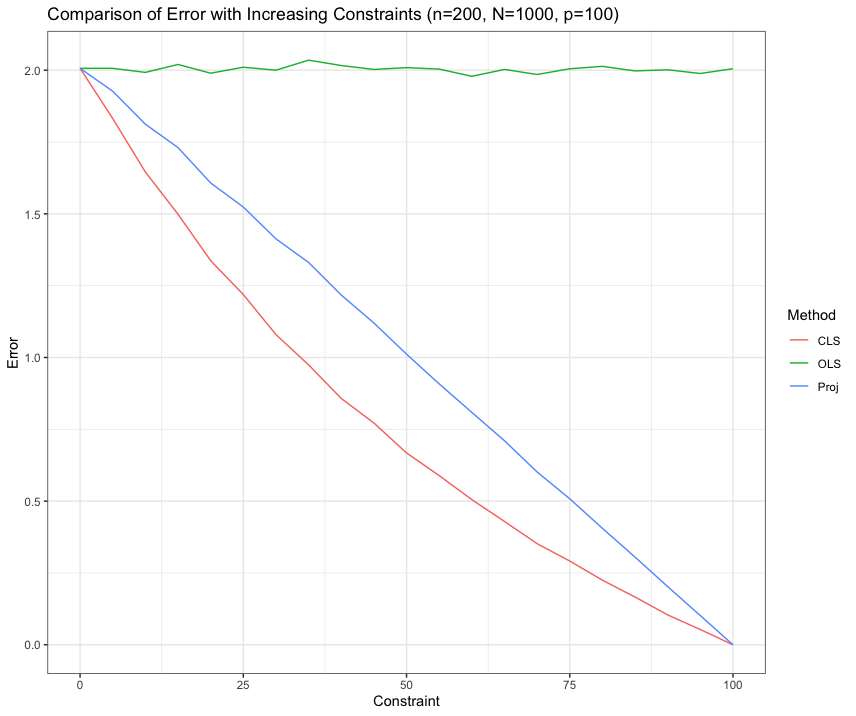}
  \caption{Error comparison ($\hat{\boldsymbol{\beta}}_{LS},\hat{\boldsymbol{\beta}}_{\mathcal{P}},\tilde{\boldsymbol{\beta}}$) [\textbf{s2,m2}]}
    \label{fig:error_comparison}
\end{minipage}%
\hfill
\begin{minipage}{.47\textwidth}
\centering
  \includegraphics[width = \textwidth, height = 13em]{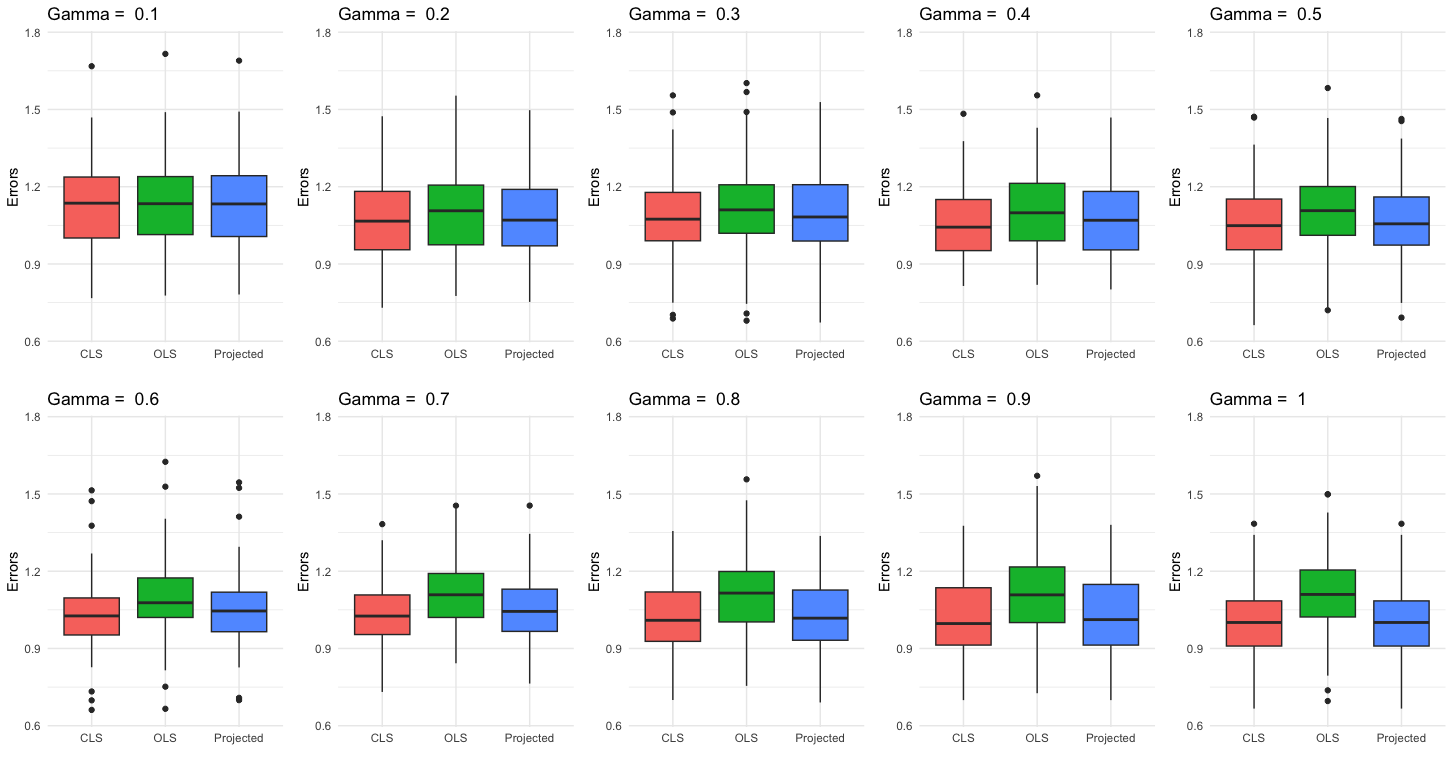}
  \caption{Prediction error ($\hat{\boldsymbol{\beta}}_{LS},\hat{\boldsymbol{\beta}}_{\mathcal{P}},\tilde{\boldsymbol{\beta}}$) [\textbf{s2,m1}]}
    \label{fig:pred_err}
\end{minipage}
\end{figure}

Fig. \ref{fig:error_comparison} [\textbf{s2,m2}] shows comparisons of mean squared error (MSE) between the OLS estimator, a standard projection estimator, and the CLS estimator, described in the moderately high dimension part of Section \ref{methods}. The error of the OLS estimator $\hat{\boldsymbol{\beta}}_{\mathrm{LS}}$ is constant since it is constraint-agnostic, while the other estimators have errors that tend to zero as the constraint level increases. As verified theoretically, the CLS estimator $\tilde{\boldsymbol{\beta}}$ has  uniformly lower error than the projected estimator $\hat{\boldsymbol{\beta}}_{\mathcal{P}}$. Next, Fig. \ref{fig:pred_err} [\textbf{s2,m1}] shows the out-of-sample prediction error for a subset of values for $q$, on $100$ new observations generated under a similar setup. We see that as the constraint ratio increases, the predictive errors of the projected and CLS estimators decrease, while the predictive error of the OLS estimator remains steady.

\begin{figure}
\centering
\begin{minipage}{.47\textwidth}
\centering
  \includegraphics[width = \textwidth, height = 13em]{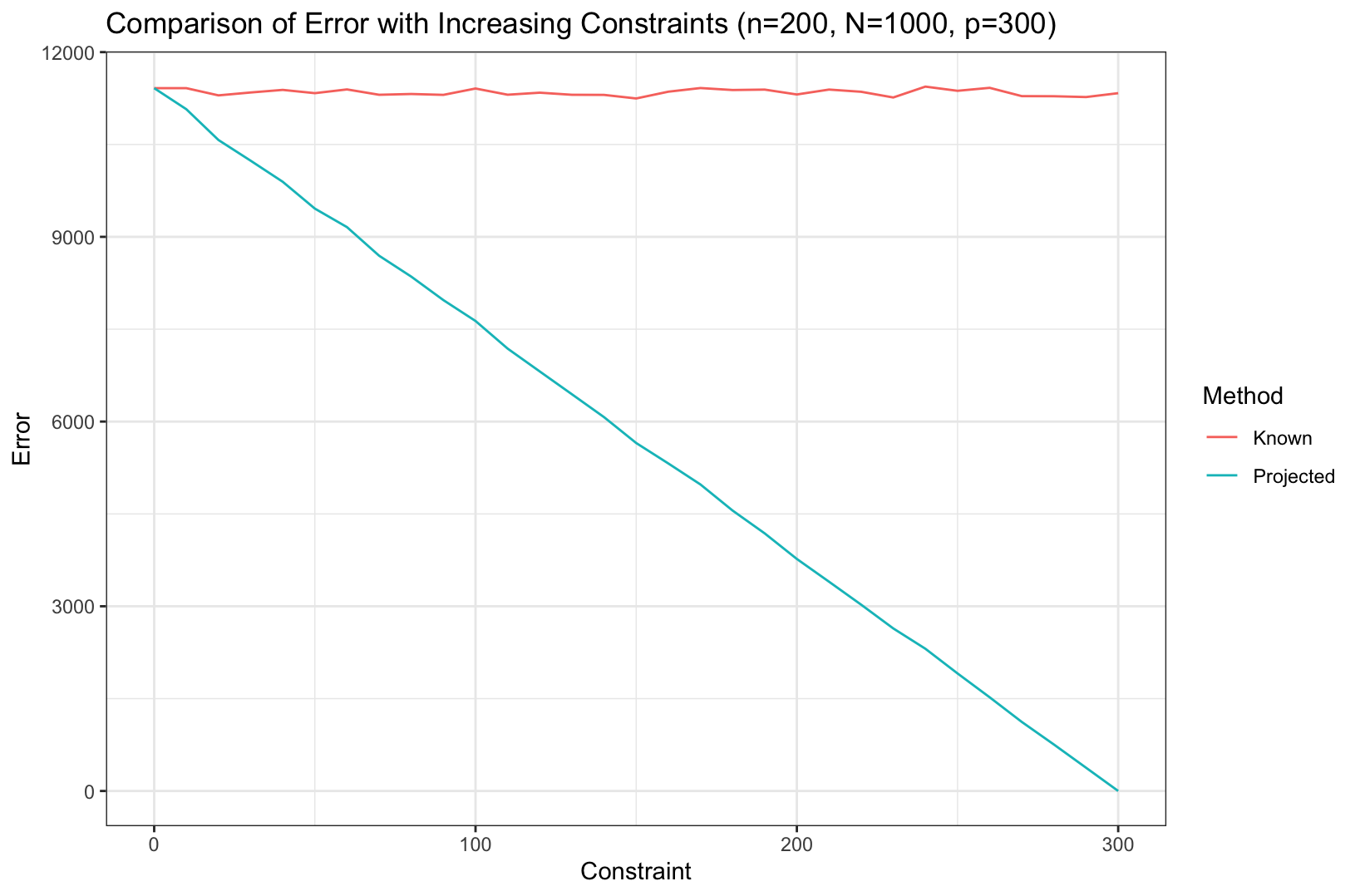}
    \caption{Error comparison ($\hat{\boldsymbol{\beta}}_{\Sigma}$ and $\hat{\boldsymbol{\beta}}_{\Sigma,\mathcal{P}}$) [\textbf{s3,m1}]}
    \label{fig:error_comparison_high}
\end{minipage}
\hfill
\begin{minipage}{.47\textwidth}
  \centering
  \includegraphics[width = \textwidth, height = 13.5em]{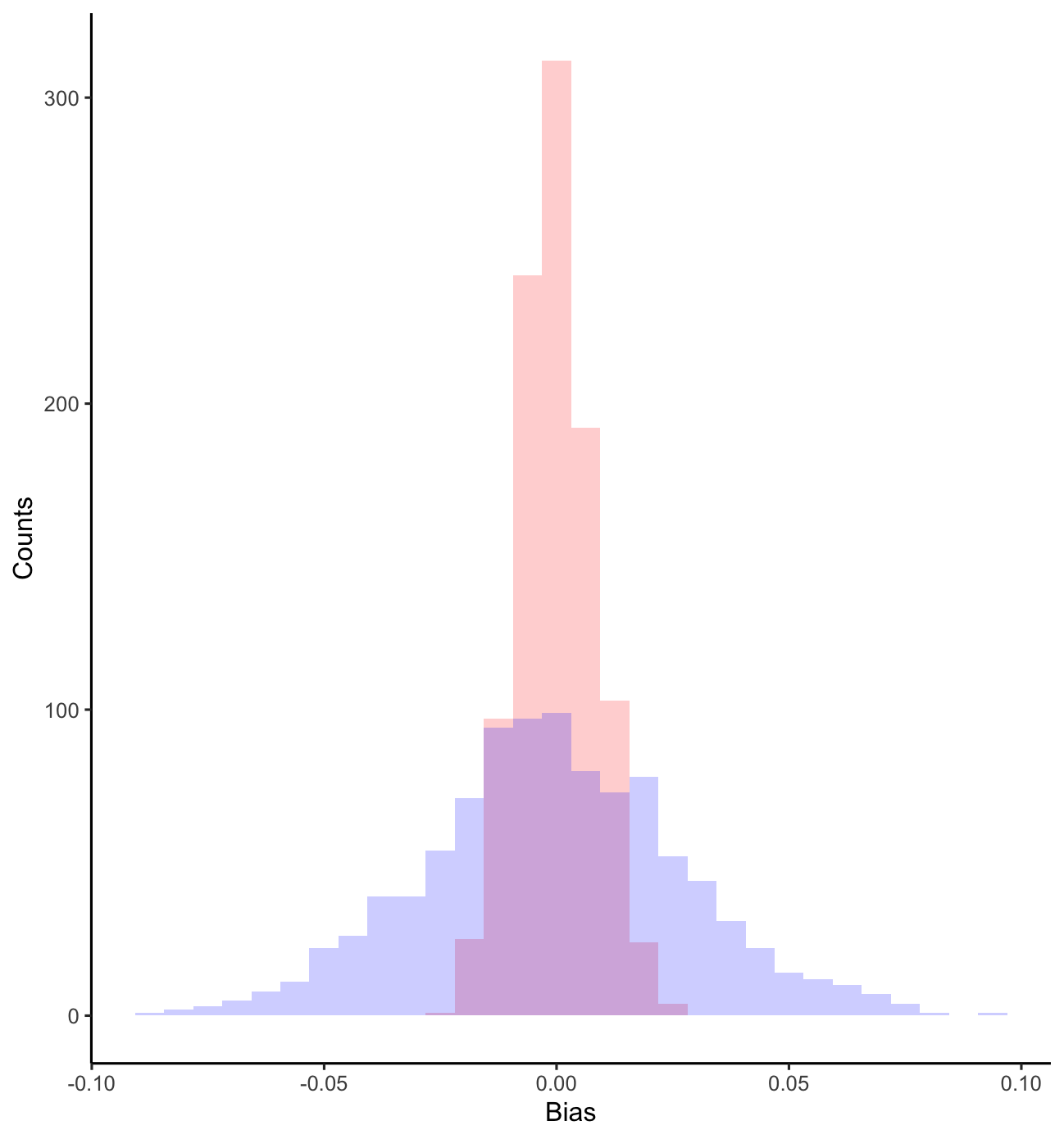}
  \caption{Fluctuations of error for a coordinate of the CLS (red) and OLS (blue) estimators with $n=3000, p = 300, q = 10$ over 1000 iterations }
    \label{fig:sec_ord_fluc}
\end{minipage}

\end{figure}

Fig. \ref{fig:error_comparison_high} [\textbf{s3,m1}] shows a similar conclusion as Fig. \ref{fig:error_comparison} but in a high dimensional case. Here, the comparison is between the oracle estimator $\hat{\boldsymbol{\beta}}_{\Sigma}$ and the projected oracle estimator $\hat{\boldsymbol{\beta}}_{\Sigma,\mathcal{P}}$ described in the high dimension part of Section \ref{methods}. 

Fig. \ref{fig:sec_ord_fluc} shows the fluctuations of a coordinate of the CLS and OLS estimators around the true value of the regression vector at this coordinate over 1000 iterations, for a dimension specification that is similar to our data. We see that the fluctuations are lesser for the CLS estimator, which means that there are still benefits to incorporating constraints in the analysis even if the number of constraints is much smaller than the number of covariates.

A suite of data-informed simulations, using the genetic datasets described in the Section \ref{subsec:data_analysis}, are available in the Appendix \ref{s:add_exp}.


\subsection{Genetic Datasets}\label{subsec:data_analysis}

\begin{figure}
  \centering
  \includegraphics[width = 0.75\textwidth]{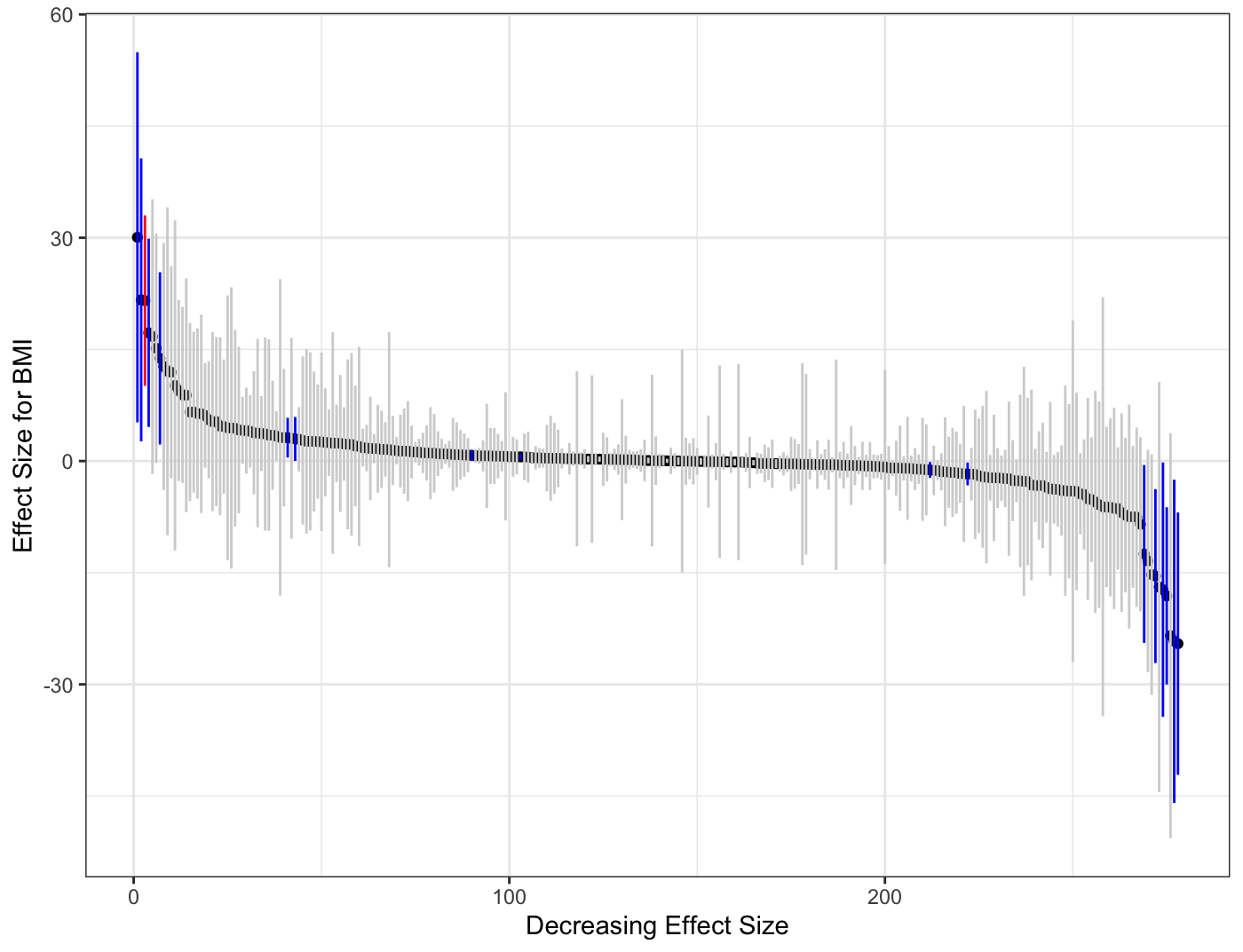}
  \caption{Estimated coefficients ordered by effect size and 95\% CIs of genetic variants in association with BMI, using $\tilde{\boldsymbol{\beta}}$. Blue and red CIs are significant before and after controlling for the family-wise error rate}
    \label{fig:ci_bmi}
\end{figure}

The remaining numerical analyses will revolve around the Jackson Heart Study (JHS) \citep{Jackson} and the Multi-Ethnic Study of Atherosclerosis (MESA) \citep{MESA} datasets. In what follows, we explain how we used the JHS dataset as our reference population, to extract information for building constraints, and the MESA dataset as our target population, to perform association analysis under these constraints.

The Jackson Heart Study (JHS) is a study with the purpose of establishing a single-site cohort study to identify the risk factors for cardiovascular disease in African American men and women in Mississippi. The total sample size is 5306 individuals, with a subsample of 2050 individuals ($N=2050$) with protein expression data, alongside metabolic traits including body-mass index (BMI), fasting insulin, and glycated hemoglobin (HbA1c). For each metabolic trait, we selected 10 proteins that are both highly heritable and have high genetic correlations to the relevant trait to define constraints \citep{Elgart_Goodman_Isasi_Chen_De_Vries_Xu_Manichaikul_Guo_Franceschini_Psaty_etal._2021, Tsai_Hrytsenko_Elgart_Tahir_Chen_Wilson_Gerszten_Sofer_2024}. For each trait, we performed GWAS of each of the proteins matched to this trait and identified genetic associations, i.e., single nucleotide polymorphisms (SNPs) associated with the proteins and their estimated effect sizes. The heritability and genetic correlations of each of these proteins, along with the processes of protein selection and selection of SNPs are detailed in the Appendix \ref{app:protein}.

The Multi-Ethnic Study of Atherosclerosis (MESA) is a longitudinal cohort study established with the objective of investigating the prevalence, correlates and progression of subclinical cardiovascular disease in an ethnically diverse population of over 6500 men and women in the United States. We used a sample of 3280 MESA individuals ($n=3280$) who participated in MESA Exam 5, and had available metabolic traits (BMI, fasting insulin, HbA1c). 

We applied the estimator from Section \ref{mod_known} to the MESA dataset, using the constraints derived from the JHS dataset, for BMI, fasting insulin and HbA1c. We explain the procedure focusing on BMI and show the results for all the traits. For BMI, after selecting protein-associated SNPs, there were $p=278$ genetic variants to consider. We construct the CLS estimator, utilizing the constraint information and covariance matrix from the JHS population, and used this estimator to make confidence intervals using the distribution described in Proposition \ref{eq:asymp_norm}. Fig. \ref{fig:ci_bmi} shows the coefficients of the CLS estimator, along with the confidence intervals. For each estimated coefficient, we calculated a p-value based on Proposition \ref{eq:asymp_norm} and used the Holm-Bonferroni correction method for multiple hypothesis testing, controlling for the family-wise error rate at the 0.05 level \citep{Holm_1979}. We rejected the null hypothesis of no association with BMI for one genetic variant (rs5510), which is associated with the Kallistatin protein. We apply a similar procedure for fasting insulin and HbA1c. For fasting insulin, out of $p=1080$ genetic variants, we identified 5 genetic variants (rs2006232, rs6740281, rs7210719, rs2277668, rs3218911), and for HbA1c, we identified 7 of the $p=539$ considered genetic variants (rs11240346, rs3901740, rs3767283, rs7137327, rs77920745, rs80228806, rs80228806). Table \ref{tab:sig_snps} lists the associated genetic variants for each metabolic trait, and the protein each variant was associated with in the JHS protein GWAS.

The selected genetic variant for BMI belongs to the gene SERPINA4 (potentially controlling gene expression \citep{Edwards_Hing_Perry_Blaisdell_Kopelman_Fathke_Plum_Newell_Allen_S._etal._2012}), which produces the protein Kallistatin, which is responsible for activity of the adrenal gland \citep{Wang_Song_Chen_Chao_Chao_1996}. This gland is responsible for steroid hormones, and thus, would have an effect on BMI \citep{Gateva_Assyov_Velikova_Kamenov_2017}. In fact, rs5510 is associated with the trait ``protein levels in obesity'' in the GWAS Catalog \citep{Cerezo_Sollis_Ji_Lewis_Abid_Bircan_Hall_Hayhurst_John_Mosaku_etal._2025}. For fasting insulin, the genetic variants belong to SARM1, MAP4K4, TMEM199, SEBOX and IL1R2 genes, all genes responsible for proteins for cellular function, which could have a potential downstream effect on pancreatic tissue. For HbA1c, the genetic variants belong to CNTN2, CHPT1 and PC genes, which are all responsible for proteins related to blood cell function. This could also have a potential downstream effect on the level of glycation of the hemoglobin. While these might be feasible etiological pathways, the variants isolated are not strongly associated with the given traits, which could be the result of relatively weaker signal in the genetic data. Additionally, there are intrinsically two levels of correction for multiplicity, one at the genome-wide level and one at the constrained level, so this procedure acts conservatively for isolating variants.

\begin{table}[]
    \centering
    \begin{tabular}{|c|c|c|}
    \hline
    BMI $(p=278)$ & Fasting Insulin $(p=1080)$& HbA1c $(p=539)$\\
    \hline
      rs5510 (KALLISTATIN)   & rs2006232 (VITRONECTIN) & rs3735169 (TIG\_2) \\
         & rs2277668 (VITRONECTIN)& rs3767283 (CNTN\_2)\\
              &  rs3218911 (IL\_1\_SRII)& rs3901740 (CNTN\_2)\\
         & rs6740281 (IL\_1\_SRII)& rs7137327 (CATF)\\
              &  rs7210719 (VITRONECTIN)& rs11240346 (CNTN\_2)\\
         & & rs77920745 (CATF)\\
         & & rs80228806 (CATF)\\
         \hline
    \end{tabular}
    \centering
    \caption{Significant associated genetic variants for each metabolic trait (with the associated protein) recovered after multiplicity corrections, along with number of genetic variants that reached genome-wide level of significance $(p)$}
    \label{tab:sig_snps}
\end{table}

\section{Discussion}
\label{discuss}

In this paper, we have developed methods for constrained linear regression, that arise when incorporating protein expression data into the pipeline of genetic effect estimation. We have shown the optimality of our methods through theory and simulations in moderately high dimensional cases, and provided $\sqrt{n}$ consistent estimators in the high dimensional case. These estimators were implemented on genetic datasets for the purpose of genetic discovery.

There are many natural extensions and pressing points to address with this work. In the high dimensional case, when $\Sigma$ is known, a oft-used framework for estimation is through the use of convex regularization and debiasing. Indeed, under the assumptions of additional structure, such as sparsity, a regularized estimator would have better performance, and knowledge of the dispersion matrix allows the construction of an estimator which minimizes asymptotic variance \citep{Bellec_Zhang_2022,bellec2023debiasing}. It can be shown that the methodology of these papers can be easily extended to the framework of constraints, warranting comparative studies. In the case where $\Sigma$ is not known, works such as \cite{kong2018estimating,Li_Sur_2023, Chen_Liu_Mukherjee_2024} take initial steps to the construction of potentially useful estimators, and there is scope for considering the impact of these works in a constrained setting. 

In terms of direct next steps, working towards a proper characterization of estimated constraint estimators and their errors would allow for accurate implementation of these estimators on real datasets. Additionally, there is a direct connection to problems in the transfer learning literature, and utilizing the tools of this paper to investigate problems of ``growing'' numbers of sources could be quite elucidating, in the age of online data. We keep these goals for future research directions.

\section*{Data Availability Statement}
Genetic data from the Jackson Heart Study is available via a data use agreement with the data base of genotypes and phenotypes (dbGaP), study accession phs000286. Whole genome sequencing data for JHS are available via data use agreement with dbGaP project “NHLBI TOPMed: The Jackson Heart Study (JHS)”, study accession phs000964, and for MESA they are available via dbGaP project “NHLBI TOPMed: MESA and MESA Family AA-CAC”, study accession phs001416. The JHS proteomics data set used in this analysis is available via a data use agreement with the JHS Data Coordinating Center (DCC), see website \url{https://www.jacksonheartstudy.org/}. 

\section*{Acknowledgements}
This work was supported by the National Institute of Diabetes and Digestive and Kidney Diseases R01DK081572. The Jackson Heart Study is supported by Contracts HHSN268201800010I, HHSN268201800011I, HHSN268201800012I, HHSN268201800013I, HHSN268201800014I, HHSN268201800015I from the National Heart Lung and Blood Institute (NHLBI) with additional support from the National Institute of Minority Health and Health Disparities (NIMHD). The authors also wish to thank the staffs and participants of the JHS. The views expressed in this manuscript are those of the authors and do not necessarily represent the views of the National Heart, Lung, and Blood Institute; the National Institute of Minority Health and Health Disparities (NIMHD); the National Institutes of Health; or the U.S. Department of Health and Human Services.  MESA and the MESA SHARe projects are conducted and supported by the National Heart, Lung, and Blood Institute (NHLBI) in collaboration with MESA investigators. Support for MESA is provided by contracts 75N92025D00022, 75N92020D00001, HHSN268201500003I, N01-HC-95159, 75N92025D00026, 75N92020D00005, N01-HC-95160, 75N92020D00002, N01-HC-95161, 75N92025D00024, 75N92020D00003, N01-HC-95162, 75N92025D00027, 75N92020D00006, N01-HC-95163, 75N92025D00025, 75N92020D00004, N01-HC-95164, 75N92025D00028, 75N92020D00007, N01-HC-95165, N01-HC-95166, N01-HC-95167, N01-HC-95168, N01-HC-95169, UL1-TR-000040, UL1-TR-001079, UL1-TR-001420, UL1TR001881, DK063491, and R01HL105756. Funding for SHARe genotyping was provided by NHLBI Contract N02-HL-64278.  Genotyping was performed at Affymetrix (Santa Clara, California, USA) and the Broad Institute of Harvard and MIT (Boston, Massachusetts, USA) using the Affymetrix Genome-Wide Human SNP Array 6.0. The authors thank the MESA participants and the MESA investigators and staff for their valuable contributions.  A full list of participating MESA investigators and 
institutions can be found at \url{http://www.mesa-nhlbi.org}.



\bibliographystyle{abbrvnat}
\bibliography{references}
\appendix
\section{Proof of Results}
\subsection{Proof of Theorem \ref{thm:minimax}}
\label{app:minimax}
We structure the proof into the following steps:
\begin{itemize}
    \item Reducing the constrained problem to an equivalent unconstrained problem
    \item Deducing the minimax risk of the unconstrained problem
    \item Showing the CLS estimator $\tilde{\boldsymbol{\beta}}$ matches the minimax risk
\end{itemize}

We will denote $\mathcal{N}(A)$ as the null space of $A$, and $\hat{\Sigma}_n = \frac{X^TX}{n}$. The loss function we will be dealing with is the squared error loss. Firstly, we will consider the constraint vector $\boldsymbol{c}$ to be equivalently equal to $0$. Subsequently, we will argue that the value of this vector does not affect the minimax error. 

Consider $\boldsymbol{\beta}^{*}$ such that $A\boldsymbol{\beta}^{*}=0$.  Clearly $\boldsymbol{\beta}^{*}\in\mathcal{N}(A)$, which is a subspace. Let $V\in\mathbb{R}^{p\times(p-q)}$ be an orthonormal basis of $\mathcal{N}(A)$. This assures the existence of a unique $\boldsymbol{\alpha}^{*}\in\mathbb{R}^{p-q}$ such that $\boldsymbol{\beta}^{*} = V\boldsymbol{\alpha}^{*}$. The mapping between $\boldsymbol{\beta}^{*}$ and $\boldsymbol{\alpha}^{*}$ is bijective, and $\boldsymbol{\alpha}^{*}$ is unconstrained in $\mathbb{R}^{p-q}$. Now, the constraint in our minimization is equivalent to the existence of such an $\boldsymbol{\alpha}^{*}$. Thus, we can write the following:
$$\min_{\boldsymbol{\beta}\in\mathcal{N}(A)}\left|\left|y-X\boldsymbol{\beta}\right|\right|^2=\min_{\boldsymbol{\alpha}\in\mathbb{R}^{p-q}}\left|\left|y-XV\boldsymbol{\alpha}\right|\right|^2$$

The second minimization is an ordinary least squares with design matrix $XV$.

Now, we can deduce the minimax risk of this unconstrained problem. We have the following lemma that follows from \cite{Mourtada_2022} and the analysis of the minimax risk for the OLS estimator

\begin{lemma}
\label{lem:ols_minimax}
    For the setup in (\ref{eq:main_setup}) with $A\boldsymbol{\beta}^{*}=0$ and $\boldsymbol{\beta}^{*}=V\boldsymbol{\alpha}^{*}$, where $V$ is an orthonormal basis of $\mathcal{N}(A)$, we have $$\inf_{\hat{\boldsymbol{\alpha}}}\sup_{\boldsymbol{\alpha}^*}\mathbb{E}\left|\left|\hat{\boldsymbol{\alpha}}-\boldsymbol{\alpha}^*\right|\right|^2=\frac{\sigma^2}{n}\mathbb{E}\left[\mathrm{Tr}\left(V^T\hat{\Sigma}_nV\right)^{-1}\right]$$
\end{lemma}


Now, let us reintroduce the constraint vector $c$. This converts the subspace $\mathcal{N}(A)$ into an affine space. Thus, if $\boldsymbol{\beta}$ satisfies $A\boldsymbol{\beta}=c$, then there exists $u_{\mathcal{N}},u_c$ such that $Au_{\mathcal{N}}=0,Au_c=c$ and $\boldsymbol{\beta}=u_{\mathcal{N}}+u_c$. If we limit $u_c$ to be of a specific form $A^{\dagger}c$, where $AA^\dagger=I_q$, then we have a bijective correspondence between $\boldsymbol{\beta}$ and $u_{\mathcal{N}}$.

$$\mathbb{E}\left|\left|\hat{\boldsymbol{\beta}}-\boldsymbol{\beta}^*\right|\right|^2=\mathbb{E}\left|\left|(\hat{\boldsymbol{\beta}}-u_c)-(\boldsymbol{\beta}^*-u_c)\right|\right|^2=\mathbb{E}\left|\left|V\left(\hat{\boldsymbol{\alpha}}-\boldsymbol{\alpha}^*\right)\right|\right|^2$$

Thus, we have the same error as the case where we have a degenerate constraint vector. 
This, along with Lemma \ref{lem:ols_minimax}, establishes the error optimality.

We will now show that the CLS estimator achieves this minimax risk. Going back to the original optimization problem, we would like to find the estimator
\begin{equation}
    \tilde{\boldsymbol{\beta}} = \underset{\boldsymbol{\beta}:A\boldsymbol{\beta}=c}{\arg\min}\left|\left|y-X\boldsymbol{\beta}\right|\right|^2
    \label{eq:opt_prob}
\end{equation}
The following lemma shows the form of the CLS estimator given in \cite{amemiya1985advanced} using the Lagrangian operator.
\begin{lemma}
\label{lem:cls_form}
    Given the optimization problem in (\ref{eq:opt_prob}) with the constraint of $A\boldsymbol{\beta}=\boldsymbol{c}$, the solution is the estimator given by $$\tilde{\boldsymbol{\beta}}=C_{A^\perp}\hat{\Sigma}_n^{-1}\left(\frac{X^Ty}{n}\right)+\hat{\Sigma}_n^{-1}A^T(A\hat{\Sigma}_n^{-1}A^T)^{-1}c$$
where $C_{A^\perp} = I-\hat{\Sigma}_n^{-1}\left(A^T\left(A\hat{\Sigma}_n^{-1}A^T\right)^{-1}A\right)$
\end{lemma}

We can evaluate the error of this estimator directly. Without loss of generality, we assume $\boldsymbol{c}=0$, which reduces the form to $\tilde{\boldsymbol{\beta}}=C_{A^\perp}\hat{\boldsymbol{\beta}}_{\mathrm{LS}}$, where $\hat{\boldsymbol{\beta}}_{\mathrm{LS}}$ is the standard OLS estimator. 

\begin{eqnarray*}
    \mathbb{E}\left|\left|\tilde{\boldsymbol{\beta}}-\boldsymbol{\beta}^*\right|\right|^2&=&\mathbb{E}\left|\left|C_{A^\perp}\hat{\boldsymbol{\beta}}_{\mathrm{LS}}-\boldsymbol{\beta}^*\right|\right|^2=\mathbb{E}\left|\left|C_{A^\perp}\hat{\boldsymbol{\beta}}_{\mathrm{LS}}-C_{A^\perp}\boldsymbol{\beta}^*\right|\right|^2\\
    &=&\frac{1}{n^2}\mathbb{E}\left|\left|\sum_{i=1}^nC_{A^\perp}\hat{\Sigma}_n^{-1}X_i\epsilon_i\right|\right|^2=\frac{1}{n^2}\mathbb{E}\left[\mathbb{E}\left[\left|\left|\sum_{i=1}^nC_{A^\perp}C_{A^\perp}^T\hat{\Sigma}_n^{-1}X_i\epsilon_i\right|\right|^2|X\right]\right]\\
    &=&\frac{\sigma^2}{n}\mathbb{E}\left[\mathrm{Tr}\left(\hat{\Sigma}_n^{-1}C_{A^\perp}C_{A^\perp}^T\right)\right]=\frac{\sigma^2}{n}\mathbb{E}\left[\mathrm{Tr}\left(C_{A^\perp}\hat{\Sigma}_n^{-1}\right)\right]
\end{eqnarray*}

Now, if we show that this error is equivalent to the error given in the minimax derivation, then the estimator $\tilde{\boldsymbol{\beta}}$ achieves the minimax error. We utilize the following lemma for this equivalence of errors

\begin{lemma}
\label{lem:lin_alg}
    Given matrices $A\in\mathbb{R}^{n\times d}$ and $B\in\mathbb{R}^{d\times n}$ such that $BA=I_d$. Then $X=AB$ uniquely satisfies the following properties: \begin{itemize}
        \item $X\in\mathbb{R}^{n\times n}$
        \item $\mathrm{rank}(X)=d$
        \item $XA=A$
        \item $BX=B$
    \end{itemize}
\end{lemma}

Given $AV=0$ and Lemma \ref{lem:lin_alg}, we have the following equations:
    $$C_{A^\perp}V = V
    \implies C_{A^\perp}V\left(V^T\hat{\Sigma}_nV\right)^{-1} = V\left(V^T\hat{\Sigma}_nV\right)^{-1}
    \implies V^T\hat{\Sigma}_nC_{A^\perp}=V^T\hat{\Sigma}_n$$

Also, we have $V^T\hat{\Sigma}_nV\left(V^T\hat{\Sigma}_nV\right)^{-1}=I_d$. Therefore, we have
$$
C_{A^\perp}=V\left(V^T\hat{\Sigma}_nV\right)^{-1}V^T\hat{\Sigma}_n\implies
\mathrm{Tr}\left(C_{A^\perp}\hat{\Sigma}_n^{-1}\right)=\mathrm{Tr}\left(V\left(V^T\hat{\Sigma}_nV\right)^{-1}V^T\right)=\mathrm{Tr}\left(V^T\hat{\Sigma}_nV\right)^{-1}
$$

This proves the equivalence of the errors from the previous two parts, which implies the theorem.

To prove Lemma \ref{lem:lin_alg}, note that $X$ and $A$ are both rank $d$ matrices. This implies that they share the same column space.

$$\exists Y\in\mathbb{R}^{d\times n}\;s.t.\; X=AY\implies
B=BX=BAY=Y
    \implies X=AB$$

\subsection{Minimax Optimality of CLS estimator (Proof of Lemma \ref{lem:ols_minimax})}
\label{app:cls_minimax}
We consider $A\boldsymbol{\beta}^{*}=0$ and $\boldsymbol{\beta}^{*}=V\boldsymbol{\alpha}^{*}$.

We denote the ordinary least squares estimator for $\boldsymbol{\alpha}^{*}$ as $\hat{\boldsymbol{\alpha}}_{\mathrm{LS}}$. 
This estimator provides an upper bound of the minimax error, which can be evaluated directly.

\begin{eqnarray*}
    \mathbb{E}\left|\left|\hat{\boldsymbol{\alpha}}_{\mathrm{LS}}-\boldsymbol{\alpha}^*\right|\right|^2&=&\mathbb{E}\left|\left|\left(V^T\hat{\Sigma}_{n}V\right)^{-1}\frac{1}{n}V^TX^T\epsilon\right|\right|^2\\
    &=&\frac{1}{n^2}\mathbb{E}\left|\left|\sum_{i=1}^n\left(V^T\hat{\Sigma}_nV\right)^{-1}\left(XV\right)_i\epsilon_i\right|\right|^2=\frac{1}{n^2}\mathbb{E}\left[\mathbb{E}\left[\left|\left|\sum_{i=1}^n\left(V^T\hat{\Sigma}_nV\right)^{-1}\left(XV\right)_i\epsilon_i\right|\right|^2|X\right]\right]\\
    &=&\frac{\sigma^2}{n^2}\mathbb{E}\left[\sum_{i=1}^n\boldsymbol{X}_i^TV\left(V^T\hat{\Sigma}_nV\right)^{-2}V^T\boldsymbol{X}_i\right]=\frac{\sigma^2}{n}\mathbb{E}\left[\mathrm{Tr}\left(V^T\hat{\Sigma}_nV\right)^{-1}\right]
\end{eqnarray*}

Hence, the minimax risk of estimation of $\boldsymbol{\alpha}^{*}$ is given by this quantity.

$$\inf_{\hat{\boldsymbol{\alpha}}}\sup_{\boldsymbol{\alpha}^*}\mathbb{E}\left|\left|\hat{\boldsymbol{\alpha}}-\boldsymbol{\alpha}^*\right|\right|^2\leq \sup_{\boldsymbol{\alpha}^*}\mathbb{E}\left|\left|\hat{\boldsymbol{\alpha}}_{\mathrm{LS}}-\boldsymbol{\alpha}^*\right|\right|^2=\frac{\sigma^2}{n}\mathbb{E}\left[\mathrm{Tr}\left(V^T\hat{\Sigma}_nV\right)^{-1}\right]$$

We will construct a lower bound for the minimax error considering the decision-theoretic problem of estimating $\boldsymbol{\alpha}^{*}$ under the squared error loss function, following the procedure given in \cite{Mourtada_2022}. Note that using the fact that $\boldsymbol{\alpha}^{*}\in\mathbb{R}^{p-q}$, we can think of the decision theoretic problem as unconstrained. 

We can consider a diffuse normal prior distribution on $\boldsymbol{\alpha}^{*}$, that is $\boldsymbol{\alpha}^{*}\sim \Pi_{\lambda}$ where $\Pi_{\lambda}=\mathcal{N}_{p-q}\left(0,\sigma^2/\lambda n\right)$. The posterior for $\boldsymbol{\alpha}^{*}$ will therefore follow the distribution $$\Pi_{\lambda}\left(\cdot\mid X,\boldsymbol{y}\right)=\mathcal{N}_{p-q}\left(\hat{\boldsymbol{\alpha}}_{\lambda,n},\left(\sigma^2/n\right)\left(V^T\hat{\Sigma}_nV+\lambda I_{p-q}\right)\right)$$where $\hat{\boldsymbol{\alpha}}_{\lambda,n}=\left(V^T\hat{\Sigma}_nV+\lambda I_{p-q}\right)^{-1}\frac{1}{n}V^TX^T\boldsymbol{y}$ is a ridge estimator. Also, for a squared error loss, the posterior mean is the minimizer of the loss. Hence, we will investigate $\mathbb{E}\mid\mid\hat{\boldsymbol{\alpha}}_{\lambda,n}-\boldsymbol{\alpha}^{*}\mid\mid^2$

\begin{eqnarray*}
    \mathbb{E}\mid\mid\hat{\boldsymbol{\alpha}}_{\lambda,n}-\boldsymbol{\alpha}^{*}\mid\mid^2&=&\mathbb{E}\left|\left|\left(V^T\hat{\Sigma}_nV+\lambda I_{p-q}\right)^{-1}\frac{1}{n}V^TX^T\boldsymbol{y}-\boldsymbol{\alpha}^{*}\right|\right|^2\\
    &=&\mathbb{E}\left|\left|\left(V^T\hat{\Sigma}_nV+\lambda I_{p-q}\right)^{-1}\frac{1}{n}\sum_{i=1}^n\epsilon_i\left(XV\right)_i -\lambda\left(V^T\hat{\Sigma}_nV+\lambda I_{p-q}\right)^{-1}\boldsymbol{\alpha}^{*}\right|\right|^2\\
    &=&\mathbb{E}\left[\mathbb{E}\left[\left|\left|\left(V^T\hat{\Sigma}_nV+\lambda I_{p-q}\right)^{-1}\frac{1}{n}\sum_{i=1}^n\epsilon_i\left(XV\right)_i -\lambda\left(V^T\hat{\Sigma}_nV+\lambda I_{p-q}\right)^{-1}\boldsymbol{\alpha}^{*}\right|\right|^2\right]\mid X\right]\\
    &=&\underbrace{\mathbb{E}\left|\left|\lambda\left(V^T\hat{\Sigma}_nV+\lambda I_{p-q}\right)^{-1}\boldsymbol{\alpha}^{*}\right|\right|^2}_{I}+\underbrace{\frac{\sigma^2}{n^2}\mathbb{E}\sum_{i=1}^n\left|\left|\left(V^T\hat{\Sigma}_nV+\lambda I_{p-q}\right)^{-1}\left(XV\right)_i\right|\right|^2}_{II}\\\\
    II&=&\frac{\sigma^2}{n^2}\mathbb{E}\sum_{i=1}^n\left|\left|\left(V^T\hat{\Sigma}_nV+\lambda I_{p-q}\right)^{-1}\left(XV\right)_i\right|\right|^2\\
    &=&\frac{\sigma^2}{n^2}\mathbb{E}\mathrm{Tr}\left(\left(V^T\hat{\Sigma}_nV+\lambda I_{p-q}\right)^{-2}\sum_{i=1}^n\left(V^T\boldsymbol{X}_i\boldsymbol{X}_i^TV\right)\right)\\
    &=&\frac{\sigma^2}{n}\mathbb{E}\mathrm{Tr}\left(\left(V^T\hat{\Sigma}_nV+\lambda I_{p-q}\right)^{-2}\left(V^T\hat{\Sigma}_nV\right)\right)\\
    I&=&\mathbb{E}\left|\left|\lambda\left(V^T\hat{\Sigma}_nV+\lambda I_{p-q}\right)^{-1}\boldsymbol{\alpha}^{*}\right|\right|^2\\
    &=&\mathbb{E}\left[\mathbb{E}_{\Pi_{\lambda}}\left|\left|\lambda\left(V^T\hat{\Sigma}_nV+\lambda I_{p-q}\right)^{-1}\boldsymbol{\alpha}^{*}\right|\right|^2\mid X\right]\text{ by Fubini's Theorem}\\
    &=& \frac{\sigma^2}{n}\mathbb{E}\mathrm{Tr}\left(\left(V^T\hat{\Sigma}_nV+\lambda I_{p-q}\right)^{-2}\right)\\
    \implies I+II&=&\frac{\sigma^2}{n}\mathbb{E}\left[\mathrm{Tr}\left(V^T\hat{\Sigma}_nV+\lambda I_{p-q}\right)^{-1}\right]
\end{eqnarray*}
This provides a lower bound on the risk. For all $\lambda>0$,
$$\inf_{\hat{\boldsymbol{\alpha}}}\sup_{\boldsymbol{\alpha}^*}\mathbb{E}\left|\left|\hat{\boldsymbol{\alpha}}-\boldsymbol{\alpha}^*\right|\right|^2\geq \inf_{\boldsymbol{\alpha}}\mathbb{E}_{\boldsymbol{\alpha}^*\sim\Pi_{\lambda}}\left|\left|\hat{\boldsymbol{\alpha}}-\boldsymbol{\alpha}^*\right|\right|^2=\mathbb{E}_{\boldsymbol{\alpha}^*\sim\Pi_{\lambda}}\left|\left|\hat{\boldsymbol{\alpha}}_{\lambda,n}-\boldsymbol{\alpha}^*\right|\right|^2=\frac{\sigma^2}{n}\mathbb{E}\left[\mathrm{Tr}\left(V^T\hat{\Sigma}_nV+\lambda I_{p-q}\right)^{-1}\right]$$

Now, this risk is well defined for all $\lambda>0$. Additionally, it is a decreasing function of $\lambda$ and it is strictly bounded below by $0$. Therefore, by monotone convergence theorem, we have $\lim_{\lambda\rightarrow0}\frac{\sigma^2}{n}\mathbb{E}\left[\mathrm{Tr}\left(V^T\hat{\Sigma}_nV+\lambda I_{p-q}\right)^{-1}\right]=\frac{\sigma^2}{n}\mathbb{E}\left[\mathrm{Tr}\left(V^T\hat{\Sigma}_nV\right)^{-1}\right]$, and this will act as a lower bound as well. Therefore, we have 
$$\inf_{\hat{\boldsymbol{\alpha}}}\sup_{\boldsymbol{\alpha}^*}\mathbb{E}\left|\left|\boldsymbol{\alpha}-\boldsymbol{\alpha}^*\right|\right|^2=\frac{\sigma^2}{n}\mathbb{E}\left[\mathrm{Tr}\left(V^T\hat{\Sigma}_nV\right)^{-1}\right]$$

\subsection{Constrained Least Squares estimator derivation (Proof of Lemma \ref{lem:cls_form})}
\label{app:cls_form}
We have the following optimization problem 
$$\min||\boldsymbol{y}-X\boldsymbol{\beta}||^2\text{ subject to }A\boldsymbol{\beta}=\boldsymbol{c}$$

We label our objective function as $L(\boldsymbol{\beta}) = \left(\boldsymbol{y}-X\boldsymbol{\beta}\right)^T\left(\boldsymbol{y}-X\boldsymbol{\beta}\right)$. We know, by definition, that $L(\boldsymbol{\beta})$ is minimized by $\hat{\boldsymbol{\beta}}_{\text{LS}}$. We can expand the objective around this minimizer as follows:
$$L(\boldsymbol{\beta})=L(\hat{\boldsymbol{\beta}})+\left(\boldsymbol{\beta}-\hat{\boldsymbol{\beta}}\right)^TX^TX\left(\boldsymbol{\beta}-\hat{\boldsymbol{\beta}}\right)$$Denote $\boldsymbol{\beta}-\hat{\boldsymbol{\beta}}$ as $\boldsymbol{\delta}$ and $A\hat{\boldsymbol{\beta}}-\boldsymbol{c}=\boldsymbol{\gamma}$. Then the optimization problem becomes 
$$\min \boldsymbol{\delta}^TX^TX\boldsymbol{\delta}\text{ subject to }A\boldsymbol{\delta}=\boldsymbol{\gamma}$$
We use a Lagrangian method to optimize this by setting the new objective to $$L(\boldsymbol{\delta},\boldsymbol{\lambda}) = \boldsymbol{\delta}^TX^TX\boldsymbol{\delta}+\boldsymbol{\lambda}^T\left(A\boldsymbol{\delta}-\boldsymbol{\gamma}\right)$$
Taking the derivative with respect to $\boldsymbol{\delta}$ and the multiplier parameter and setting to zero, we get 

\begin{eqnarray*}
\frac{\partial L(\boldsymbol{\delta},\boldsymbol{\lambda})}{\partial\boldsymbol{\lambda}}&=& A\boldsymbol{\delta}-\boldsymbol{\gamma}:=0\\
    \frac{\partial L(\boldsymbol{\delta},\boldsymbol{\lambda})}{\partial\boldsymbol{\delta}}&=&2\boldsymbol{\delta}^TX^TX+\lambda^TA:=0\\
   2\boldsymbol{\delta}=-\left(X^TX\right)^{-1}A^T\lambda
    &\implies& \lambda =-2\left(A\left(X^TX\right)^{-1}A^T\right)^{-1}\gamma\\
    \implies \boldsymbol{\delta}&=&\left(X^TX\right)^{-1}A^T\left(A\left(X^TX\right)^{-1}A^T\right)^{-1}\gamma\\
\end{eqnarray*}
Translating this solution into the original formulation, we get 
$$\tilde{\boldsymbol{\beta}}=\hat{\boldsymbol{\beta}}+\left(X^TX\right)^{-1}A^T\left(A\left(X^TX\right)^{-1}A^T\right)^{-1}\left(A\hat{\boldsymbol{\beta}}-c\right)$$which is our desired estimator.
\subsection{Asymptotics of Minimax Error (Proof of Corollary \ref{thm:asymp_minimax})}
\label{app:asymp_minimax}
We define $D\in\mathbb{R}^{p\times p}$ as the deterministic equivalent of a random matrix $\tilde{D}\in\mathbb{R}^{p\times p}$, if, for any unit vectors $\boldsymbol{a},\boldsymbol{b}\in\mathbb{R}^p$ we have 
$$\frac{1}{p}\mathrm{Tr}\left(D-\tilde{D}\right)=o_{\mathbb{P}}(1)\text{ and }\boldsymbol{a}^T\left(D-\tilde{D}\right)\boldsymbol{b}=o_{\mathbb{P}}(1)$$

We will be using Theorem 4.1 and Corollary 4.2 from \cite{Dobriban_Sheng_2022}, a consequence of \cite{Rubio_Mestre_2011}, which gives the notion of asymptotic equivalence for the inverse of a covariance matrix using deterministic equivalents. 
\begin{lemma}[Corollary 4.2 in \cite{Dobriban_Sheng_2022}]
\label{lem:det_eq}
    $$\frac{1}{p}\mathrm{Tr}\left(\hat{\Sigma}^{-1}-\frac{\Sigma^{-1}}{1-\alpha}\right)=o_{\mathbb{P}}(1)\text{ and }\boldsymbol{a}^T\left(\hat{\Sigma}^{-1}-\frac{\Sigma^{-1}}{1-\alpha}\right)\boldsymbol{b}=o_{\mathbb{P}}(1)$$ for vectors $\boldsymbol{a},\boldsymbol{b}$ bounded in $\ell_2$ norm.
\end{lemma}

In the given setup, we have $\boldsymbol{X}_i$ being drawn from $N_p(0,\Sigma)$, and our estimate for the covariance matrix is given by $\hat{\Sigma}_n=\frac{1}{n}\sum_{i=1}^n\boldsymbol{X}_i\boldsymbol{X}_i^T$.
We define $\tilde{X}_i=V^T\boldsymbol{X}_i\sim N_{p-q}\left(0,V^T\Sigma V\right)$. Thus, we have 
$$\frac{1}{n}\sum_{i=1}^n\tilde{X}_i\tilde{X}_i^T=V^T\hat{\Sigma}_nV$$

\textbf{Remark:} We appeal to this theory of deterministic equivalents, put forth in \cite{Hachem_Loubaton_Najim_2007} to deal with the asymptotics associated with these sample covariance matrices and their inverses. We utilize assumptions from \cite{Hachem_Loubaton_Najim_2007}, that is $X$ is equivalent in distribution to $\Sigma^{1/2}Z$, where entries of $Z$ are mean zero, unit variance and have finite $8+c$th moment where $c>0$. Additionally, in either formulation of Assumption \ref{as:rd}, we have $\mu_{\hat{\Sigma}_n}$ converges to $\mathrm{MP}(\alpha) \boxtimes \mu_{\infty}$, where $\mu_{\hat{\Sigma}_n}$ is the empirical spectral distribution of $\hat{\Sigma}_n$, $\mu_{\infty}$ is the limiting empirical spectral distribution of $\Sigma$, and $\boxtimes$ denotes the free multiplicative convolution function \cite{Bai_Zhou_2008, Chouard_2022}. Recent works \citep{Louart_Couillet_2021,Chouard_2022} have allowed us to utilize the deterministic equivalents for weaker assumptions, which we have utilized as Assumption \ref{as:rd}(a). Namely, we assume that every row of $X$ are i.i.d. distributed from a centered distribution, any contrast of the form $a^TX$ is subgaussian with variance proxy $a^T\Sigma a$, and $\Sigma$ has bounded spectral norm.

By Lemma \ref{lem:det_eq}, we have the following result, by definition of deterministic equivalent.
$$\frac{1}{n}\mathrm{Tr}\left(V^T\hat{\Sigma}_nV\right)^{-1}=\frac{\mathrm{Tr}\left(V^T\Sigma V\right)^{-1}}{n(1-(1-\gamma)\alpha)}+o_{\mathbb{P}}(1)$$

This gives a convergence in probability. In order to show a convergence of the expectation, we must show uniform integrability of $T_n = \frac{1}{n}\mathrm{Tr}\left(V^T\hat{\Sigma}_nV\right)^{-1}$ for all $n$. We leave this as a claim and then prove it at the end of this section.


Also, by the equivalence proven by the theorem, we know that 
$$\lim_{n\rightarrow\infty}\frac{1}{n}\mathrm{Tr}\left(V^T\Sigma V\right)^{-1}=\lim_{n\rightarrow\infty}\frac{1}{n}\mathrm{Tr}\left(C_\Sigma\Sigma^{-1}\right)$$
where $C_\Sigma =I_p-\Sigma^{-1}A\left(A^T\Sigma^{-1}A\right)^{-1}A^T$, which are all positive random variables. 

Thus, we have $$\lim_{n\rightarrow\infty}\mathbb{E}\left|\left|\tilde{\boldsymbol{\beta}}-\boldsymbol{\beta}^*\right|\right|^2=\frac{\sigma^2}{(1-(1-\gamma)\alpha)}\lim_{n\rightarrow\infty}\frac{1}{n}\mathrm{Tr}\left(\Sigma^{-1}-\Sigma^{-1}\left(A^T\left(A\Sigma^{-1}A^T\right)^{-1}A\right)\Sigma^{-1}\right)$$

We assume the existence of the limit on the right hand side. This is reasonable since the limiting spectrum of $\Sigma$ is bounded away from $0$ and $\infty$, allowing for invertibility. 
Plugging in $\Sigma = I_p$ for the isotropic case gives us the corresponding result.

Now, we prove uniform integrability of $T_n$. Since we are operating in a regime where $n$ is bounded below by $p$, we will show the result for all $n$ large enough, leaving a finite complement, which allows for uniform integrability. We will utilize the De la Vallée Poussin's condition for uniform integrability. That is, if we can show that $T_n^{1+\delta}$ is bounded for some $\delta>0$, this is an equivalent definition of uniform integrability. 

Note that for $A_n$ a standard Wishart matrix, we can write $T_n = \mathrm{Tr}\left(\left(V^T\Sigma V\right)^{-1/2}A_n\left(V^T\Sigma V\right)^{-1/2}\right)$, which means we can bound $$\frac{1}{\lambda_{\max}(V^T\Sigma V)}\mathrm{Tr}(A_n^{-1})\leq T_n\leq \frac{1}{\lambda_{\min}(V^T\Sigma V)}\mathrm{Tr}(A_n^{-1})$$

This shows that it is enough to show UI for $\tilde{T}_n:=\mathrm{Tr}(A_n^{-1})$, since we have constant bounds on either side. 
For a standard Wishart matrix, using the Schur complement, we can show that $(A_n)_{jj}\sim\chi^2_{n-p+q+1}$. For large enough $n$ and $\delta>0$, we have 
\begin{eqnarray*}
    \mathbb{E}\left[\left(\mathrm{Tr}\left(A_n^{-1}\right)\right)^{1+\delta}\right]&\leq& \left(p-q\right)^{\delta}\sum_{j=1}^{p-q}\mathbb{E}\left[\left(A_n^{-1}\right)_{jj}^{1+\delta}\right]\\
    &=&(p-q)^{1+\delta}\mathbb{E}\left[\left(\chi^2_{n-p+q+1}\right)^{-1-\delta}\right]\\
    &=&(p-q)^{1+\delta}2^{-1-\delta}\frac{\Gamma\left(\frac{n-p+q+1}{2}-1-\delta\right)}{\Gamma\left(\frac{n-p+q+1}{2}\right)}
\end{eqnarray*}

The last equality holds if $\frac{n-p+q+1}{2}>1+\delta$

Using Gautschi's inequality \citep{Gautschi_1959}, we can show that for a large enough constant $C$, we have $$\mathbb{E}\left[\left(\mathrm{Tr}\left(A_n^{-1}\right)\right)^{1+\delta}\right]\leq C\left(p-q\right)^{1+\delta}n^{-1-\delta}<\infty$$

This shows that $\tilde{T}_n^{1+\delta}$ is bounded for all $n$ such that $\frac{n-p+q+1}{2}>1+\delta$. Note that uniform integrability holds for any finite set of $\mathcal{O}(1)$ variables. Thus, we have shown uniform integrability of $\tilde{T}_n$ for all $n>p$, which implies uniform integrability of $T_n$ for all $n>p$.
\subsection{``Gain'' of using Constraints when $q\ll p$ (Proof of Proposition 1)}
\label{app:low_q}
We are analyzing the ``gain'' \begin{eqnarray*}
    G_n &=& ||\hat{\boldsymbol{\beta}}_{LS}-\boldsymbol{\beta}^{*}||^2-||\hat{\boldsymbol{\beta}}_{\mathcal{P}}-\boldsymbol{\beta}^{*}||^2\\
    &=&||\left(X^TX\right)^{-1}X^T\boldsymbol{\epsilon}||^2-||\mathcal{P}_A^{\perp}\left(X^TX\right)^{-1}X^T\boldsymbol{\epsilon}||^2\\
    &=&||\mathcal{P}_A\left(X^TX\right)^{-1}X^T\boldsymbol{\epsilon}||^2\\
    &=&\boldsymbol{\epsilon}^TX\left(X^TX\right)^{-1}\mathcal{P}_A\left(X^TX\right)^{-1}X^T\boldsymbol{\epsilon}
\end{eqnarray*}
We have $\mathbb{E}\left[G_n\mid X\right] = \sigma^2\mathrm{Tr}\left(\mathcal{P}_{A}\left(X^TX\right)^{-1}\right) = \frac{\sigma^2}{n}\mathrm{Tr}\left(\mathcal{P}_{A}\hat{\Sigma}_n^{-1}\right)$

We will use deterministic equivalents as given in Appendix \ref{app:asymp_minimax} to evaluate this expectation.
\begin{eqnarray*}
    \mathbb{E}\left[G_n\mid X\right] &=& \frac{\sigma^2}{n}\mathrm{Tr}\left(\mathcal{P}_{A}\hat{\Sigma}_n^{-1}\right)\\
     &=& \sigma^2\underbrace{\frac{1}{n}\left[\mathrm{Tr}\left(\mathcal{P}_{A}\hat{\Sigma}_n^{-1}\right)-\frac{1}{1-\alpha}\mathrm{Tr}\left(\mathcal{P}_{A}\Sigma^{-1}\right)\right]}_{g_1}+\frac{\sigma^2}{n(1-\alpha)}\mathrm{Tr}\left(\mathcal{P}_{A}\Sigma^{-1}\right)\\
\end{eqnarray*}
Given Corollary 4.2 from \cite{Dobriban_Sheng_2022}, we see that $g_1 = \mathcal{O}_{\mathbb{P}}\left(\frac{||\mathcal{P}_A||_{op}}{n\sqrt{n}}\right)$. Considering a rescaling, we see that $$\mathbb{E}\left[\frac{n}{q}G_n\mid X\right] = \underbrace{\frac{1}{q(1-\alpha)}\mathrm{Tr}\left(\mathcal{P}_A\Sigma^{-1}\right)}_{g_2}+\mathcal{O}_{\mathbb{P}}\left(\frac{||\mathcal{P}_A||_{op}}{q\sqrt{n}}\right)$$
Under isotropic design, we have $g_2 = \frac{\sigma^2}{q(1-\alpha)}\mathrm{Tr}\left(\mathcal{P}_{A}\Sigma^{-1}\right) = \frac{\sigma^2}{q(1-\alpha)}\mathrm{Tr}\left(\mathcal{P}_{A}\right) = \frac{\sigma^2}{1-\alpha}$.
The proof of uniform integrability of $\left\{\mathbb{E}\left[\frac{n}{q}G_n\mid X\right]\right\}_{n\geq 1}$ follows quite directly from the analysis of $T_n$ in the previous section. Indeed, for some $\delta>0$, we have $$|\mathbb{E}\left[G_n\mid X\right]|^{1+\delta}=\left|\frac{\sigma^2}{n}\mathrm{Tr}\left(\mathcal{P}_A\hat{\Sigma}_n^{-1}\right)\right|^{1+\delta}\leq \sigma^{2+2\delta}\left|\frac{1}{n}\mathrm{Tr}\left(\hat{\Sigma}_n^{-1}\right)\right|^{1+\delta}=\sigma^{2+2\delta}\left|\mathrm{Tr}\left(A_n^{-1}\right)\right|^{1+\delta}$$where $A_n$ is a standard Wishart matrix. From the previous section, we have shown the uniform integrability of these trace objects, which in turn gives us the uniform integrability of $\left\{\mathbb{E}\left[\frac{n}{q}G_n\mid X\right]\right\}_{n\geq 1}$. Given this uniform integrability, we can assert that $\underset{n\rightarrow\infty}{\lim}\mathbb{E}\left[\frac{n}{q}G_n\right] = \frac{\sigma^2}{1-\alpha}$. Therefore, we have $\mathbb{E}\left[G_n\right] = \frac{q}{n(1-\alpha)}(1+o(1))$. We will evaluate the variance of $G_n$ now.
\begin{eqnarray*}
    \mathrm{Var}\left(G_n\right)&=&2\sigma^4\mathrm{Tr}\left(X\left(X^TX\right)^{-1}\mathcal{P}_{A}\left(X^TX\right)^{-1}X^TX\left(X^TX\right)^{-1}\mathcal{P}_{A}\left(X^TX\right)^{-1}X^T\right)\\
    &=&\frac{2\sigma^4}{n^2}\underbrace{\mathrm{Tr}\left(\left(\mathcal{P}_{A}\hat{\Sigma}_n^{-1}\right)^2\right)}_{\mathcal{O}(q)}
\end{eqnarray*}
Thus, we have \begin{eqnarray*}
    \mathbb{P}\left(|G_n|>\varepsilon\mathbb{E}|G_n|\right)\leq\frac{\mathrm{Var}(G_n)}{\varepsilon^2\mathbb{E}^2|G_n|}=\frac{1}{2\sigma^2\varepsilon^2q}\rightarrow 0 
\end{eqnarray*}
if $q\rightarrow0$. Therefore, $G_n\stackrel{\mathbb{P}}{\rightarrow}\frac{\sigma^2\gamma\alpha}{1-\alpha}$.

Now, we consider the case where $q=\mathcal{O}(1)$. We additionally assume Assumption \ref{as:err}. Therefore, $G_n$ is a quadratic form. Let us denote $Q = X\left(X^TX\right)^{-1}\mathcal{P}_A\left(X^TX\right)^{-1}X^T$. Let us diagonalize $Q = UW U^T$, where $U$ is an orthogonal matrix and $W$ is a diagonal matrix of eigenvalues $w_1\geq\ldots\geq w_n$ of $Q$. Note that $w_{q+1}=\ldots=w_n=0$. Since $\boldsymbol{\epsilon}/\sigma\sim N_n(0,I_n)$, we have $U^T\boldsymbol{\epsilon}/\sigma\sim N_n(0,I_n)$. Therefore, we have $\frac{G_n}{\sigma^2}\stackrel{d}{=}\boldsymbol{v}^TW\boldsymbol{v}$ where $\boldsymbol{v}\sim N_n(0,I_n)$. Therefore, $$\frac{G_n}{\sigma^2}\stackrel{d}{=}\sum_{i = 1}^qw_iv_i^2$$which gives us the necessary result. Also, in order to have non-vanishing eigenvalues, we scale $G_n$ by $n$.
\subsection{Asymptotic Normality of $\tilde{\boldsymbol{\beta}}$ (Proof of Proposition \ref{eq:asymp_norm})}
\label{app:asymp_norm}
We can establish the asymptotic normality of the CLS estimator using the independence and identical distribution of each of the errors. We operate under the assumption that the errors are independent of the draws of the design matrix. The form of the CLS estimator for the $j$th coordinate is 
$$\tilde{\beta}_j=\beta_j+\boldsymbol{e}_j^TC_{A^\perp}\hat{\Sigma}_n^{-1}\left(\frac{1}{n}X^T\boldsymbol{\epsilon}\right)$$
Rearranging, we have 
$$\sqrt{n}\left(\tilde{\beta}_j-\beta_j\right)=\boldsymbol{e}_j^TC_{A^\perp}\hat{\Sigma}_n^{-1}\left(\frac{1}{\sqrt{n}}X^T\boldsymbol{\epsilon}\right)=\sum_{i=1}^na_{n,i}\epsilon_i$$

where $a_{n,i}=\frac{1}{\sqrt{n}}\boldsymbol{e}_i^TX\hat{\Sigma}_n^{-1}C_{A^\perp}^T\boldsymbol{e}_j$, $\epsilon_i$ are iid with $\mathbb{E}[\epsilon_i]=0$, $\mathrm{Var}\left(\epsilon_i\right)=\sigma^2$ and $\mathbb{E}[\epsilon^{2+\delta}]<\infty$ for $\delta$ given in (\ref{eq:main_setup}). 
We can invoke the CLT if this quantity satisfies the Lyapunov condition \citep{Billingsley_1986}. Note that $a_{n,i}$ are random variables independent of $\epsilon_i$, therefore, we will invoke a conditional version of the Lyapunov CLT and then finally resolve the distribution of the $a_{n,i}$'s. This will provide sufficient conditions on $X$ that would allow for asymptotic normality. Let us denote $S_n=\sum_{i = 1}^na_{n,i}\epsilon_i$ and $v_n^2 = \sigma^2\sum_{i=1}^na_{n,i}^2$. We have the following lemma that provides conditions on $a_{n,i}$.

\begin{lemma}
\label{lem:rand_wts}
\begin{itemize}
    \item[(i)]$$\boldsymbol{e}_j^TC_{A^\perp}\hat{\Sigma}_n^{-1} C_{A^\perp}^T\boldsymbol{e}_j>0$$ with probability $1$
    \item[(ii)]$$\lim_{n\rightarrow\infty}\left(\boldsymbol{e}_j^TC_{A^\perp}\hat{\Sigma}_n^{-1} C_{A^\perp}^T\boldsymbol{e}_j-\frac{1}{1-(1-\gamma)\alpha}\boldsymbol{e}_j^TC_\Sigma\Sigma^{-1} C_\Sigma^T\boldsymbol{e}_j\right)=0$$
    
    \item[(iii)]$$\frac{\stackrel{n}{\underset{i=1}{\sum}} \left|\frac{1}{\sqrt{n}}\boldsymbol{e}_i^TX\hat{\Sigma}_n^{-1}C_{A^\perp}^T\boldsymbol{e}_j\right|^{2+\delta}}{\left(\boldsymbol{e}_j^TC_{A^\perp}\hat{\Sigma}_n^{-1}C_{A^\perp}^T\boldsymbol{e}_j\right)^{1+\delta/2}}\stackrel{\mathbb{P}}{\rightarrow}0$$for $\delta$ given in (\ref{eq:main_setup})
\end{itemize}
\end{lemma}

We can see that given Lemma \ref{lem:rand_wts}, we have $v_n>0$ w.p. 1 and $v_n^2$ converges to a constant (in probability) as $n\rightarrow\infty$. Let $\mathcal{A}_n$ denote the $\sigma$-field generated from $\{a_{n,i}\}$. Considering a conditional version of the scaled sum
\begin{eqnarray*}
    \Lambda_n:=\frac{1}{v_n^{2+\delta}}\sum_{i=1}^n\mathbb{E}\left[|a_{n,i}\epsilon_i|^{2+\delta}\mid\mathcal{A}_n\right]&=&\frac{1}{v_n^{2+\delta}}\sum_{i=1}^n|a_{n,i}|^{2+\delta}\mathbb{E}\left[|\epsilon_i|^{2+\delta}\right]\\
    &=&\underbrace{\frac{\mathbb{E}\left[|\epsilon_1|^{2+\delta}\right]}{\sigma_n^{2+\delta}}}_{<\infty}\frac{\sum_{i=1}^n|a_{n,i}|^{2+\delta}}{\left(\sum_{i=1}^na_{n,i}^2\right)^{1+\delta/2}}\\
    &\stackrel{\mathbb{P}}{\rightarrow}& 0\text{ by Lemma \ref{lem:rand_wts}(iii)}
\end{eqnarray*}
We will use the standard Lyapunov CLT on the event that $\{\Lambda_n\leq\eta\}$ for any $\eta>0$. We have $$\frac{S_n}{v_n}\mid\mathcal{A}_n\stackrel{d}{\rightarrow}N(0,1)\text{ on }\{\Lambda_n\leq\eta\}$$by Lyapunov CLT. Since $\Lambda_n\stackrel{\mathbb{P}}{\rightarrow}0$, we have uniform convergence $$\sup_{x\in\mathbb{R}}\left|\mathbb{P}\left(\frac{S_n}{v_n}\mid\mathcal{A}_n\right)-\Phi(x)\right|\stackrel{\mathbb{P}}{\rightarrow}0$$

Now, for any bounded, uniformly continuous $f$, 
$$\mathbb{E}\left[f\left(\frac{S_n}{v_n}\right)\right]=\mathbb{E}\left[\mathbb{E}\left(f\left(\frac{S_n}{v_n}\right)\mid X\right)\right]\rightarrow\mathbb{E}[f(Z)]$$where $Z\sim N(0,1)$ and $\Phi$ is the CDF of the standard normal distribution. This gives us the asymptotic normality. Now we need to prove Lemma \ref{lem:rand_wts}. Lemma \ref{lem:rand_wts}(ii) follows from Lemma \ref{lem:det_eq} using deterministic equivalents. We will prove Lemma \ref{lem:rand_wts}(iii) and in turn show (i).

Note that we have the following inequality
$$\sum_{i=1}^n|a_{n,i}|^{2+\delta}\leq \underset{1\leq i\leq n}{\max}|a_{n,i}|^\delta\sum_{i=1}^na_{n,i}^2$$Therefore, it is enough for us to verify $$\left(\frac{\underset{1\leq i\leq n}{\max} |a_{n,i}|}{\left(\sum_{j=1}^na_j^2\right)^{1/2}}\right)^\delta\stackrel{\mathbb{P}}{\rightarrow}0\text{ as }n\rightarrow\infty$$

We have $$\frac{\underset{1\leq i\leq n}{\max} \left|\frac{1}{\sqrt{n}}\boldsymbol{e}_i^TX\hat{\Sigma}_n^{-1}C_{A^\perp}^T\boldsymbol{e}_j\right|}{\left(\boldsymbol{e}_j^TC_{A^\perp}\hat{\Sigma}_n^{-1}C_{A^\perp}^T\boldsymbol{e}_j\right)^{1/2}}=\frac{A_1}{A_2}$$

If we show $A_1\stackrel{\mathbb{P}}{\rightarrow} 0$ as $n \rightarrow\infty$ and show that $A_2$ is bounded away from $0$ with high probability, then we would have shown that the Lindeberg condition holds. 
$$A_1=\frac{\underset{1\leq i\leq n}{\max} \left|\boldsymbol{e}_i^TX\hat{\Sigma}_n^{-1}C_{A^\perp}^T\boldsymbol{e}_j\right|}{\sqrt{n}}\leq\frac{\lambda_{\max}\left(\hat{\Sigma}_n^{-1}C_{A^\perp}^T\right)}{\sqrt{n}\underset{1\leq i\leq n}{\min}||\boldsymbol{e}_i^TX||_1}=\frac{B_1}{B_2}$$

We use the variational form of the largest eigenvalue and the properties of the deterministic equivalent from Lemma \ref{lem:det_eq}. Define $\boldsymbol{a}=\underset{||\boldsymbol{v}||=1}{\text{argmax }} \boldsymbol{v}^T\hat{\Sigma}_nC_{A^{\perp}}^T\boldsymbol{v}$.

We have
\begin{eqnarray*}
    B_1&=&\boldsymbol{a}^T\hat{\Sigma}_nC_{A^{\perp}}^T\boldsymbol{a}\\
    &=&\frac{\boldsymbol{a}^TC_\Sigma\Sigma^{-1}\boldsymbol{a}}{1-(1-\gamma)\alpha}+o_{\mathbb{P}}(1)\text{ from Lemma \ref{lem:det_eq}}\\
    &\leq&\frac{\lambda_{\max}\left(C_\Sigma\Sigma^{-1}\right)}{1-(1-\gamma)\alpha}+o_{\mathbb{P}}(1)\\
    &=&\mathcal{O}(1)+o_{\mathbb{P}}(1)
\end{eqnarray*}

Each of the $\boldsymbol{X}_i$ are drawn independently from a multivariate normal with dispersion matrix $\Sigma$. Thus, we have $B_2=\sqrt{n}\underset{1\leq i\leq n}{\min}||\boldsymbol{Z}_i||_1$, where $\boldsymbol{Z}_i\stackrel{i.i.d.}{\sim} \mathcal{N}(0,\Sigma)$.

If $\mathbb{P}\left(\underset{1\leq i\leq n}{\min}||\boldsymbol{Z}_i||_1>\frac{\kappa}{\sqrt{n}}\right)\rightarrow1$ for any fixed $\kappa$, then we have $\frac{1}{B_2}\stackrel{\mathbb{P}}{\rightarrow}0$
\begin{eqnarray*}
    \mathbb{P}\left(\underset{1\leq i\leq n}{\min}||\boldsymbol{Z}_i||_1>\frac{\kappa}{\sqrt{n}}\right)&=&\mathbb{P}\left(||\boldsymbol{Z}_1||_1>\frac{\kappa}{\sqrt{n}}\right)^n\geq\mathbb{P}\left(|\boldsymbol{Z}_1^T\boldsymbol{\mathbbm{1}}|>\frac{\kappa}{\sqrt{n}}\right)^n\\
    &=&\mathbb{P}\left(|\boldsymbol{Z}|>\frac{\kappa}{\sqrt{n\boldsymbol{\mathbbm{1}}^T\Sigma\boldsymbol{\mathbbm{1}}}}\right)^n=\left(1-\mathrm{erf}\left(\frac{\kappa}{\sqrt{2n\boldsymbol{\mathbbm{1}}^T\Sigma\boldsymbol{\mathbbm{1}}}}\right)\right)^n
\end{eqnarray*}

where $\boldsymbol{Z}\sim \mathcal{N}(0,1)$ and $\mathrm{erf}$ is the error function. We can see that $\mathrm{erf}\left(\frac{\kappa}{\sqrt{2n\boldsymbol{\mathbbm{1}}^T\Sigma\boldsymbol{\mathbbm{1}}}}\right)=o\left(\frac{1}{n}\right)$, since $\boldsymbol{\mathbbm{1}}^T\Sigma\boldsymbol{\mathbbm{1}}=\Omega(p)$. Thus, we have 
$$\left(1-\mathrm{erf}\left(\frac{\kappa}{\sqrt{2n\boldsymbol{\mathbbm{1}}^T\Sigma\boldsymbol{\mathbbm{1}}}}\right)\right)^n\rightarrow1$$
which proves that $\frac{1}{B_2}\stackrel{\mathbb{P}}{\rightarrow}0$


Finally, we need to find a condition such that we can bound $A_2$ away from zero.
Relabelling $C_{A^\perp}^T\boldsymbol{e}_j=:\boldsymbol{v}_j$, we have 
$$\boldsymbol{v}_j^T\hat{\Sigma}^{-1}\boldsymbol{v}_j\geq ||\boldsymbol{v}_j||^2\lambda_{\min}\left(\hat{\Sigma}^{-1}\right)$$
Now, this is bounded away from $0$ with high probability if $||\boldsymbol{v}_j||^2$ is bounded away from $0$ with high probability.
\begin{eqnarray*}
    ||\boldsymbol{v}_j||^2&=&\boldsymbol{e}_j^TC_{A^\perp}C_{A^\perp}^T\boldsymbol{e}_j\\
    &=&\boldsymbol{e}_j^TV\left(V^T\hat{\Sigma}_nV\right)^{-1}\left(V^T\hat{\Sigma}_n^2V\right)\left(V^T\hat{\Sigma}_nV\right)^{-1}V^T\boldsymbol{e}_j\text{ using the notation from the proof of Theorem \ref{thm:minimax}}\\
    &\geq&||V^T\boldsymbol{e}_j||^2\left(\lambda_{\max}\left(V^T\hat{\Sigma}_nV\right)\right)^{-2}\lambda_{\min}\left(V^T\hat{\Sigma}_n^2V\right)
\end{eqnarray*}
The second line follows from the following alternate form of the projection matrix
$$C_{A^\perp}=V\left(V^T\hat{\Sigma}_nV\right)^{-1}V^T\hat{\Sigma}_n$$ where $V$ is an orthogonal basis of $\mathcal{N}(A)$.

By Assumption \ref{as:rd}, $\lambda_{\max}\left(V^T\Sigma V\right)<\infty$ and $\lambda_{\min}\left(V^T\Sigma^2V\right)>0$. Define $\boldsymbol{a} = \underset{||\boldsymbol{v}||=1}{\mathrm{argmax}}\;\boldsymbol{v}^TV^T\hat{\Sigma}_nV\boldsymbol{v}$ and $\boldsymbol{b} = \underset{||\boldsymbol{v}||=1}{\mathrm{argmin}}\;\boldsymbol{v}^TV^T\hat{\Sigma}^2_nV\boldsymbol{v}$. Using a similar line of argument using deterministic equivalents, we can conclude the following:

\begin{eqnarray*}
    \lambda_{\max}\left(V^T\hat{\Sigma}_nV\right)&=&\boldsymbol{a}^TV^T\hat{\Sigma}_nV\boldsymbol{a}\\
    &=&\boldsymbol{a}^TV^T\Sigma V\boldsymbol{a}+o_{\mathbb{P}}(1)\text{ from Lemma \ref{lem:det_eq}}\\
    &\leq&\lambda_{\max}\left(V^T\Sigma V\right)+o_{\mathbb{P}}(1)\\
\end{eqnarray*}
\begin{eqnarray*}
\lambda_{\min}\left(V^T\hat{\Sigma}^2_nV\right)&=&\boldsymbol{b}^TV^T\hat{\Sigma}_n^2V\boldsymbol{b}\\
    &\geq&\boldsymbol{b}^TV^T\Sigma^2 V\boldsymbol{b}+o_{\mathbb{P}}(1)\text{ from Lemma \ref{lem:det_eq} and }\Sigma\text{ p.s.d.}\\
    &\geq&\lambda_{\min}\left(V^T\Sigma^2 V\right)+o_{\mathbb{P}}(1)\\
\end{eqnarray*}

Therefore, we can conclude that asymptotically, $\lambda_{\max}\left(V^T\Sigma V\right)<\infty$ and $\lambda_{\min}\left(V^T\Sigma^2V\right)>0$ with high probability. Additionally, if $V$ is an orthonormal basis, we have $V^T\boldsymbol{e}_j$ is non-zero with high probability. Therefore, we have $A_2>0$ with high probability, proving Lemma \ref{lem:rand_wts}(i) and (iii).



Therefore, by Lyapunov CLT, we have 
\begin{eqnarray*}
    \left(\frac{\sqrt{n}\left(\boldsymbol{e}_j^TC_{A^\perp}\hat{\Sigma}_n^{-1}\left(\frac{1}{n}X^T\epsilon\right)\right)}{\sqrt{\sigma^2\boldsymbol{e}_j^TC_{A^\perp}\hat{\Sigma}_n^{-1} C_{A^\perp}^T\boldsymbol{e}_j}}\right)\mid X\stackrel{d}{\rightarrow}N\left(0,1\right)
\end{eqnarray*}

Applying Lemma \ref{lem:rand_wts}(ii) with Slutsky's Theorem and plugging in the matrices for $C_{\Sigma}$ gives us the result.

\subsection{Jackknife Variance (Proof of Proposition \ref{thm:jackknife})}
\label{app:jackknife}
Proposition \ref{thm:jackknife} gave a scaling that allows us to adjust the variance of the jackknife estimator and in turn, the lengths of the confidence intervals. We will show that this adjustment is accurate. Without loss of generality, we will study the case where $\Sigma = I_p$ and $\boldsymbol{\beta} = 0$. The proof techniques are similar to the calculations in \cite{Karoui_Purdom_2016a}. 

We have defined $\tilde{\boldsymbol{\beta}}$ as the CLS estimator, and similarly $\tilde{\boldsymbol{\beta}}_{(i)}$ as the leave-one-out CLS estimator. That is, we have $\hat{\Sigma}_{(i)}=\frac{1}{n-1}X_{(i)}X_{(i)}^T$ and thus $$C_{A^\perp(i)}=I_p-\hat{\Sigma}_{(i)}^{-1}A^T\left(A\hat{\Sigma}_{(i)}^{-1}A^T\right)^{-1}A$$ We define $\hat{\epsilon}_i$ as the $i$th residual from the CLS estimator, and similarly $\hat{\epsilon}_{i(i)}$ as the $i$th residual from the leave-one-out CLS estimator, with the $i$th observation removed. We have the following quantification of the leave-one-out error, following from standard results involving the sample precision matrix:

\begin{eqnarray*}
    \tilde{\boldsymbol{\beta}}-\tilde{\boldsymbol{\beta}}_{(i)}&=& C_{A^\perp}\frac{1}{n}\hat{\Sigma}_{n}^{-1}X^T\hat{\boldsymbol{\epsilon}} - C_{A^\perp(i)}\frac{1}{n}\hat{\Sigma}_{(i)}^{-1}X_{(i)}^T\hat{\boldsymbol{\epsilon}}_{(i)} \\
    &=& \left(C_{A^\perp(i)}\frac{1}{n}\hat{\Sigma}_{(i)}^{-1}X^T\hat{\boldsymbol{\epsilon}} - C_{A^\perp(i)}\frac{1}{n}\hat{\Sigma}_{(i)}^{-1}X_{(i)}\hat{\boldsymbol{\epsilon}}_{(i)}\right) + \left(C_{A^\perp}\frac{1}{n}\hat{\Sigma}_{n}^{-1}X^T\hat{\boldsymbol{\epsilon}} - C_{A^\perp(i)}\frac{1}{n}\hat{\Sigma}_{(i)}^{-1}X^T\hat{\boldsymbol{\epsilon}}\right) \\
    &=& C_{A^\perp(i)}\frac{1}{n}\hat{\Sigma}_{(i)}^{-1}\boldsymbol{X}_i\hat{\epsilon}_i + \underbrace{\left(C_{A^\perp}\hat{\Sigma}_{n}^{-1}-C_{A^\perp(i)}\hat{\Sigma}_{(i)}^{-1}\right)X^T\hat{\boldsymbol{\epsilon}}}_{E}\\
    \boldsymbol{e}_j^TE&=&\boldsymbol{e}_j^T\left(C_{A^\perp}\hat{\Sigma}_{n}^{-1}-C_{A^\perp(i)}\hat{\Sigma}_{(i)}^{-1}\right)X^T\hat{\boldsymbol{\epsilon}}\\
    &=&\boldsymbol{e}_j^TV\left(\left(V^T\hat{\Sigma}_nV\right)^{-1}-\left(V^T\hat{\Sigma}_{(i)}V\right)^{-1}\right)V^TX^T\hat{\boldsymbol{\epsilon}}\\
    &=&\boldsymbol{e}_j^TV\left(\left(\sum_{j\neq i}\left(V^T\boldsymbol{X}_j\right)\left(V^T\boldsymbol{X}_j\right)^T+\left(V^T\boldsymbol{X}_i\right)\left(V^T\boldsymbol{X}_i\right)^T\right)^{-1}-\left(V^T\hat{\Sigma}_{(i)}V\right)^{-1}\right)V^TX^T\hat{\boldsymbol{\epsilon}}\\
    &=&\boldsymbol{e}_j^T\underbrace{V\left(\frac{\left(V^T\hat{\Sigma}_{(i)}V\right)^{-1}\left(V^T\boldsymbol{X}_i\boldsymbol{X}_i^TV\right)\left(V^T\hat{\Sigma}_{(i)}V\right)^{-1}}{1+\boldsymbol{X}_i^TV\left(V^T\hat{\Sigma}_{(i)}V\right)^{-1}V^T\boldsymbol{X}_i}\right)V^T}_{M}X^T\hat{\boldsymbol{\epsilon}}\text{ by Sherman-Morrison}\\
\end{eqnarray*}
In order to show this object is $o_{\mathbb{P}}(1)$ for all $i$, we need to show that $M$ has a uniform bound in operator norm for all $i$, and Proposition \ref{thm:asymp_norm} gives us the necessary result. We have $1+\boldsymbol{X}_i^TV\left(V^T\hat{\Sigma}_{(i)}V\right)^{-1}V^T\boldsymbol{X}_i = \mathcal{O}_{\mathbb{P}}(p-q)$. 

\begin{eqnarray*}
    ||M||_{op}&\leq&\frac{1}{\left(1+\boldsymbol{X}_i^TV\left(V^T\hat{\Sigma}_{(i)}V\right)^{-1}V^T\boldsymbol{X}_i\right)^2}\underbrace{\lambda_{\max}\left(\left(V^T\hat{\Sigma}_{(i)}V\right)^{-2}\right)^2}_{L}\underbrace{\lambda_{\max}\left(V^T\boldsymbol{X}_i\boldsymbol{X}_i^TV\right)^2}_{(p-q)^2}=\mathcal{O}_{\mathbb{P}}(1)
\end{eqnarray*}
\begin{eqnarray*}
L&=&\lambda_{\max}\left(\left(V^T\hat{\Sigma}_{(i)}V\right)^{-2}\right)^2=\left(\frac{1}{\lambda_{\min}\left(V^T\hat{\Sigma}_{(i)}V\right)}\right)^4\leq\left(\frac{1}{\lambda_{\min}\left(V^T\Sigma V\right)}\frac{n-1}{\left(\sqrt{n-1}-\sqrt{q}-t\right)^2}\right)^4\\
\end{eqnarray*}
with probability $1-2\exp\left(-t^2/2\right)$ \citep{vershynin2010introduction}. Note that this bound is independent of $i$, as it only depends on the Gaussianity of $\boldsymbol{X}_i$. This gives $L=\mathcal{O}_{\mathbb{P}}(1)$.
Therefore, we have $\boldsymbol{e}_j^TE=o_{\mathbb{P}}(1)$. 
Using a similar line of reasoning from \cite{Karoui_Purdom_2016a}, we can show that $\hat{\epsilon}_i=\frac{\hat{\epsilon}_{i(i)}}{1+\frac{1}{n}\boldsymbol{X}_i^TC_{A^\perp(i)}\hat{\Sigma}_{(i)}^{-1}\boldsymbol{X}_i}$. Thus, for a given contrast $\boldsymbol{v}$, we have 
$$\boldsymbol{v}^T\left(\tilde{\boldsymbol{\beta}}-\tilde{\boldsymbol{\beta}}_{(i)}\right)=\frac{1}{n}\boldsymbol{v}^TC_{A^\perp(i)}\hat{\Sigma}_{(i)}^{-1}\boldsymbol{X}_i\frac{\hat{\epsilon}_{i(i)}}{1+\frac{1}{n}\boldsymbol{X}_i^TC_{A^\perp(i)}\hat{\Sigma}_{(i)}^{-1}\boldsymbol{X}_i}+o_{\mathbb{P}}(1)$$
Therefore, we have 
$$n\sum_{i=1}^n\left[\boldsymbol{v}^T\left(\tilde{\boldsymbol{\beta}}-\tilde{\boldsymbol{\beta}}_{(i)}\right)\right]^2=\frac{1}{n}\sum_{i=1}^n\frac{\left[\boldsymbol{v}^TC_{A^\perp(i)}\hat{\Sigma}_{(i)}^{-1}\boldsymbol{X}_i\hat{\epsilon}_{i(i)}\right]^2}{\left[1+\frac{1}{n}\boldsymbol{X}_i^TC_{A^\perp(i)}\hat{\Sigma}_{(i)}^{-1}\boldsymbol{X}_i\right]^2}+o_{\mathbb{P}}(1)$$

We can look at the asymptotics of the denominator separately.
\begin{eqnarray*}
    1+\frac{1}{n}\boldsymbol{X}_i^TC_{A^\perp(i)}\hat{\Sigma}_{(i)}^{-1}\boldsymbol{X}_i&=&1+\frac{1}{n}\mathrm{Tr}\left(C_{A^\perp(i)}\hat{\Sigma}_{(i)}^{-1}\right)+o_{\mathbb{P}}(1)\\
    &=&1+\frac{1}{n}\mathrm{Tr}\left(\left(V^T\hat{\Sigma}_{(i)}V\right)^{-1}\right)+o_{\mathbb{P}}(1)\\
     &=&1+\frac{1}{n-1}\frac{\mathrm{Tr}\left(\left(V^T\Sigma V\right)^{-1}\right)}{1-(1-\gamma)\alpha}+o_{\mathbb{P}}(1)\text{ from Lemma \ref{lem:det_eq}}\\
    &=&1+\frac{(1-\gamma)\alpha}{1-(1-\gamma)\alpha}+o_{\mathbb{P}}(1)\\
    &=&\frac{1}{1+(1-\gamma)\alpha}+o_{\mathbb{P}}(1)
\end{eqnarray*}

Using the deterministic equivalent, we can show this uniform control of the error terms for all $i$. Therefore, we can essentially factor out the denominator, and only focus on the sum of the numerators.

We will use notation from \cite{Karoui_Purdom_2016a} to investigate the numerator. Let $$T_i = \boldsymbol{v}^TC_{A^\perp(i)}\hat{\Sigma}^{-1}_{(i)}\boldsymbol{X}_i\hat{\epsilon}_{i(i)}=\boldsymbol{v}^TC_{A^\perp(i)}\hat{\Sigma}^{-1}_{(i)}\boldsymbol{X}_i\left(\epsilon_i-\boldsymbol{X}_i^T\left(\tilde{\boldsymbol{\beta}}_{(i)}-\boldsymbol{\beta}\right)\right)$$

Using the independence of the errors, we have 
$$\mathbb{E}\left[T_i^2\right]=\underbrace{\mathbb{E}\left(\epsilon_i^2\right)}_{=\sigma^2}\underbrace{\mathbb{E}\left(\left(\boldsymbol{v}^TC_{A^\perp(i)}\hat{\Sigma}^{-1}_{(i)}\boldsymbol{X}_i\right)^2\right)}_{I}+\underbrace{\mathbb{E}\left(\left[\boldsymbol{X}_i^T\left(\tilde{\boldsymbol{\beta}}_{(i)}-\boldsymbol{\beta}\right)\right]^2\left[\boldsymbol{v}^TC_{A^\perp(i)}\hat{\Sigma}^{-1}_{(i)}\boldsymbol{X}_i\right]^2\right)}_{II}$$
\begin{eqnarray*}
    I&=&\mathbb{E}\left(\left(\boldsymbol{v}^TC_{A^\perp(i)}\hat{\Sigma}^{-1}_{(i)}\boldsymbol{X}_i\right)^2\right)\\
    &=&\mathbb{E}\left(\boldsymbol{v}^TC_{A^\perp(i)}\hat{\Sigma}^{-1}_{(i)}\boldsymbol{X}_i\boldsymbol{X}_i^T\hat{\Sigma}^{-1}_{(i)}C_{A^\perp(i)}^T\boldsymbol{v}\right)\\
    &=&\mathbb{E}\left(\mathbb{E}\left[\boldsymbol{v}^TC_{A^\perp(i)}\hat{\Sigma}^{-1}_{(i)}\boldsymbol{X}_i\boldsymbol{X}_i^T\hat{\Sigma}^{-1}_{(i)}C_{A^\perp(i)}^T\boldsymbol{v}|X_{(i)}\right]\right)\\
    &=&\mathbb{E}\left(\boldsymbol{v}^TC_{A^\perp(i)}\hat{\Sigma}^{-1}_{(i)}\mathbb{E}\left[\boldsymbol{X}_i\boldsymbol{X}_i^T|X_{(i)}\right]\hat{\Sigma}^{-1}_{(i)}C_{A^\perp(i)}^T\boldsymbol{v}\right)\\
    &=&\mathbb{E}\left(\boldsymbol{v}^T\left(C_{A^\perp(i)}\hat{\Sigma}_{(i)}^{-1}\right)^2\boldsymbol{v}\right) 
\end{eqnarray*}

For $II$, we use the fact that this can be expressed as the second moment of the covariance between two Gaussian vectors. That is 
$$\mathbb{E}\left(\left(\boldsymbol{a}^T\boldsymbol{X}_i\right)^2\left(\boldsymbol{b}^T\boldsymbol{X}_i\right)^2\right)=||\boldsymbol{a}||_2^2||\boldsymbol{b}||_2^2+2\left(\boldsymbol{a}^T\boldsymbol{b}\right)^2$$
\begin{eqnarray*}
    II&=&\mathbb{E}\left(\left[\boldsymbol{X}_i^T\left(\tilde{\boldsymbol{\beta}}_{(i)}-\boldsymbol{\beta}\right)\right]^2\left[\boldsymbol{v}^TC_{A^\perp(i)}\hat{\Sigma}^{-1}_{(i)}\boldsymbol{X}_i\right]^2\right)\\
    &=&\mathbb{E}\left|\left|\tilde{\boldsymbol{\beta}}_{(i)}-\boldsymbol{\beta}\right|\right|^2_2\mathbb{E}\left|\left|\boldsymbol{v}^TC_{A^\perp(i)}\hat{\Sigma}_{(i)}^{-1}\right|\right|^2_2+2\mathbb{E}\left(\boldsymbol{v}^TC_{A^\perp(i)}\hat{\Sigma}_{(i)}^{-1}\left(\tilde{\boldsymbol{\beta}}_{(i)}-\boldsymbol{\beta}\right)\right)^2\\
    &=&\frac{\sigma^2||\boldsymbol{v}||^2_2\left((1-\gamma)\alpha\right)}{\left(1-\left((1-\gamma)\alpha\right)\right)}I+o(1)
\end{eqnarray*}
The first term follows from Theorem \ref{thm:minimax}, and the second term has an additional $n$ scaling and therefore vanishes asymptotically. Putting this together, we have $$\mathbb{E}\left[T_i^2\right]=\left(1+\frac{(1-\gamma)\alpha}{1-(1-\gamma)\alpha}\right)I||\boldsymbol{v}||^2_2\sigma^2+o(1)=\frac{||\boldsymbol{v}||^2_2\sigma^2I}{\left(1-(1-\gamma)\alpha\right)}+o(1)$$

We calculate $I$ based on computations in \cite{Haff_1979,Holgersson_Pielaszkiewicz_2020}. Using the formulation given in Theorem \ref{thm:minimax}, we have 
$$C_{A^\perp(i)}\hat{\Sigma}_{(i)}^{-1}=V\underbrace{\left(V^T\hat{\Sigma}_{(i)}^{-1}V\right)^{-1}}_{\sim W(n-1,I_{p-q})}V^T$$

Based on Theorem 3.2 in \cite{Haff_1979} for the transformed Wishart matrix, we have,
$$I=\mathbb{E}\left[(V^T\boldsymbol{v})^T\left(C_{A^\perp(i)}\hat{\Sigma}_{(i)}^{-1}\right)^2(V^T\boldsymbol{v})\right]\rightarrow\left(\frac{1}{\left(1-(p-q)/n\right)^3}+o(1)\right)\boldsymbol{v}^T\underbrace{VV^T}_{\mathcal{P}_{A^\perp}}\boldsymbol{v}$$

Note that for the general $\Sigma$ case, this form will hold, with the inner matrix would be of the form $$\left(\Sigma^{-1}-\Sigma^{-1}\left(A^T\left(A\Sigma^{-1}A^T\right)^{-1}A\right)\Sigma^{-1}\right)$$and for the subsequent parts of the proof, we will use this form

Therefore, we have
$$\mathbb{E}\left[n\sum_{i=1}^n\left[\boldsymbol{v}^T\left(\tilde{\boldsymbol{\beta}}-\tilde{\boldsymbol{\beta}}_{(i)}\right)\right]^2\right]=\sigma^2\frac{\boldsymbol{v}^T\left(\Sigma^{-1}-\Sigma^{-1}\left(A^T\left(A\Sigma^{-1}A^T\right)^{-1}A\right)\Sigma^{-1}\right)\boldsymbol{v}}{(1-(1-\gamma)\alpha)^2}+o(1)$$
Plugging in the variance of our CLS estimator from Proposition \ref{thm:asymp_norm}, we have 
$$\mathbb{E}\left[\sum_{i=1}^n\left[\boldsymbol{v}^T\left(\tilde{\boldsymbol{\beta}}-\tilde{\boldsymbol{\beta}}_{(i)}\right)\right]^2\right]=\left[\frac{1}{(1-(1-\gamma)\alpha) }+o(1)\right]\mathrm{Var}\left(\boldsymbol{v}^T\tilde{\boldsymbol{\beta}}\right)$$

Now, we just need to show that the centering does not affect the final scaling. We have proved the result centered around the CLS estimator, and now, we must show the same holds around the mean of the leave-one-out estimators. Let us define $\tilde{\boldsymbol{\beta}}_{(\cdot)}=\frac{1}{n}\sum_{i=1}^n\tilde{\boldsymbol{\beta}}_{(i)}$. We will consider the quantity $n^2\left[\boldsymbol{v}^T\left(\tilde{\boldsymbol{\beta}}-\tilde{\boldsymbol{\beta}}_{(\cdot)}\right)\right]^2$ and if we can show that this quantity vanishes with $n$, then the results will still hold for this new centering. 

Since $\tilde{\boldsymbol{\beta}}-\tilde{\boldsymbol{\beta}}_{(i)}= \frac{1}{n}C_{A^\perp(i)}\hat{\Sigma}_{(i)}^{-1}\boldsymbol{X}_i\hat{\epsilon}_i+o_{\mathbb{P}}(1)$, we have 
$$\tilde{\boldsymbol{\beta}}-\tilde{\boldsymbol{\beta}}_{(\cdot)}= \frac{1}{n^2}\sum_{i=1}^nC_{A^\perp(i)}\hat{\Sigma}_{(i)}^{-1}\boldsymbol{X}_i\hat{\epsilon}_i+o_{\mathbb{P}}(1)$$

Therefore, we have 
$$n^2\left[\boldsymbol{v}^T\left(\tilde{\boldsymbol{\beta}}-\tilde{\boldsymbol{\beta}}_{(\cdot)}\right)\right]^2=\left[\frac{1}{n}\sum_{i=1}^n\boldsymbol{v}^TC_{A^\perp(i)}\hat{\Sigma}_{(i)}^{-1}\boldsymbol{X}_i\left(\epsilon_i-\boldsymbol{X}_i^T\left(\tilde{\boldsymbol{\beta}}-\boldsymbol{\beta}\right)\right)\right]^2+o_{\mathbb{P}}(1)$$

The $\epsilon_i$ term can be interpreted as a weighted mean of errors, which has mean $0$ and variance going to $0$ as well. Therefore, that term goes to $0$ in $\ell_2$ norm. Note that $C_{A^\perp}\hat{\Sigma}^{-1}\boldsymbol{X}_i=C_{A^\perp(i)}\hat{\Sigma}^{-1}_{(i)}\boldsymbol{X}_i(1-\gamma)\alpha$. We can use this to show that the second term also goes to $0$. Therefore, we have $n^2\left[\boldsymbol{v}^T\left(\tilde{\boldsymbol{\beta}}-\tilde{\boldsymbol{\beta}}_{(\cdot)}\right)\right]^2\stackrel{\mathbb{P}}{\rightarrow
}0$. This proves the proposition.

\subsection{Asymptotic Normality of $\hat{\boldsymbol{\beta}}_{\Sigma,\mathcal{P}}$ (Proof of Proposition \ref{thm:asymp_norm_high})}
\label{app:asymp_norm_high}
The high-dimensional projection estimator, when centered, can be viewed as a linear combination of independent errors, similar to the formulation in the Proof of Proposition \ref{thm:asymp_norm}. Under Assumption \ref{as:err}, we have $\hat{\boldsymbol{\beta}}_{\Sigma,\mathcal{P}}$ following a normal distribution. Therefore, we will show coordinate-wise $\sqrt{n}$-consistency, and show the variance of this estimator to prove the proposition, using the method of moments.

$\sqrt{n}$\textbf{-consistency of  }$\hat{\beta}_{\Sigma,\mathcal{P},j}$\textbf{ to }$\beta^{*}_j$

We need to evaluate the centering of $\sqrt{n}\left(\hat{\beta}_{\Sigma,\mathcal{P},j}-\beta^{*}_j\right)$, that is, if we show $\mathbb{E}\left[\sqrt{n}\left(\hat{\beta}_{\Sigma,\mathcal{P},j}-\beta^{*}_j\right)\right]=0$

We have defined $\hat{\boldsymbol{\beta}}_{\Sigma,\mathcal{P}}=\frac{1}{n}\mathcal{P}_{A^\perp}\Sigma^{-1}X^T\boldsymbol{y}=\frac{1}{n}\mathcal{P}_{A^\perp}\Sigma^{-1}X^TX\boldsymbol{\beta}^{*}+\frac{1}{n}\mathcal{P}_{A^\perp}\Sigma^{-1}X^T\boldsymbol{\epsilon}$

\begin{eqnarray*}
    \mathbb{E}\left[\sqrt{n}\left(\hat{\beta}_{\Sigma,\mathcal{P},j}-\beta^{*}_j\right)\right]&=&\mathbb{E}\left[\mathbb{E}\left[\sqrt{n}\left(\hat{\beta}_{\Sigma,\mathcal{P},j}-\beta^{*}_j\right)\right]\mid X\right]\\
    &=&\mathbb{E}\left[\mathbb{E}\left[\sqrt{n}\left(\boldsymbol{e}_j^T\left(\frac{1}{n}\mathcal{P}_{A^\perp}\Sigma^{-1}X^TX\boldsymbol-I_{p}\right){\beta}^{*}+\frac{1}{n}\boldsymbol{e}_j^T\mathcal{P}_{A^\perp}\Sigma^{-1}X^T\boldsymbol{\epsilon}\right)\right]\mid X\right]\\
    &=&\mathbb{E}\left[\sqrt{n}\left(\boldsymbol{e}_j^T\left(\frac{1}{n}\mathcal{P}_{A^\perp}\Sigma^{-1}X^TX\boldsymbol-I_{p}\right){\beta}^{*}\right)\right]\text{ by the Cramer-Wold device}\\
    &=&\sqrt{n}\left(\boldsymbol{e}_j^T\left(\frac{1}{n}\mathcal{P}_{A^\perp}\Sigma^{-1}\mathbb{E}\left[X^TX\right]\boldsymbol-I_{p}\right){\beta}^{*}\right)\\
    &=&\sqrt{n}\left(\boldsymbol{e}_j^T\left(\mathcal{P}_{A^\perp}\boldsymbol-I_{p}\right){\beta}^{*}\right)\\
    &=&0\\
\end{eqnarray*}
Therefore, the projected estimator is $\sqrt{n}$-consistent. 

\textbf{Variance of }$\hat{\beta}_{\Sigma,\mathcal{P},j}$

\begin{eqnarray*}
    \mathrm{Var}\left(\hat{\beta}_{\Sigma,\mathcal{P},j}\right)&=&\mathbb{E}\left[\mathrm{Var}\left(\hat{\beta}_{\Sigma,\mathcal{P},j}\mid X\right)\right]+\mathrm{Var}\left[\mathbb{E}\left(\hat{\beta}_{\Sigma,\mathcal{P},j}\mid X\right)\right]\\
    &=&\underbrace{\mathbb{E}\left[\mathrm{Var}\left(\frac{1}{n}\boldsymbol{e}_j^T\mathcal{P}_{A^\perp}\Sigma^{-1}X^T\epsilon\mid X\right)\right]}_{I}+\underbrace{\mathrm{Var}\left[\frac{1}{n}\boldsymbol{e}_j^T\Sigma^{-1}X^TX\boldsymbol{\beta}^{*}\right]}_{II}\text{ by the Cramer-Wold device}\\
\end{eqnarray*}
By the independence of $X$ and $\epsilon$, we have $$I=\frac{\sigma^2}{n}\boldsymbol{e}_j^T\mathcal{P}_{A^\perp}\Sigma^{-1}\mathcal{P}_{A^\perp}\boldsymbol{e}_j$$
As a distinction from the analysis in the low dimensional case, the use of the inverse covariance matrix introduces variance in the form of the second term $II$.
\begin{eqnarray*}
    II&=&\mathrm{Var}\left[\frac{1}{n}\boldsymbol{e}_j^T\mathcal{P}_{A^\perp}\Sigma^{-1}\left(X^TX\right)\boldsymbol{\beta}^{*}\right]\\
    &=&\mathbb{E}\left[\left(\frac{1}{n}\boldsymbol{e}_j^T\mathcal{P}_{A^\perp}\Sigma^{-1}\left(X^TX\right)\boldsymbol{\beta}^{*}\right)^2\right]-\left(\boldsymbol{e}_j^T\mathcal{P}_{A^\perp}\boldsymbol{\beta}^{*}\right)^2\\
    &=&\mathbb{E}\left[\left(\frac{1}{n}\boldsymbol{e}_j^T\mathcal{P}_{A^\perp}\Sigma^{-1}\left(X^TX\right)\boldsymbol{\beta}^{*}\right)^2\right]-\left(\beta^{*}_j\right)^2\\
    \mathbb{E}\left[\left(\frac{1}{n}\boldsymbol{e}_j^T\mathcal{P}_{A^\perp}\Sigma^{-1}\left(X^TX\right)\boldsymbol{\beta}^{*}\right)^2\right]&=&\mathbb{E}\left[\frac{1}{n^2}\boldsymbol{e}_j^T\mathcal{P}_{A^\perp}\Sigma^{-1}\left(X^TX\right)\boldsymbol{\beta}^{*}\boldsymbol{e}_j^T\mathcal{P}_{A^\perp}\Sigma^{-1}\left(X^TX\right)\boldsymbol{\beta}^{*}\right]\\
    &=&\frac{1}{n^2}\boldsymbol{e}_j^T\mathcal{P}_{A^\perp}\Sigma^{-1}\mathbb{E}\left[\left(X^TX\right)\boldsymbol{\beta}^{*}\boldsymbol{e}_j^T\mathcal{P}_{A^\perp}\Sigma^{-1}\left(X^TX\right)\right]\boldsymbol{\beta}^{*}\\
\end{eqnarray*}
We can evaluate this using Wishart calculations from \cite{Holgersson_Pielaszkiewicz_2020}
\begin{eqnarray*}
    \mathbb{E}\left[\left(X^TX\right)\boldsymbol{\beta}^{*}\boldsymbol{e}_j^T\mathcal{P}_{A^\perp}\Sigma^{-1}\left(X^TX\right)\right]&=&n\mathcal{P}_{A^\perp}\boldsymbol{e}_j\boldsymbol{\beta}^{*^T}\Sigma+n^2\Sigma\boldsymbol{\beta}^{*}\boldsymbol{e}_j^T\mathcal{P}_{A^\perp}+n\mathrm{Tr}\left(\Sigma\boldsymbol{\beta}^{*}\boldsymbol{e}_j^T\mathcal{P}_{A^\perp}\Sigma^{-1}\right)\Sigma\\
    \mathbb{E}\left[\left(\frac{1}{n}\boldsymbol{e}_j^T\mathcal{P}_{A^\perp}\Sigma^{-1}\left(X^TX\right)\boldsymbol{\beta}^{*}\right)^2\right]&=&\frac{1}{n}\boldsymbol{e}_j^T\mathcal{P}_{A^\perp}\Sigma^{-1}\mathcal{P}_{A^\perp}\boldsymbol{e}_j||\Sigma^{1/2}\boldsymbol{\beta}^{*}||^2+\left(\beta^*_j\right)^2+\frac{1}{n}\left(\beta^*_j\right)^2\\
    II&=&\frac{1}{n}\boldsymbol{e}_j^T\mathcal{P}_{A^\perp}\Sigma^{-1}\mathcal{P}_{A^\perp}\boldsymbol{e}_j||\Sigma^{1/2}\boldsymbol{\beta}^{*}||^2+\frac{1}{n}\left(\beta^*_j\right)^2\\
    \mathrm{Var}\left(\hat{\beta}_{\Sigma,\mathcal{P},j}\right)&=&\frac{1}{n}\left(\beta^*_j\right)^2+\frac{\left(\sigma^2+||\Sigma^{1/2}\boldsymbol{\beta}^{*}||^2\right)}{n}\boldsymbol{e}_j^T\mathcal{P}_{A^\perp}\Sigma^{-1}\mathcal{P}_{A^\perp}\boldsymbol{e}_j\\
\end{eqnarray*}
This gives us the variance in Proposition \ref{thm:asymp_norm_high}. 
Therefore, by matching moments, we see that $\sqrt{n}\left(\hat{\beta}_{\Sigma,\mathcal{P},j}-\beta^{*}_j\right)$ follows a normal distribution, giving the asymptotic distribution we desire.
\section{Additional Information}

\subsection{Biological Interpretation of Constraints}
\label{app:bio_info}
Consider the levels of two traits $\boldsymbol{y}_1, \boldsymbol{y}_2$ observed in an individual. We can decompose these trait levels into a genetic component and a non-genetic component
$$\boldsymbol{y}_1=\boldsymbol{g}_1+\boldsymbol{e}_1$$
$$\boldsymbol{y}_2=\boldsymbol{g}_2+\boldsymbol{e}_2$$
Here $\boldsymbol{g}_1,\boldsymbol{g}_2$ are referred to as the genetic values, and $\boldsymbol{e}_1,\boldsymbol{e}_2$ are referred to as the residual values. If we consider $\boldsymbol{\beta}_i$ to be the vector of additive allelic effects for each of the SNPs for trait $i$, and $X$ is the genotype vector for the individual (takes values 0,1,2), we have 
$$\boldsymbol{g}_i=X\boldsymbol{\beta}_i$$
as each effect allele contributes the corresponding value to the total genetic value. 
Using this genetic effects, taken for the whole population, we define the genetic correlation between the two traits as 
$$\rho_g=\frac{\sigma_{\boldsymbol{g}_{1},\boldsymbol{g}_{2}}}{\sqrt{\sigma_{\boldsymbol{g}_{1}}^2\sigma_{\boldsymbol{g}_{2}}^2}}$$where $\sigma_{\boldsymbol{g}_{1},\boldsymbol{g}_{2}}$ is the population level covariance of the genetic effects, and $\sigma_{\boldsymbol{g}_{1}}^2,\sigma_{\boldsymbol{g}_{2}}^2$ are the population level variances of the genetic effects. Now, if we are to take the covariance between two genetic values for the population, that have been normalized, we see that 
$$\sigma_{\boldsymbol{g}_{1},\boldsymbol{g}_{2}}=\frac{1}{N}\boldsymbol{\beta}_1^TX^TX\boldsymbol{\beta}_2=\boldsymbol{\beta}_1\tilde{\Sigma}_N\boldsymbol{\beta}_2$$
This gives the constraint that we are using in our analysis. Clearly, this is contingent on the allelic effects acting additively.

\subsection{Construction of Estimators in High Dimensions, Unknown $\Sigma$ regime}
\label{app:high_dim_est}
Our target of interest is the coordinate of a high dimensional regression vector $$\beta_k = \mathbb{E}\left[\boldsymbol{y}^TX\right]\Sigma^{-1}\boldsymbol{e}_k$$
We use the approximation $\beta_{k,J}:=\sum_{\ell=0}^Jc_{\ell}\beta_{k}^{(\ell)}$  where $\beta_{k}^{(\ell)}:=\mathbb{E}\left[\boldsymbol{y}^TX\right]\Sigma^{\ell}\boldsymbol{e}_k$. Here $c_\ell$ represent the Chebyshev polynomial coefficients, the selection of which is detailed in \cite{DeVore_Lorentz_1993, Orecchia_Sachdeva_Vishnoi_2011, Walczyk_Moroz_Samotyy_2025}, and $J\asymp (\log n)^c$ for some $c<1$. We construct an unbiased estimator of $\beta_{k}^{(\ell)}$ using higher order U-statistics through terms defined as such:
$$\hat{\beta}_{k}^{(\ell)}:=\frac{\left(n-\left(\ell+1\right)\right)!}{n!}\sum_{1\leq i_1\neq\ldots\neq i_{\ell+1}\leq n}y_{i_1}\boldsymbol{X}_{i_1}^T\left\{\prod_{s=2}^{\ell+1}\boldsymbol{X}_{i_s}\boldsymbol{X}_{i_s}^T\right\}\boldsymbol{e}_k$$
Thus, we have the following algorithm for constructing the coordinate of our regression vector:
\begin{enumerate}
    \item Construct $\hat{\beta}_k^{(\ell)}$ for $\ell = 0,\ldots,J$ and $k=1,\ldots,p$
    \item Estimate $\hat{\beta}_k:=\sum_{\ell=0}^Jc_{\ell}\hat{\beta}_{k}^{(\ell)}$ for $k=1,\ldots,p$ to make the vector $\hat{\boldsymbol{\beta}}$
    \item Project $\hat{\beta}_{est,j}:=\boldsymbol{e}_j^T\mathcal{P}_{A^\perp}\hat{\boldsymbol{\beta}}+\boldsymbol{e}_j^TA^T\left(AA^T\right)\boldsymbol{c}$
\end{enumerate}

This procedure provides us with an unbiased estimator for the coordinates of $\boldsymbol{\beta}$. \cite{Orecchia_Sachdeva_Vishnoi_2011} provide a bound on the error of estimation of the inverse as a function of $J$, and a similar computation can be extended to quantify the bias of estimating a coordinate of the effect vector using this method. However, we do not have guarantees on the rate of consistency, as mentioned in \cite{Chen_Liu_Mukherjee_2024}. Additionally, the algorithm is computationally expensive. While \cite{kong2018estimating} provide a polynomial-time algorithm for evaluating these U-statistics, these estimations have to be constructed over a large grid of indices, making this computationally infeasible. We demonstrate this in a simulation setup in Appendix \ref{s:add_exp}.

\subsection{Construction of Estimators for GLMs}
\label{app:glm_est}
We elucidate a procedure for constructing such estimators under the framework of generalized linear models (GLM). Note that we do not have optimality guarantees for these estimators, and look forward to working on the derivation of such estimators in the future. This methodology is based on the procedure given in \cite{Chen_Liu_Mukherjee_2024}, assuming $\Sigma, A, \boldsymbol{c}$ are known. We refer to that paper for further discussions when $\Sigma$ is not known and the next subsection for a discussion when $A, \boldsymbol{c}$ are not known. 

We have a GLM setup as follows:
$$\mathbb{E}\left[\boldsymbol{y}|X=X\right]=g\left(X^T\boldsymbol{\beta}^{*}\right)$$
We would like to estimate $\beta^{*}_j$ for all $j$ in $1,\ldots,p$. 

From Lemma 1 of \cite{Chen_Liu_Mukherjee_2024}, we have $$U:=\mathbb{E}\left[\boldsymbol{y}^TX\right]\Sigma^{-1}\mathbb{E}\left[X^T\boldsymbol{y}\right] = f^2(\boldsymbol{\beta^{*}}^T\Sigma\boldsymbol{\beta^{*}})\boldsymbol{\beta^{*}}^T\Sigma\boldsymbol{\beta^{*}}$$
$$V_j:=\mathbb{E}\left[\boldsymbol{y}^TX\right]\Sigma^{-1}\boldsymbol{e}_j = f(\boldsymbol{\beta^{*}}^T\Sigma\boldsymbol{\beta^{*}})\beta^{*}_j$$
where $f(t):=\mathbb{E}\left[g^{'}(Z)\right]$ with $Z\sim N(0,t)$ for $t\geq 0$.

Therefore, for a given link function $g$, there exists a continuous function $h$ such that $Q:=\boldsymbol{\beta^{*}}^T\Sigma\boldsymbol{\beta^{*}}=h\left(U\right)$, which we can use for estimating $\beta^{*}_j$.

Thus, we have the following algorithm to construct an estimator in the GLM case, given $\Sigma$ and this inverse function $h$ and derivative function $f$ given $g$:
\begin{enumerate}
    \item Compute $\hat{U}=\frac{1}{n(n-1)}\underset{1\leq i_1\neq i_2\leq n}{\sum}y_{i_1}y_{i_2}\boldsymbol{X}_{i_1}^T\Sigma^{-1}\boldsymbol{X}_{i_2}$
    \item Compute $\hat{V_j}=\frac{1}{n}\underset{1\leq i\leq n}{\sum}y_{i}\boldsymbol{X}_{i}^T\Sigma^{-1}\boldsymbol{e}_{j}$ for $j=1,\ldots,p$
    \item Find $\hat{Q}=h(\hat{U})$ and get $\hat{\beta}_{GLM,j}:=\hat{V}_j/f(\hat{Q})$
    \item Project $\hat{\boldsymbol{\beta}}_{\mathcal{P},GLM}:=\mathcal{P}_{A^\perp}\hat{\boldsymbol{\beta}}_{GLM}+A^T\left(AA^T\right)\boldsymbol{c}$
\end{enumerate}

\textbf{Remark:} Since
$\left(I-\mathcal{P}_{A^\perp}\right)V$ is positive semi-definite for any positive semi-definite matrix $V$, the total of the variances of each coordinate of the projected GLM estimator is lower than that of the coordinate-wise GLM estimator. Here $V$ represents a diagonal matrix where $V_{ii}=\nu_i^2$, which is defined as the asymptotic variance of the coordinate-wise estimator given in Proposition 2 of \cite{Chen_Liu_Mukherjee_2024}. We cannot make any direct comparison of the asymptotic variance for a given coordinate.

We implement this estimator in the case of logistic regression ($g(\cdot) = \mathrm{expit(\cdot)}$) and demonstrate its numerical performance in Appendix \ref{s:add_exp}.

\subsection{Details on Protein Selection}
\label{app:protein}
For our analysis, we consider $10$ proteins as the anchors of our analysis, for each of the glycemic traits. A concept necessary to choosing the proteins is heritability, which is the proportion of the trait variability attributable to variance in genetic factors. It is a population level statistic between 0 and 1 which will be useful for determining proteins whose expression levels are predominantly determined by genetics. We have genotypic data at pQTLs for the individuals in the JHS dataset. \cite{Elgart_Goodman_Isasi_Chen_De_Vries_Xu_Manichaikul_Guo_Franceschini_Psaty_etal._2021} and \cite{Sofer_2017} elucidate the procedure for estimation of genetic correlations and heritability for proteins, from individual level data in the JHS dataset. From these proteins, we choose the top 10 proteins with the greatest absolute genetic correlations, that have high heritability ($>0.4$). We do this to pick proteins that will lend the most information to the analysis. The genetic correlations for these 10 proteins are extracted as $c$. From the GWAS of each of the selected proteins, we select genetic variants with a p-value below a threshold of $5\times 10^{-8}$, accounting for collinearity of SNPs through pruning. We subset the effect vectors of each of the proteins at the selected SNPs to construct the matrix $A$. Tables \ref{tab:sig_proteins_bmi},\ref{tab:sig_proteins_fi},\ref{tab:sig_proteins_hb} shows the proteins that we use for each of the glycemic traits in our analysis.

\begin{table}[]
    \centering
    \begin{tabular}{|c|c|c|c|}
    \hline
    Protein Name  & Heritability & Genetic Correlation & Genetic Covariance\\
    \hline
      ANTITHROMBIN\_III &   0.4240401 &-0.4778983 &-0.1904203\\
C5A   & 0.5589884 & 0.5547244 & 0.2537773\\
CHL1  &  0.6229589 &-0.4617222 &-0.2229896\\
EDA   & 0.4689197 &-0.5524264 &-0.2314717\\
KALLISTATIN  &  0.5523707 &-0.4978723 &-0.2264162\\
M\_CSF\_R  &  0.4290930 & 0.4718710 & 0.1891356\\
RET  &  0.4126752 & 0.4804708 & 0.1888624\\
SCF\_SR  &  0.5021367 &-0.4785794 &-0.2075101\\
TGF\_B3  &  0.4665368 & 0.4866873 & 0.2034076\\
TRKC  &  0.6085098 &-0.4903304 &-0.2340436\\
         \hline
    \end{tabular}
    \caption{Significant proteins for BMI with their heritabilities and genetic correlations}
    \label{tab:sig_proteins_bmi}
\end{table}
\begin{table}[]
    \centering
    \begin{tabular}{|c|c|c|c|}
\hline
    Protein Name  & Heritability & Genetic Correlation & Genetic Covariance\\
    \hline
ANGIOPOIETIN\_2  &  0.4331282   &          0.6657746      &       0.1895192 \\
ANTITHROMBIN\_III  &  0.4240401  &          -0.5810212      &      -0.1636489\\
CHL1  &  0.6229589      &      -0.4808414    &        -0.1641532\\
ERP29  &  0.4087223      &       0.5870843    &         0.1623425\\
IL\_1\_SRII  &  0.6033166      &      -0.4782250 &           -0.1606655\\
TPSB2  &  0.5180600      &      0.6821772    &         0.2123757\\
TSG\_6  &  0.6053944      &      -0.9503767   &         -0.3198400\\
TYK2  &  0.4019710       &      0.5809450    &         0.1593126\\
UB2G2  &  0.4204848       &      0.5513349    &         0.1546352\\
VITRONECTIN  &  0.7120431      &      -0.5510545&            -0.2011247\\
         \hline
    \end{tabular}
    \caption{Significant proteins for fasting insulin with their heritabilities and genetic correlations}
    \label{tab:sig_proteins_fi}
\end{table}
\begin{table}[]
    \centering
    \begin{tabular}{|c|c|c|c|}
    \hline
    Protein Name  & Heritability & Genetic Correlation & Genetic Covariance\\
    \hline
C2  &  0.4469158  &  0.5761554  &  0.1576280 \\
CATF  &  0.4772143  & -0.6253038  & -0.1767782\\
CNTN2  &  0.4540917  & -0.6146444 &  -0.1695027\\
ERP29  &  0.4087223  &  0.7550191  &  0.1975391\\
PARC  &  0.5050623  &  0.5825308  &  0.1694229\\
PGCB  &  0.5056586  & -0.7173123  & -0.2087458\\
TIG2  &  0.6312174  &  0.7301262  &  0.2373932\\
TRKC  &  0.6085098  & -0.5512258  & -0.1759722\\
TYK2  &  0.4019710  &  0.8336819  &  0.2163110\\
UB2G2  &  0.4204848  &  0.8414359  &  0.2232940\\
         \hline
    \end{tabular}
    \caption{Significant proteins for HbA1c with their heritabilities and genetic correlations}
    \label{tab:sig_proteins_hb}
\end{table}
\section{Additional Experimental Results}
\label{s:add_exp}
We catalogue the simulation results that have not been included in the main body of the paper. 
\subsection{Synthetic Simulations}
\begin{figure}
\centering
\begin{minipage}{.47\textwidth}
  \centering
  \includegraphics[width=\textwidth, height = 25em]{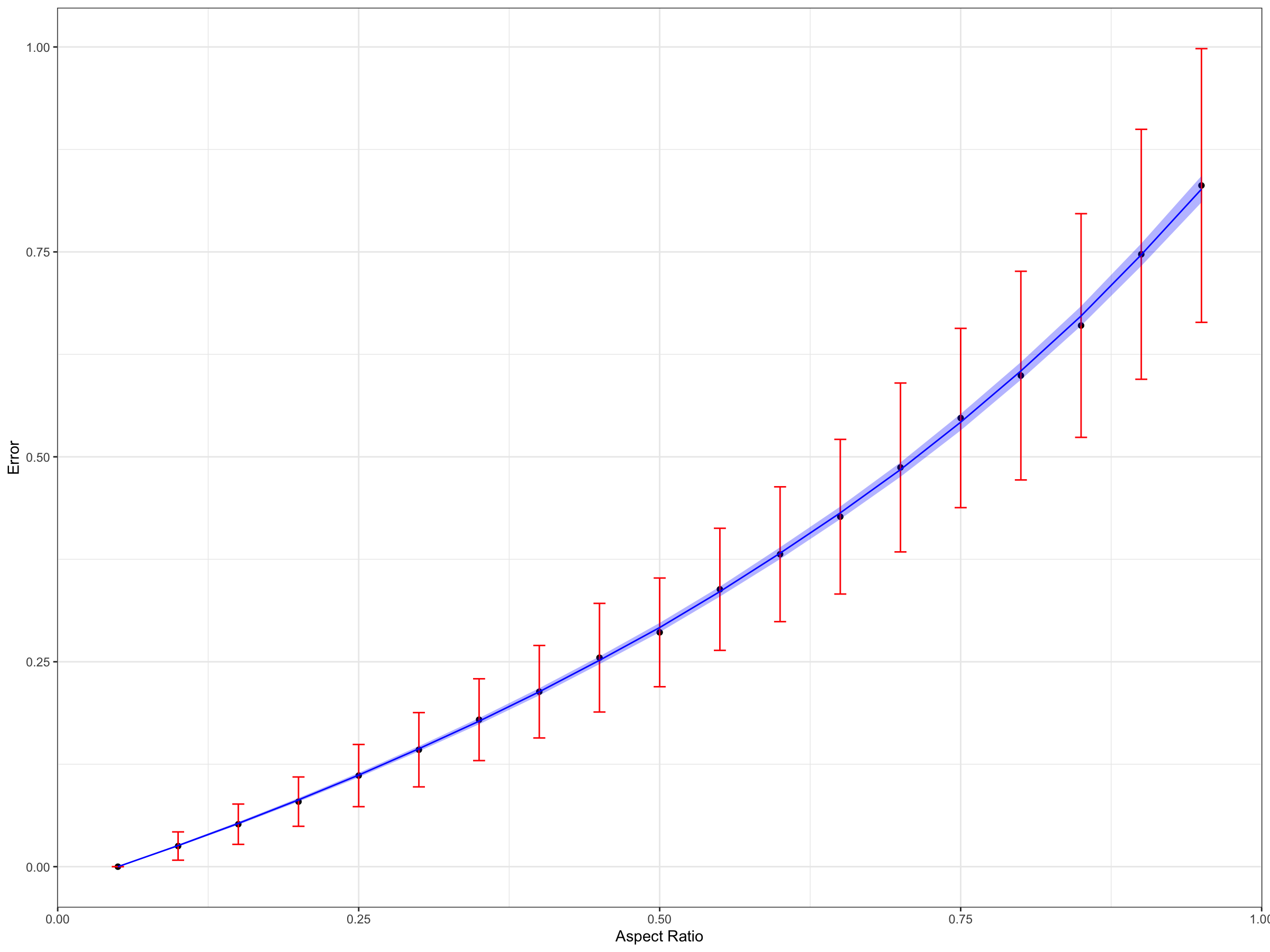}
		\caption{Isotropic dispersion [\textbf{m1}]}
		\label{fig:err_formula_1}
\end{minipage}%
\hfill
\begin{minipage}{.47\textwidth}
  \centering
  \includegraphics[width=\textwidth, height = 25em]{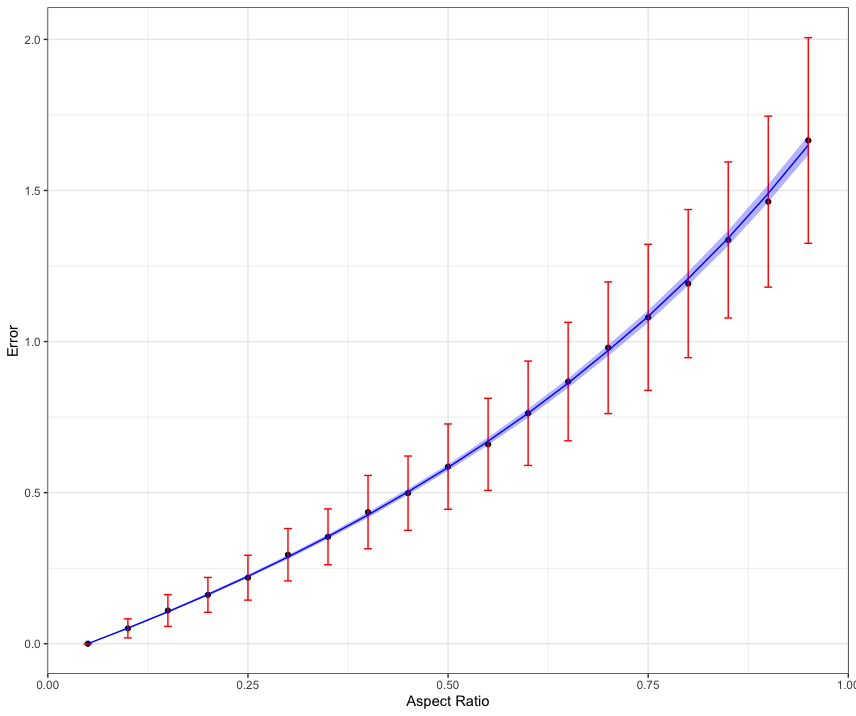}
		\caption{Anisotropic dispersion [\textbf{m2}]}
		\label{fig:err_formula_2}
\end{minipage}
	\caption{Adherence of MSE from simulation (black dots) to exact derived formula (blue line) for $\tilde{\beta}$ [\textbf{s1}]}
	\label{fig:err_formula}
\end{figure}

Figures \ref{fig:err_formula_1} and \ref{fig:err_formula_2} [\textbf{s1}] provide credence to the asymptotic formulae of the MSE derived in Theorem \ref{thm:minimax}, showing that it can be applied to small sample calculations as well. The black dots indicate the mean of the MSEs of the CLS estimator over all the iterations, and the error bars indicate the standard deviation of these MSEs. The blue line indicates the mean of theoretical asymptotic error values over all the iterations, and the blue ribbon indicates the standard deviation of these theoretical values. 

\begin{figure}
\centering
\begin{minipage}{.47\textwidth}
  \centering
  \includegraphics[width = \textwidth, height = 25em]{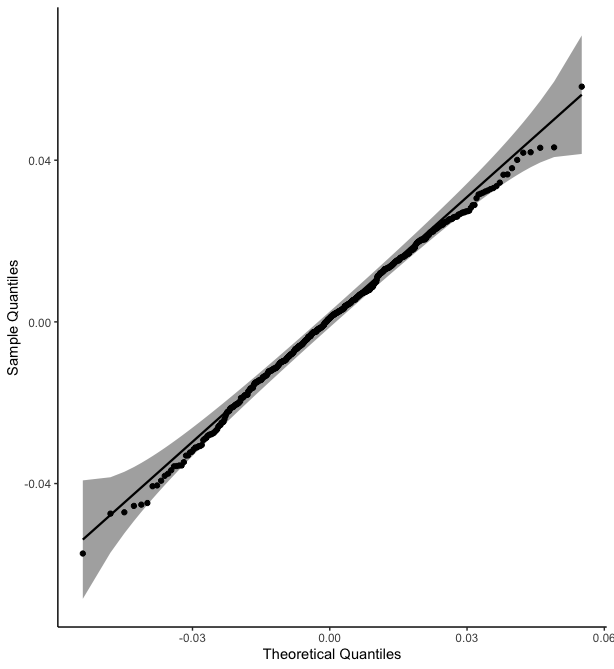}
    \caption{QQ plot of first coordinate of CLS vector [\textbf{s2,m1}]}
    \label{fig:lowdim-qqplot}
\end{minipage}%
\hfill
\begin{minipage}{.47\textwidth}
  \centering
  \includegraphics[width = \textwidth, height = 25em]{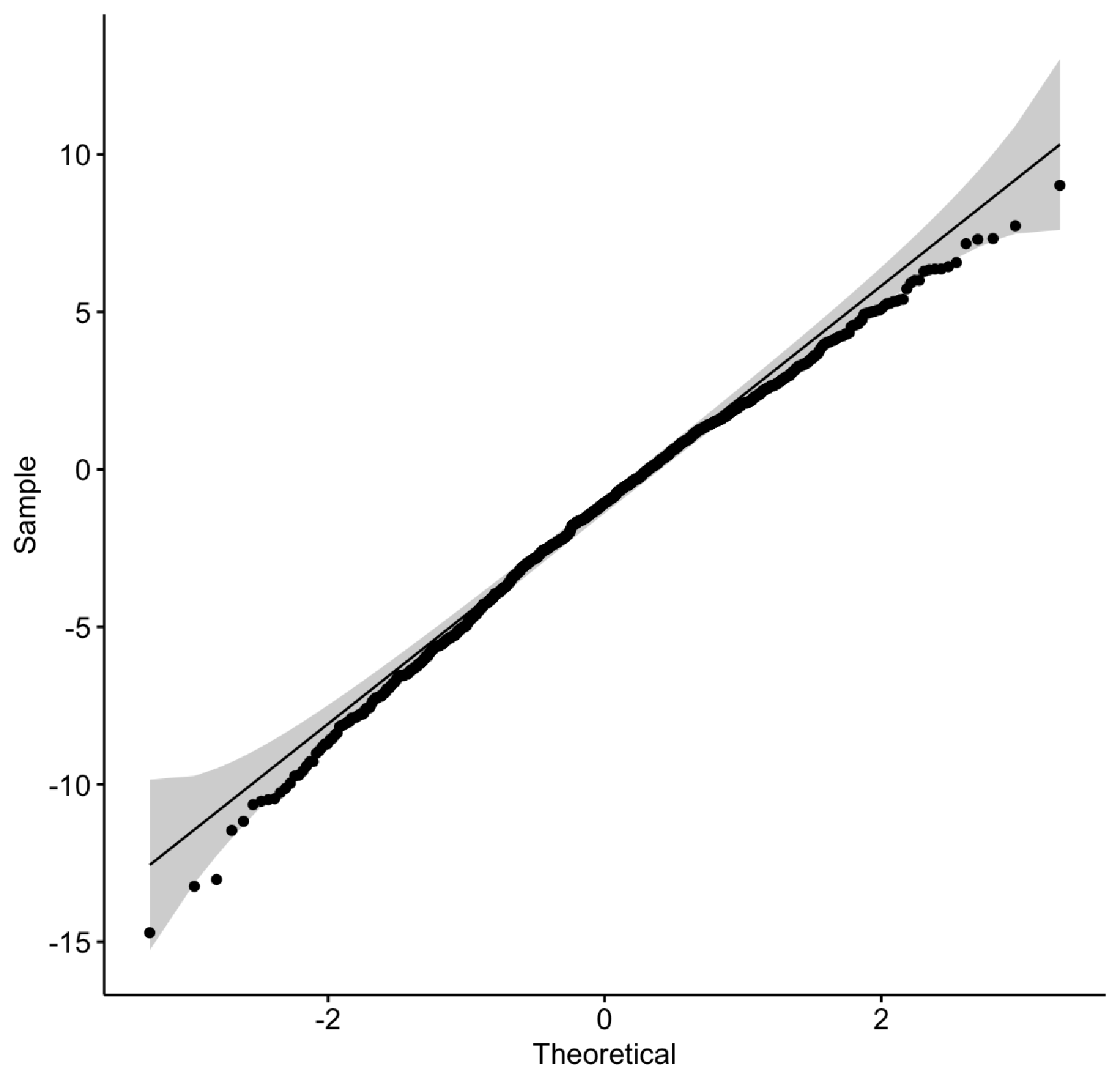}
    \caption{QQ plot of first coordinate of CLS vector [\textbf{s3,m1}]}
    \label{fig:highdim-qqplot}
\end{minipage}
\end{figure}

Figure \ref{fig:lowdim-qqplot} [\textbf{s2,m1}] shows the asymptotic normality of the first coordinate of the CLS vector for a fixed $q=50$. This demonstrates the property given in Proposition \ref{thm:asymp_norm}. 

Similarly, Figure \ref{fig:highdim-qqplot} [\textbf{s3,m1}] shows the asymptotic normality of the first coordinate of the CLS vector for a fixed $q=150$. Therefore, we can work towards the high dimensional corollary of Proposition \ref{thm:asymp_norm}. 

\begin{figure}
\centering
\begin{minipage}{.49\textwidth}
  \centering
  \includegraphics[width = \textwidth]{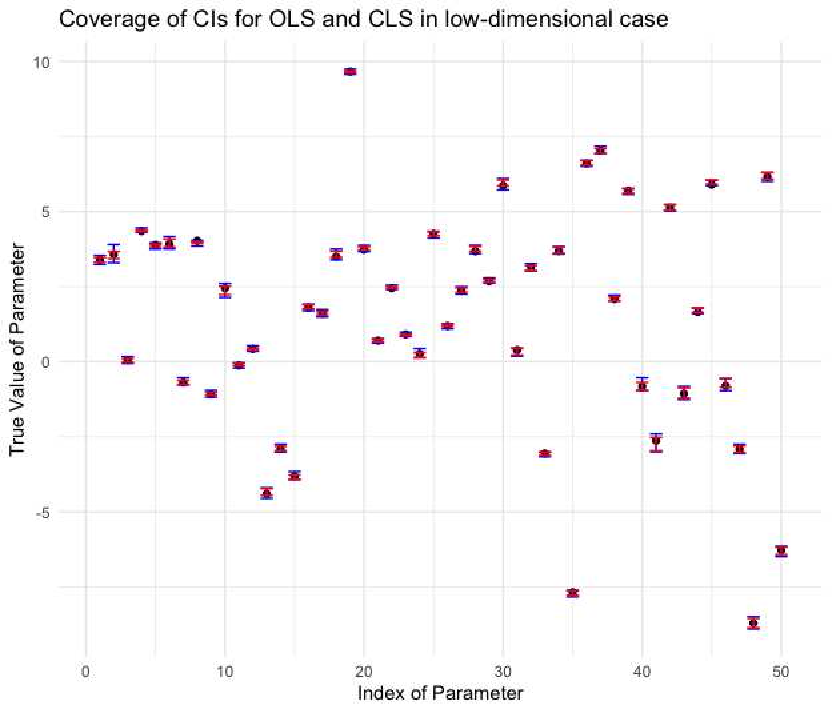}
    \caption{CIs at each coordinate for $\hat{\beta}_{LS}$ and $\tilde{\beta}$ [\textbf{s2,m1}]}
    \label{fig:ci_cover}
\end{minipage}%
\hfill
\begin{minipage}{.49\textwidth}
  \centering
  \includegraphics[width = \textwidth]{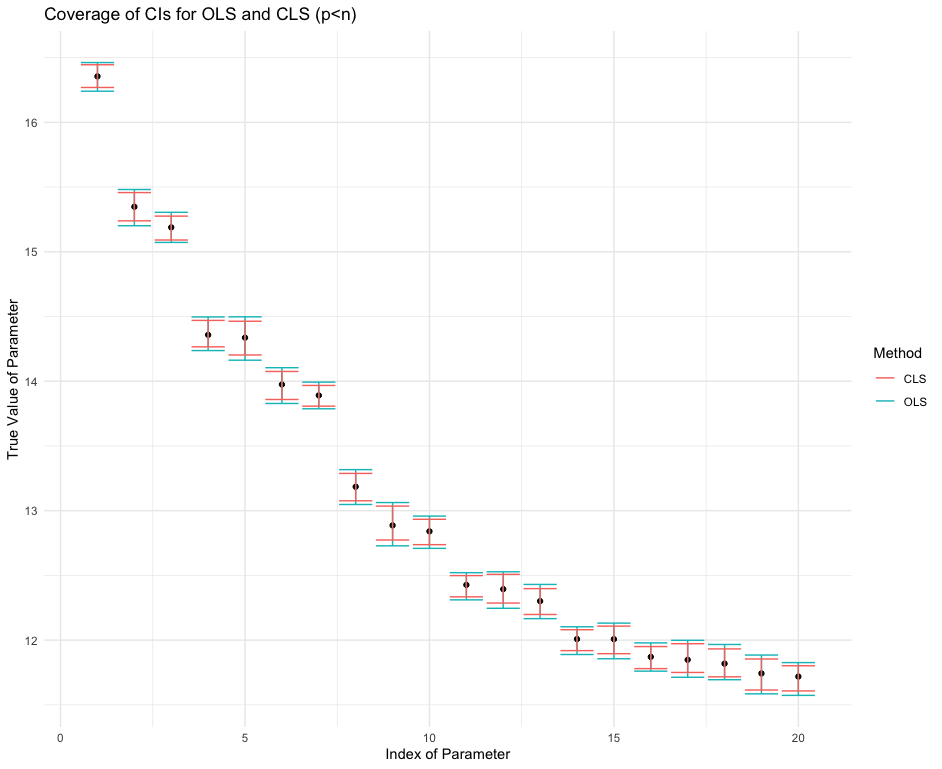}
    \caption{CIs at selected coordinates for $\hat{\beta}_{LS}$ and $\tilde{\beta}$ [\textbf{s2,m1}]}
    \label{fig:ci_cover_20}
\end{minipage}
\end{figure}

\begin{figure}
\centering
\begin{minipage}{.47\textwidth}
  \centering
  \includegraphics[width = \textwidth]{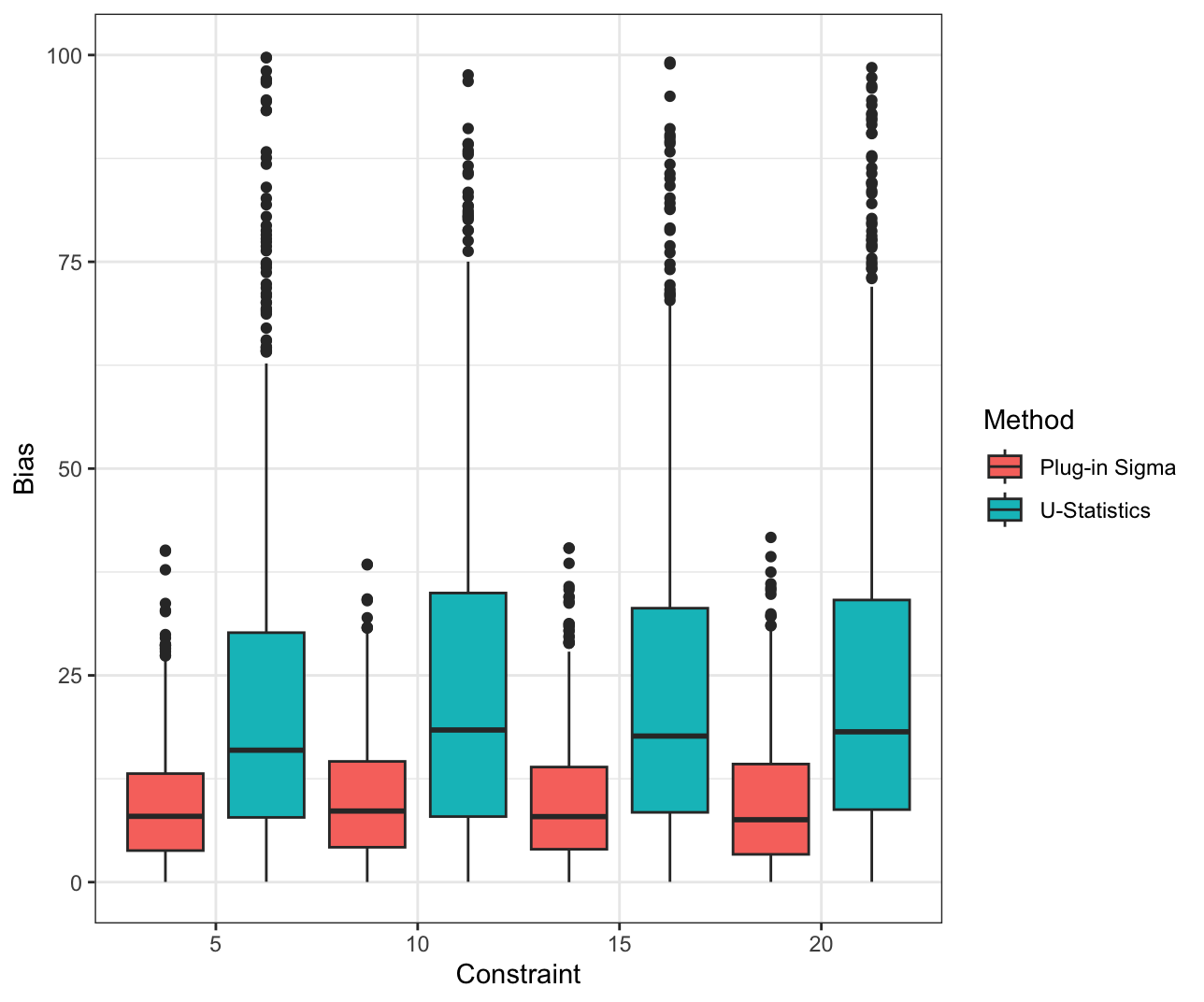}
  \caption{Comparison of absolute bias in one coordinate ($\hat{\beta}_{\Sigma,\mathcal{P},j}$ and  $\hat{\beta}_{est,j}$) with $n=10, N = 100, p = 20$ [\textbf{m2}]}
    \label{fig:bias_high_dim}
\end{minipage}
\hfill
\begin{minipage}{.47\textwidth}
\centering
  \includegraphics[width = \textwidth]{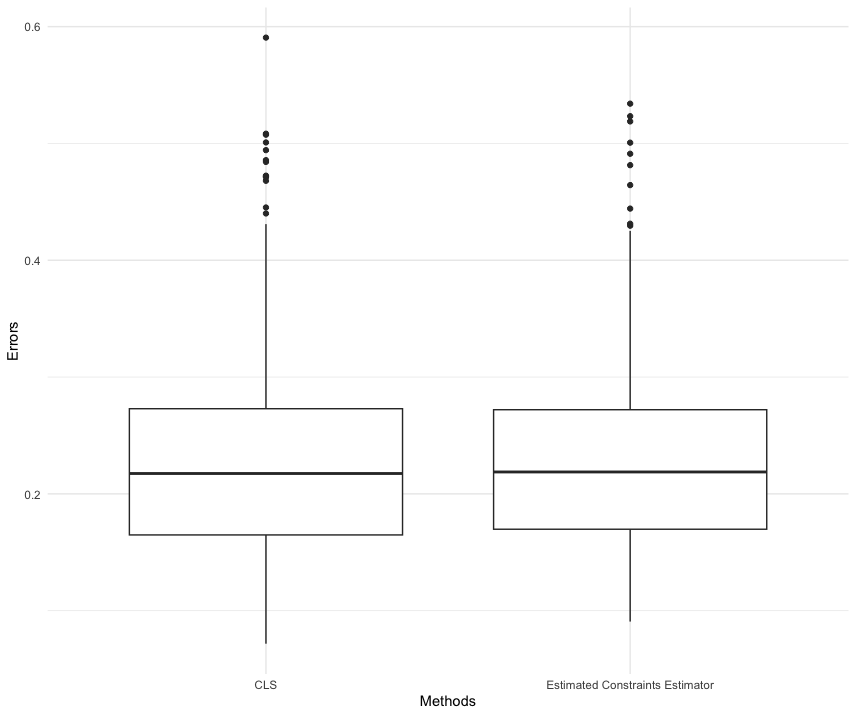}
    \caption{Comparison of errors between $\tilde{\boldsymbol{\beta}}$ and $\hat{\boldsymbol{\beta}}_{semi}$ for $q=5$ [\textbf{s2,m2}]}
    \label{fig:cls_estimated}
\end{minipage}
\end{figure}

\begin{figure}
  \centering
  \includegraphics[width = 0.6\textwidth]{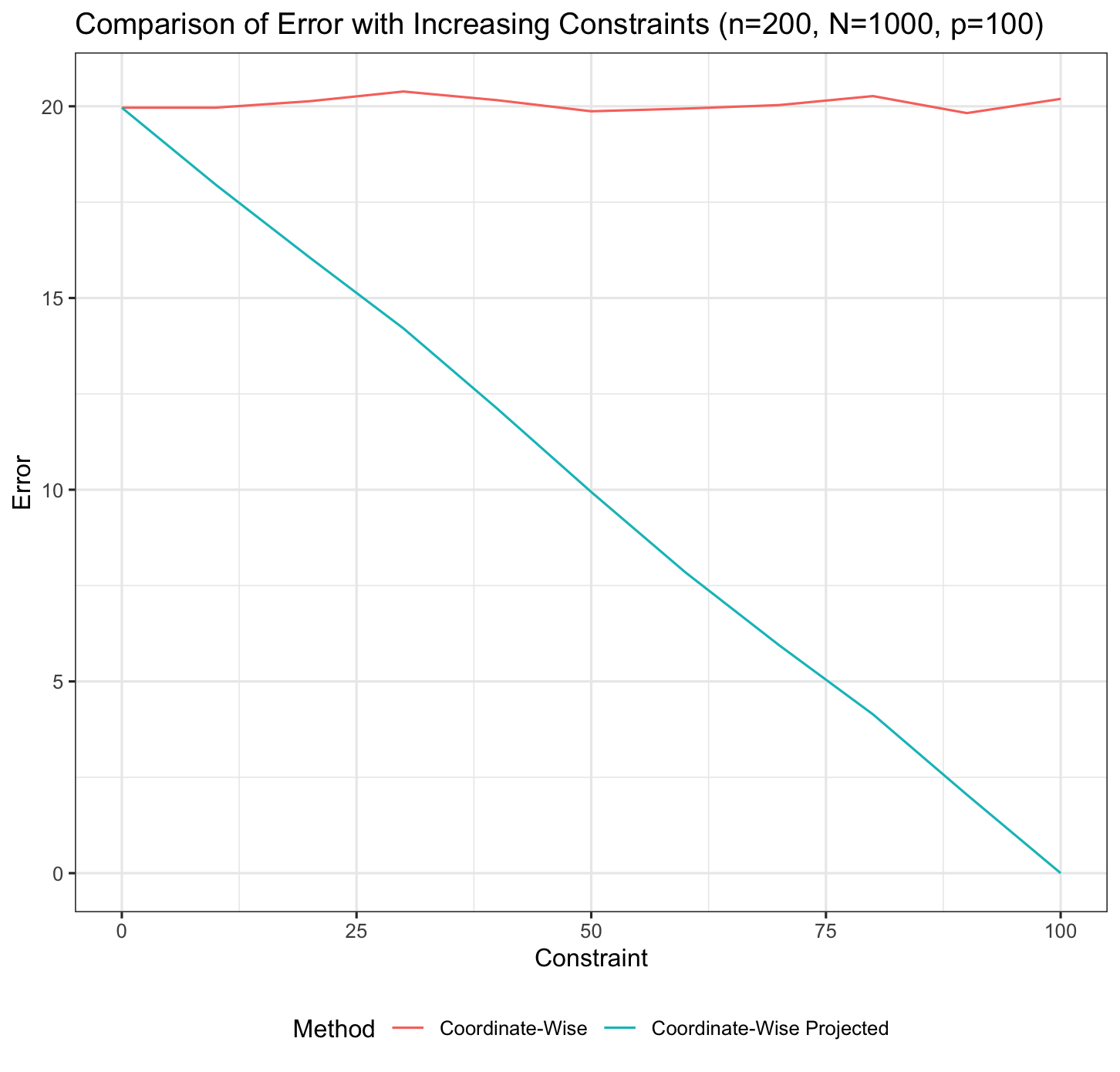}
  \caption{Error comparison in logistic regression ($\hat{\beta}_{\mathcal{P},GLM}$ and  $\hat{\beta}_{GLM}$) [\textbf{s2,m1}]}
    \label{fig:glm_errors}
\end{figure}
Figure \ref{fig:ci_cover} [\textbf{s2,m1}] shows the confidence intervals constructed from the OLS and CLS estimators around each coordinate of the true vector $\boldsymbol{\beta}^*$. Figure \ref{fig:ci_cover_20} zooms in on 20 of the coordinates, which shows that the confidence intervals constructed from the CLS estimators are thinner, due to the smaller asymptotic variance. Figure \ref{fig:bias_high_dim} [\textbf{m2}] shows the bias of 1000 iterations of the U-statistics based method-of-moments estimator $\hat{\boldsymbol{\beta}}_{est,j}$, along with the projected oracle estimator $\hat{\boldsymbol{\beta}}_{\Sigma,\mathcal{P},j}$, on a single coordinate of the regression vector. $\hat{\boldsymbol{\beta}}_{est,j}$ was computed using Chebyshev polynomials up to the order of $J=3$. There is a minimal improvement as the number of constraints increase, and a variance is quite large. Additionally, the computational time increases exponentially with the value of $J$. 

Figure \ref{fig:glm_errors} [\textbf{s2,m1}] compares the MSEs of the coordinate-wise MoM estimator and the projected estimator in a logistic regression setup over a range of constraint ratios.

Finally, in our data exploration, we employ the CLS estimator, presuming the constraints are known. This is because we have an exact formulation of the confidence intervals for this estimator. However, Fig. \ref{fig:cls_estimated} [\textbf{s2,m2}] shows that under a low level of constraint ($q=5$), we can implement the CLS estimator directly, as opposed to the semiparametric estimator, with a minimal loss of accuracy. This allows us to simply use summary level data from our reference population.

\subsection{Data-Informed Simulations}
We perform a series of simulations incorporating the partial data from the JHS and MESA datasets.
We divide our analysis into the following cases based on the glycemic traits of interest; (\textbf{g1}) BMI, (\textbf{g2}) Fasting Insulin, (\textbf{g3}) HbA1c; and the data source used; (\textbf{d1}) $A,c$ from JHS ($p,q$ fixed, varying sample size), (\textbf{d2}) $X,y$ from MESA ($n,p$ fixed, varying constraint ratio). 

\begin{figure}
\centering
\begin{minipage}{.45\textwidth}
  \centering
    \includegraphics[width = \textwidth, height = 22em]{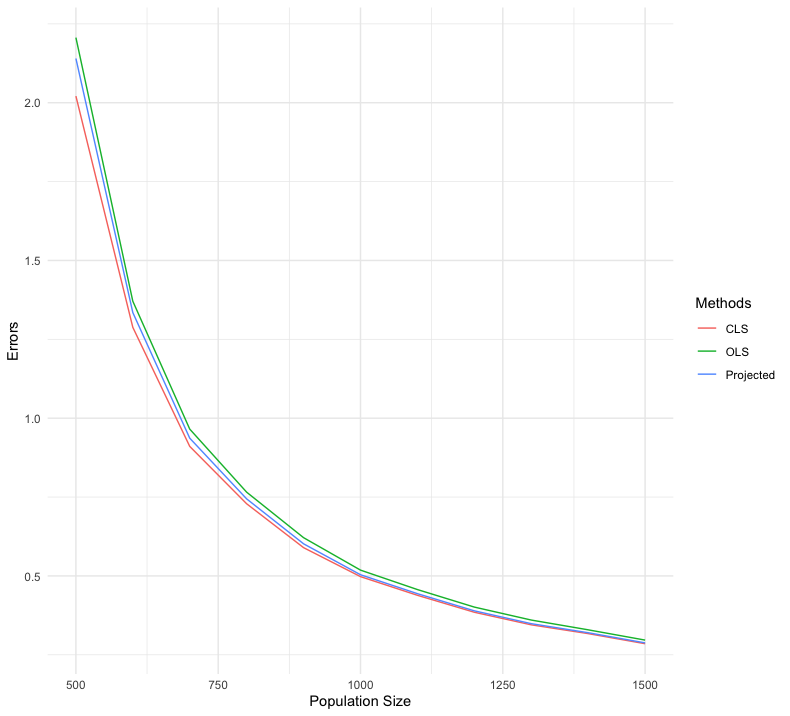}
    \caption{Comparison of MSEs between $\hat{\beta}_{LS},\hat{\beta}_{\mathcal{P}}$ and $\tilde{\beta}$ [\textbf{g1,d1}]}
\label{fig:err_ss}
\end{minipage}%
\hfill
\begin{minipage}{.45\textwidth}
  \centering
  \includegraphics[width = \textwidth, height = 22em]{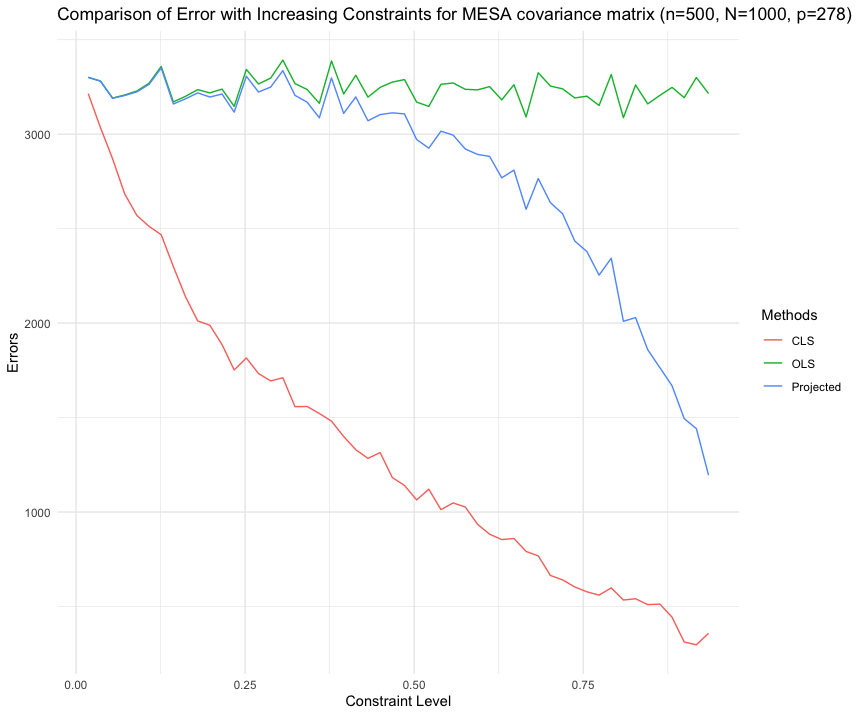}
    \caption{Error comparison ($\hat{\beta}_{LS},\hat{\beta}_\mathcal{P},\tilde{\beta}$) [\textbf{g1,d2}]}
    \label{fig:error_comparison_mesa}
\end{minipage}
\end{figure}

Akin to Figures 2 and 4, Figure \ref{fig:error_comparison_mesa} [\textbf{g1,d2}] compares the MSEs of the OLS, projected and CLS estimators over a range of constraint ratios. Additional simulations further demonstrate the theoretical properties of our proposed estimators. 
We also look at the error comparison under varying sample size. Figure \ref{fig:err_ss} shows the MSE over varying sample sizes. While the CLS estimator does perform better than the other two estimators, the gap is not large. This is indicative that the signal we are recovering from the JHS dataset is not very strong. 
\begin{figure}
\centering
\begin{minipage}{.45\textwidth}
  \centering
    \includegraphics[width = \textwidth, height = 15em]{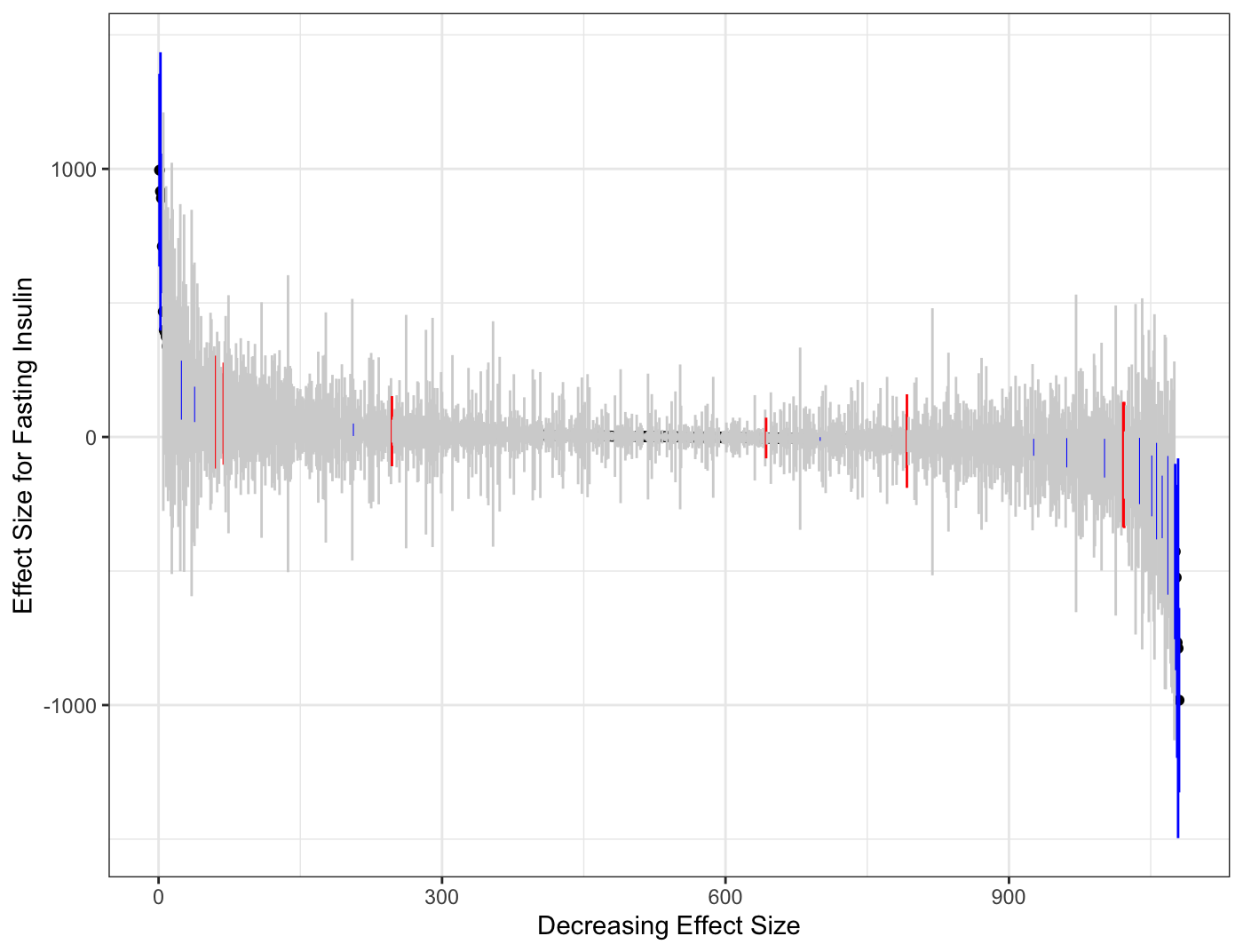}
    \caption{Estimated coefficients ordered by effect size and 95\% CIs of genetic variants in association with fasting insulin, using $\tilde{\boldsymbol{\beta}}$. Blue and red CIs are significant before and after controlling for the family-wise error rate}
    \label{fig:cis_fi}
\end{minipage}%
\hfill
\begin{minipage}{.45\textwidth}
  \centering
  \includegraphics[width = \textwidth, height = 15em]{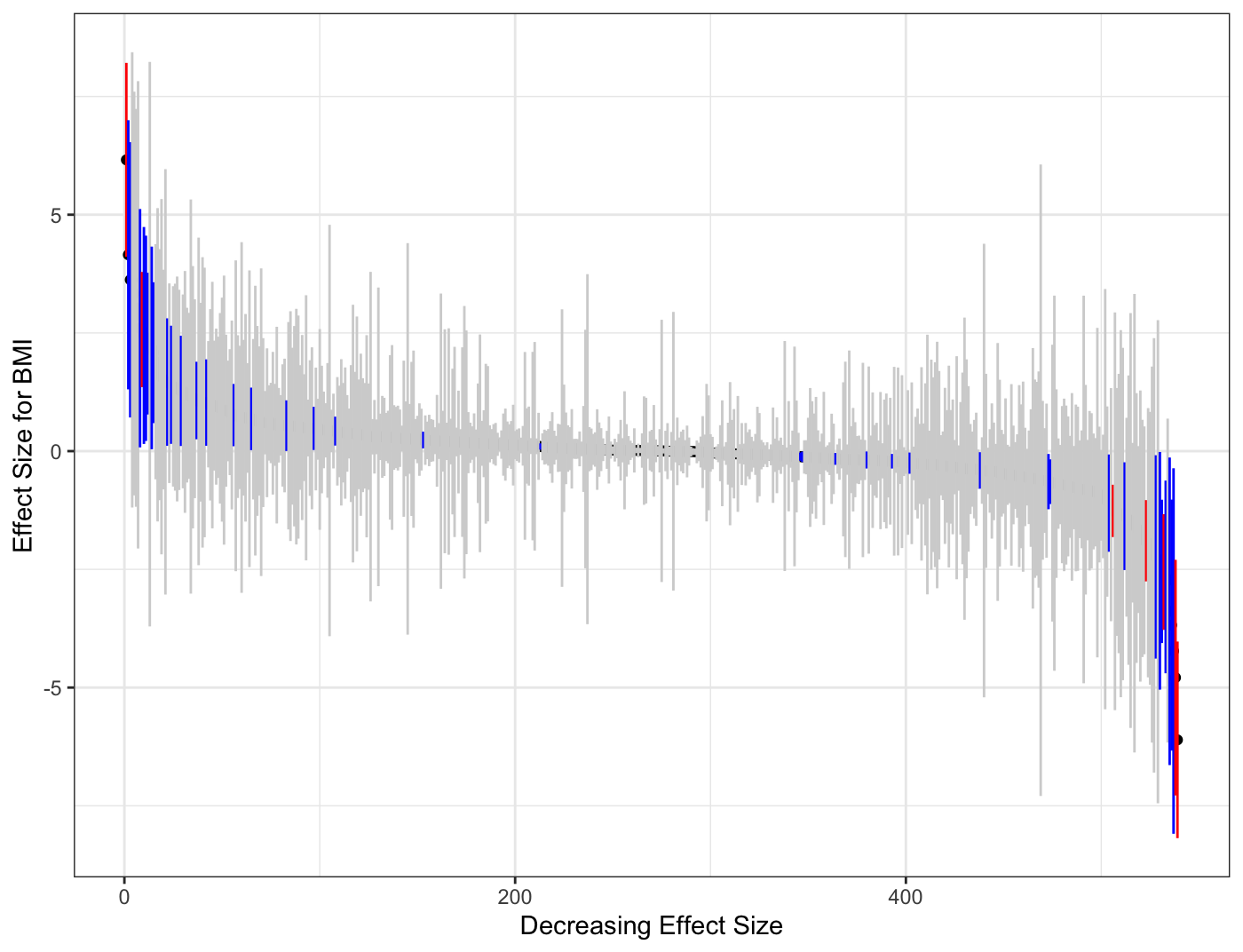}
    \caption{Estimated coefficients ordered by effect size and 95\% CIs of genetic variants in association with HbA1c, using $\tilde{\boldsymbol{\beta}}$. Blue and red CIs are significant before and after controlling for the family-wise error rate}
    \label{fig:cis_hb}
\end{minipage}
\end{figure}
\begin{figure}
\centering
\includegraphics[width = \textwidth, height = 15em]{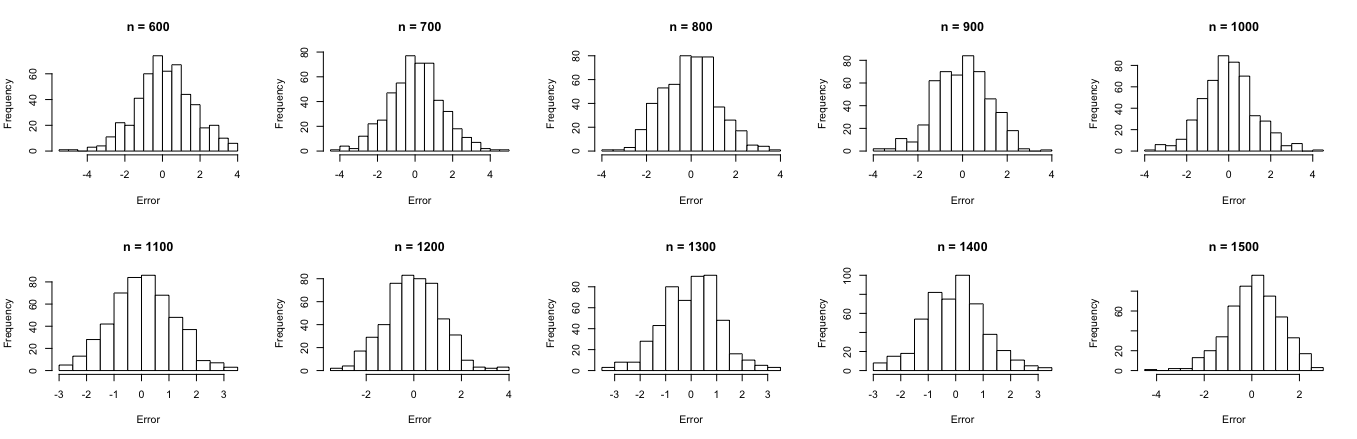}
    \caption{Normality of coordinate over varying sample sizes [\textbf{g1,d1}]}
\label{fig:normal_jhs}
\end{figure}
\begin{figure}
\centering
  \includegraphics[width = 0.6\textwidth, height = 18em]{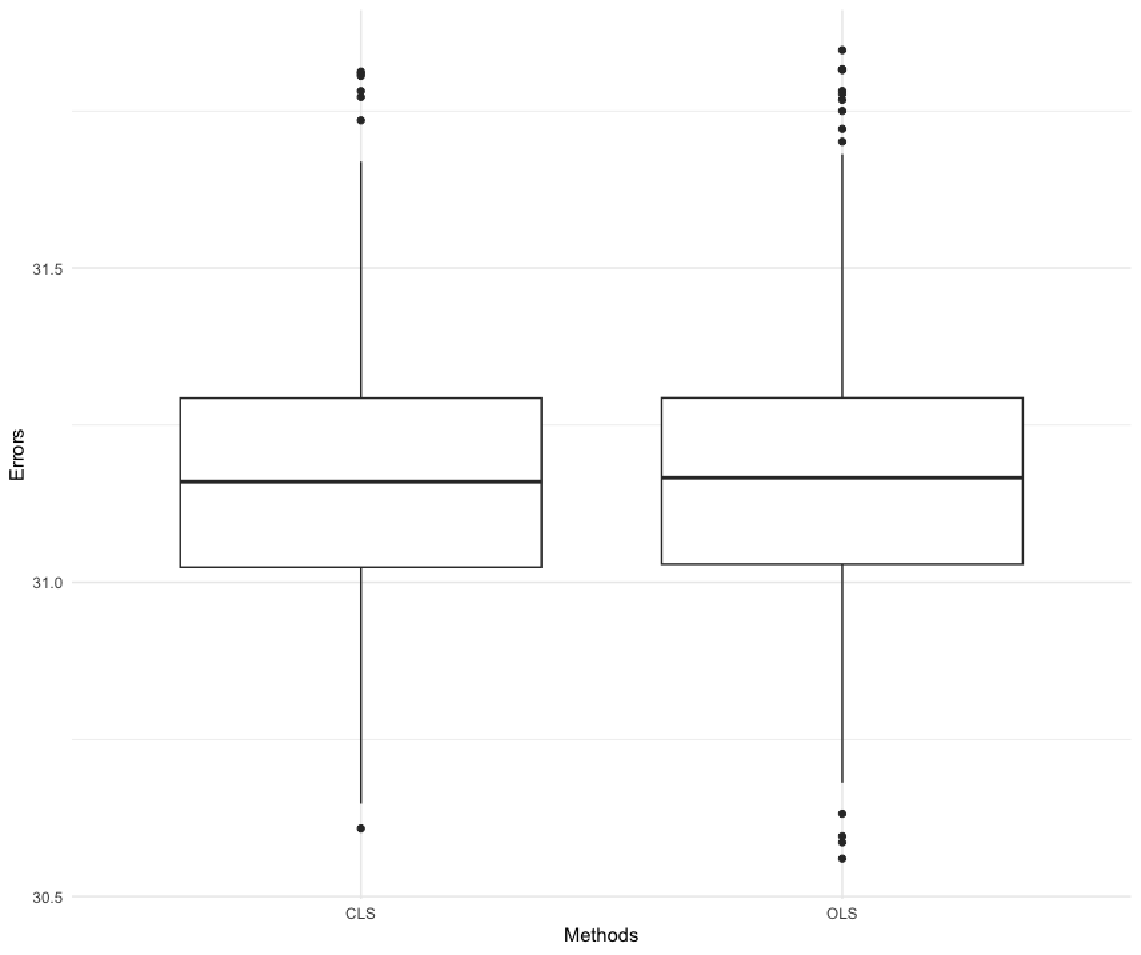}
    \caption{Comparison of cross-validation error between $\hat{\beta}_{LS}$ and $\tilde{\beta}$}
    \label{fig:cv_error}
\end{figure}

Figure \ref{fig:normal_jhs} [\textbf{g1,d1}] shows the histogram of the first estimated coordinate over all the iterations, with the appropriate normalization, as we increase the sample size. The coordinate appears to follow a normal distribution even in the small sample case, which gives credence to the usage of asymptotic confidence intervals in the small sample case. 

\subsection{Data Analysis}
For the data exploration example, we apply a similar procedure for fasting insulin and HbA1c. We tabulated the genetic variants recovered for each of the glycemic traits in Table \ref{tab:sig_snps}. Figures \ref{fig:cis_fi} and \ref{fig:cis_hb} shows the coordinates of the CLS vector, along with the confidence intervals constructed using the naïve variance estimates, for fasting insulin and HbA1c, respectively. For fasting insulin, out of $p=1080$ genetic variants, we reject the null hypothesis for 5 genetic variants (rs2006232, rs6740281, rs7210719, rs2277668, rs3218911), and for HbA1c, out of $p=539$ genetic variants, we reject the null hypothesis for 6 genetic variants (rs11240346, rs3901740, rs3767283, rs7137327, rs77920745, rs80228806). 

For fasting insulin, the genetic variants belong to SARM1, MAP4K4, TMEM199, SEBOX and IL1R2 genes, all genes responsible for proteins for cellular function, which could have a potential downstream effect on pancreatic tissue. For HbA1c, the genetic variants belong to CNTN2, CHPT1 and PC genes, which are all responsible for proteins related to blood cell function. This could also have a potential downstream effect on the level of glycation of the hemoglobin.
\begin{table}[]
    \centering
    \begin{tabular}{|c|c|c|}
    \hline
    BMI $(p=278)$ & Fasting Insulin $(p=1080)$& HbA1c $(p=539)$\\
    \hline
      rs5510   & rs2006232 & rs11240346\\
         & rs6740281& rs3901740\\
              &  rs7210719& rs3767283\\
         & rs2277668& rs7137327\\
              &  rs3218911& rs77920745\\
         & & rs80228806\\
         \hline
    \end{tabular}
    \caption{Significant genetic variants for each glycemic traits}
    \label{tab:sig_snps}
\end{table}

In addition to our data exploration, we perform a cross validation analysis to test the efficacy of our estimator in prediction. We split the dataset into 5 subpopulations and perform 5-fold cross validation to get an estimate of estimation error on a real-life dataset. Figure \ref{fig:cv_error} shows the mean-squared errors recovered over the iterations. The difference between the errors is negligible, most probably due to the strength of the signal from the JHS dataset. This concludes the extent of our data analysis.
\end{document}